\documentstyle[epsf,times]{mn2e}
\newcommand{\simgt}{\lower.5ex\hbox{$\; \buildrel > \over \sim \;$}}
\newcommand{\simlt}{\lower.5ex\hbox{$\; \buildrel < \over \sim \;$}}
\newcommand{\bm}[1]{\mbox{{\it \boldmath$#1$}}}

\newcommand{\skaco}[1]{\langle{#1}\rangle}

\newcommand{\apj}{ApJ}
\newcommand{\prd}{Phys.~ Rev.~ D.}
\newcommand{\mnras}{MNRAS}
\newcommand{\LCDM}{$\Lambda$CDM~}

\newcommand{\rvir}{r_{\rm vir}}

\newcommand{\baredth}{\;\overline{\raise1.0pt\hbox{$'$}\hskip-6pt
\partial}\;}
\newcommand{\edth}{\;\raise1.0pt\hbox{$'$}\hskip-6pt\partial\;}
\begin{document}
\title[Three-Point Correlations in Lensing]
{Three-Point Correlations in Weak Lensing Surveys: 
Model Predictions and Applications}

\author[M. Takada \& B. Jain]
{Masahiro Takada \thanks{E-mail: mtakada@hep.upenn.edu}
and Bhuvnesh Jain \thanks{E-mail: bjain@physics.upenn.edu} \\
 Department of Physics
and Astronomy, University of Pennsylvania, 209 S. 33rd Street,
Philadelphia, PA 19104, USA
}

\onecolumn
\pagerange{\pageref{firstpage}--\pageref{lastpage}}

\maketitle

\label{firstpage}
\begin{abstract}

We use the halo model of clustering to compute two- and three-point 
correlation functions for weak lensing, and apply them in a new 
statistical technique to measure properties of massive halos. 
We present analytical results on the eight shear three-point 
correlation functions constructed using combination of the two shear 
components at each vertex of a triangle.  We compare the amplitude and 
configuration dependence of the functions with ray-tracing simulations 
and find excellent agreement for different scales and models. These 
results are promising, since shear statistics are easier
to measure than the convergence.  In addition, the symmetry
properties of the shear three-point functions provide a new and precise
way of disentangling the lensing $E$-mode from 
the $B$-mode due to possible systematic errors. 

We develop an approach based on correlation functions 
to measure the properties of galaxy-group and cluster
halos from lensing surveys. Shear correlations
on small scales arise from the lensing matter within halos
of mass $M \simgt 10^{13}M_\odot$. Thus, the measurement of  two- and 
three-point correlations can be used to extract information on halo density 
profiles, primarily the inner slope and halo concentration. 
We demonstrate the feasibility of such an analysis for forthcoming 
surveys. We include covariances in the correlation functions due to 
sample variance and intrinsic ellipticity noise to show that 10\% 
accuracy on profile
parameters is achievable with surveys like the CFHT Legacy
survey, and significantly better with future surveys. Our statistical 
approach is complementary to the standard approach of identifying individual 
objects in survey data and measuring their properties. 
It can be extended to perform a combined analysis of the large-scale, 
perturbative regime down to small, sub-arcminute scales, to obtain 
consistent measurements of cosmological parameters and halo properties. 
\end{abstract}
%%%%%%%%%%%%%%%%%%%%%%%%%%%%%%%%%%%%%%%%%%%%%%%%%%%%%%%%%%%%%%%%%%%
\begin{keywords}
 cosmology: theory --- gravitational lensing --- 
large-scale structure of universe
\end{keywords}
%%%%%%%%%%%%%%%%%%%%%%%%%%%%%%%%%%%%%%%%%%%%%%%%%%%%%%%%%%%%%%%%%%%

\section{Introduction}

The gravitational lensing of distant galaxy images due to large-scale
structure, known as cosmic shear, has been well established as a cosmological probe (e.g,
Mellier 1999, Bartelmann \& Schneider 2001 and Wittman 2002 for reviews).
Many independent groups have reported significant
detections of two-point shear correlations, 
providing constraints on the mass density of
the universe ($\Omega_{\rm m0}$) and the mass power spectrum amplitude
($\sigma_8$) (e.g., Van Waerbeke et al. 2001b; Bacon et al. 2002;
Refregier, Rhodes \& Groth 2002; Hoekstra et al. 2002; Brown et
al. 2002; Hamana et al. 2002; Jarvis et al. 2003).  Non-linear
gravitational clustering substantially enhances the cosmic shear signal
on angular scales $\simlt 10'$ (Jain \& Seljak
1997). An accurate description of non-linear effects on shear correlations
is important to interpret the measurements.  

The most effective way of studying non-linear structure formation has been
$N$-body simulations, which are accurate on scales
larger than the numerical resolution limit, since the relevant physics is only
gravity on scales of interest. However, current survey data require
accurate models of large-scale structure over a huge dynamic range of
length scales. A simulation needs to sample
cosmological scale $(\sim 100$Mpc) in order to have a fair sample.
On the other hand, lensing statistics at relevant angular
scales are affected by highly non-linear structures, 
dark matter halos with a size $\simlt $Mpc, and structure that needs
to be resolved down to scales approaching $ 0.01 $Mpc.  The required dynamic
range is prohibitive for current computational resources.  In addition,
to perform multiple evaluations in model parameter space requires thousands
of simulations runs, which is also prohibitive.  While some numerical short-cuts can
be used to get around these constraints,  having an accurate
analytic model would be a valuable complementary, and sometimes essential, tool. 
It could also provide physical insights into the complex non-linear
phenomena involved in gravitational clustering. 

The dark matter halo model of clustering fulfills such a niche. 
The method was first constructed in real space to understand
the correlation functions of the galaxy distribution in terms of halos
(Neyman \& Scott 1952; Peebles 1974; McClelland \& Silk 1977;
Scherrer \& Bertschinger 1991; Sheth \& Jain 1997; Yano \& Gouda 1999;
Ma \& Fry 2000a,b; Sheth et al. 2001; Berlind \& Weinberg 2002; Takada
\& Jain 2003b, hereafter TJ03b). The model was also
formulated in Fourier space, since this leads to
simpler expressions for the Fourier-transformed counterparts of the
higher-order moments (Seljak 2000; Peacock \& Smith 2000; Ma \& Fry
2000c; Scoccimarro et al. 2001; Cooray \& Hu 2001a,b; Takada \& Jain
2002 hereafter TJ02; also see Cooray \& Sheth 2002 for a review).  Given
the model ingredients: the halo density profile, mass
function and bias, each of which has been well investigated
in the literature, the halo model can be used to
compute the statistics of cosmic fields. There are several advantages in
using the halo model. First, it allows us to easily implement multiple
model predictions in parameter space. Second, the measured signals are
explicitly understood in terms of the halo model ingredients. This will
be relevant for the comparison with other observations such as the
$X$-ray or the Sunyaev-Zel'dovich (SZ) effect for clusters of galaxies.
Third, the model accuracy can be easily refined by incorporating results
from $N$-body simulations with higher resolution. Interestingly, so far
the halo model has led to consistent predictions to interpret
observational results of galaxy clustering as well as reproduce 
simulation results (e.g., Seljak 2000; Ma \& Fry 2000c; Scoccimarro et
al. 2001; TJ02; Guzik \& Seljak 2002; Sheth et 2001; Takada \& Jain
2003a hereafter TJ03a; TJ03b; Zehavi et al. 2003).

Recently, we have extended the  halo approach 
to analytically compute the real-space three-point correlation functions
(3PCF) of the mass, galaxies and  weak lensing
fields with reasonable computational expense (TJ03b).  The 3PCF is the
lowest-order statistical quantity to probe non-Gaussianity,
generated by non-linear structure formation from primordial Gaussian fluctuations.  
It also allows one to explore the shapes of clustered mass
distributions via its configuration dependence, which is not 
contained in  the two-point
correlation function (2PCF). Hence, measurements of the 3PCF can provide
additional cosmological information. 
Our recent work described above and this paper provide the methods and
formalism to compute the real space 3PCF.  On large scales, the real space
3PCF provides no advantage over the well studied bispectrum. On small scales,
however, it is easier to measure the 3PCF in real space 
because the bispectrum
requires taking the Fourier transform of survey data, which typically
involves dealing with the complex survey geometry (e.g. in lensing surveys
there are several areas that are masked out due to bright
stars, and thus the transform is likely noisy).  
In addition,
the different wavenumber modes are highly
correlated with each other on small scales, and thus one merit of 
working in Fourier space is lost (in contrast to the
CMB, where nonlinearties are minimal on scales of interest). Indeed 
so far most measurements of the 2PCF have also been in real-space for cosmic shear
(though Pen, Van Waerbeke \& Mellier 2002; Brown et al. 2002; Pen et
al. 2003 have estimated the shear power spectrum as well).

In lensing surveys, the 3PCF of the shear is easier to measure than
the 3PCF of the convergence field which requires a non-local 
reconstruction from the data (Bernardeau, Van Waerbeke \& Mellier 2003).  
An exciting recent development has been the detection of 
statistical measures based on the shear 3PCF from the
Virmos-Descart Survey 
(Bernardeau, Mellier \& Van Waerbeke 2002; Pen et al. 2003). 
Theoretical study of the shear 3PCF has just begun: 
Schneider \& Lombardi (2003, hereafter SL03) and
Zaldarriaga \& Scoccimarro (2003, hereafter ZS03) analyzed how to
construct the eight shear 3PCFs from combinations of the $+/\times$
shear components at each vertex of a given triangle. 
Using the ray-tracing simulations, in TJ03a we
verified that the eight shear 3PCFs display characteristic
configuration dependences, and pointed out that the characteristics can
be used to separate the lensing $E$-mode from the measured signals that
are generally contaminated by the $B$-mode due to systematic errors and 
other effects.

The purpose of this paper is to develop an accurate, analytic
model to predict the 3PCFs of the shear field, extending the
halo model for the convergence field (TJ03b). 
We will carefully examine the accuracy of the model 
predictions by comparing with ray-tracing simulations. In particular, we
will focus on whether or not the halo model can reproduce the complex
configuration dependences  seen in the
simulations, since the halo model relies on simplified
assumptions of spherical halos. 

We develop a new application of higher order shear correlations to measure
the properties of massive halos. These are relevant
for upcoming surveys such as the CFHT
Legacy Survey\footnote{\tt www.cfht.hawaii.edu/Science/CFHLS/} and the
Deep Lens Survey\footnote{\tt dls.bell-labs.com/} and proposed projects
such as DML/LSST\footnote{\tt www.dmtelescope.org/dark\_home.html},
Pan-STARRS\footnote{\tt www.ifa.hawaii.edu/pan-starrs/} and
SNAP\footnote{\tt snap.lbl.gov}.  These surveys
promise to measure $n$-point correlation functions of the cosmic shear
fields even on sub-arcminutes scales with high significance.
Therefore, with the
halo model, these small-scale signals can be used to probe properties of
the halo density profile such as its inner slope and concentration. 
We will work with a generalized
universal profile for halos (Navarro, Frenk \& White 1996; 1997, hereafter NFW). 
These properties remain uncertain theoretically as
well as observationally.  To implement this approach, 
we will develop a method to compute the covariance for the 2PCF and 3PCF
measurements, extending the method of Schneider et al. (2002b).  We will estimate
the accuracy with which forthcoming surveys can 
constrain halo profile parameters from combined measurements of the 2PCF and 3PCF.  

The plan of this paper is as follows. In \S \ref{model}, we develop the
real-space halo approach for computing the two- and three-point correlation
functions of the lensing convergence and shear fields.  In \S \ref{parity},
we summarize the triangle configuration
dependences of the shear 3PCFs. In \S \ref{results}, we compare the halo model
predictions with ray-tracing simulation results. Then, in \S \ref{profile}, we 
show how measurements of the 2PCF and 3PCF on sub-arcminute scales are
feasible for ongoing and future lensing surveys, taking into account 
statistical sources of error. We then address how these
small-scale measurements can be used to constrain halo
profile properties. \S \ref{disc} is devoted to a summary and
discussion. We will consider mainly two cosmological models. One is the
$\Lambda$CDM model with $\Omega_{m0}=0.3$, $\Omega_{\lambda0}=0.7$,
$\Omega_{\rm b0}$, $h=0.7$ and $\sigma_8=0.9$.  The other is the SCDM
model with $\Omega_{m0}=1.0$, $h=0.5$ and $\sigma_8=0.6$.  Here
$\Omega_{\rm m0}$, $\Omega_{b0}$ and $\Omega_{\lambda0}$ are the
present-day density parameters of matter, baryons and the cosmological
constant, $h$ is the Hubble parameter, and $\sigma_8$ is the rms mass
fluctuation in a sphere of radius $8h^{-1}$Mpc.

\section{Formalism:
Real-space Halo Approach to Cosmic shear statistics}
\label{model}

In this section we develop an analytic method for calculating the 3PCFs
of shear fields by extending the real-space halo approach developed in TJ03b.

\subsection{Convergence and shear fields}
\label{lens}

Weak gravitational lensing can be separated into two effects:
magnification (described by the convergence) and shear (e.g., Bartelmann \& Schneider 2001).

The lensing convergence field is a scalar quantity and simply expressed
as a weighted projection of the density fluctuation field between source
galaxy and observer:
\begin{equation}
\kappa(\bm{\theta})=\frac{1}{2}\nabla^2\Psi(\bm{\theta})
=\int_0^{\chi_H}\!\!d\chi W(\chi) 
\delta[\chi, d_A(\chi)\bm{\theta}],
\label{eqn:kappa}
\end{equation}
where we have introduced the two-dimensional lensing potential $\Psi$,
the Laplacian operator $\nabla^2$ defined as $\nabla^2\equiv
\partial^2/\partial \theta_i\partial\theta_i$, $\chi$ is the comoving
distance, and $\chi_H$ is the distance to the horizon.  Note that $\chi$
is related to redshift $z$ via the relation $d\chi=dz/H(z)$, where $H(z)$ is
the Hubble parameter at epoch $z$.  Following the early work of
Blandford et al. (1991), Miralda-Escude (1991) and Kaiser (1992),
we used two key simplifications to derive the equation above; the
flat-sky approximation, which is valid on angular scales of
interest, and the Born approximation, where the convergence field is
computed along the unperturbed path. Using ray-tracing simulations,
Jain et al. (2000) showed that the Born approximation is an excellent
approximation for the two-point statistics.  We
will assume that it also holds for the higher-order statistics we are
interested in.  The function $W$ is the lensing projection defined by
\begin{equation}
W(\chi)=\frac{3}{2}\Omega_{m0}H_0^2 a^{-1}(\chi)
d_A(\chi)
\int^{\chi_H}_0\!\!d\chi_s~ 
n_s(\chi_s) \frac{d_A(\chi_{\rm s}-\chi)}{d_A(\chi_s)},
\label{eqn:weightgl}
\end{equation}
where $n_s(\chi_s)$ is the redshift selection function of source galaxies.  
Here $H_0$ is the Hubble constant ($H_0=100h{~\rm km~s}^{-1}{\rm Mpc}^{-1}$)
and $d_A(\chi)$ is the comoving angular diameter distance. 
In this paper we assume all source galaxies are at a single redshift
$z_s$ for simplicity; $n_s(\chi)=\delta_D(\chi-\chi_s)$.  Note that
$d_A=\chi$ for a flat universe. 

A more direct observable of weak lensing is the shearing of
images of source galaxies. Since this effect
is of order 1\% for large-scale structures lensing in a CDM model,
it is measurable only in a statistical sense.  
The shear field is described by the two components,
$\gamma_1$ and $\gamma_2$, which correspond to elongation or
compressions along the x-axis, or at $45^\circ$ to it, respectively 
(given Cartesian coordinates on the sky).  The shear field is expressed
in terms of the lensing potential as
\begin{equation}
\gamma_{1}=\frac{1}{2}(\Psi_{,11}-\Psi_{,22}),\hspace{1em}
\gamma_2=\Psi_{,12},
\end{equation}
where $\Psi_{,ij}\equiv\partial^2 \Psi/{\partial \theta_i}{\partial
\theta_j}$.  In Fourier space, these fields are simply related to
the convergence field via the relation
\begin{equation}
 \tilde{\gamma}_1(\bm{l})=\tilde{\kappa}(\bm{l})\cos(2\varphi_{l}),
\hspace{1em}
\tilde{\gamma}_2(\bm{l})=\tilde{\kappa}(\bm{l})\sin(2\varphi_{l}), 
\label{eqn:shear}
\end{equation}
where $\bm{l}=l(\cos\varphi_{l},\sin\varphi_{l})$ and, in the following,
quantities with tilde symbol denote Fourier components. 
Equation (\ref{eqn:shear}) shows that $\gamma_i$ behaves
like a spin-2 field:  if at a given point one rotates the coordinate
system by an angle $\alpha$ in the anti-clockwise direction, the shear
fields are transformed as
\begin{eqnarray}
\gamma'_1&=&\cos2\alpha~ \gamma_1+\sin2\alpha~ \gamma_2, \nonumber\\
\gamma'_2&=&-\sin2\alpha~ \gamma_1+\cos2\alpha~ \gamma_2. 
\label{eqn:spattern}
\end{eqnarray}
We will  often use the vector notation
$\bm{\gamma}=\gamma_1+i\gamma_2$.  A general two-dimensional spin-2
field can be decomposed into an $E$-mode derivable from a scalar
potential and a pseudo-scalar $B$-mode (Kamionkowski, Kosowski \&
Stebbins 1997; Zaldarriaga \& Seljak 1997; Hu \& White 1997 for the CMB
polarization and Stebbins 1996; Kamionkowski et al. 1998; Crittenden et
al. 2002; Schneider et al. 2002a for the cosmic shear). Gravitational
lensing induces a pure $E$-mode in the weak lensing regime, while source
galaxy clustering, intrinsic alignments and observational systematics
induce both $E$ and $B$-modes in general (Crittenden et al. 2002;
Schneider et al. 2002a). The clean separation of $E/B$ modes
from survey data is of great importance in extracting cosmological
information. 

\subsection{Halo mass function and halo bias}

To develop the halo model, we begin by describing models of the mass
function and the halo bias that are used in this paper.

Following TJ02 and TJ03b, we employ an analytical fitting formula proposed by
Sheth \& Tormen (1999), which is more accurate than the original
Press-Schechter mass function \cite{PS74}.  The number density of halos with
mass in the range between $M$ and $M+dM$ is given by
\begin{eqnarray}
n(M)dM&=&\frac{\bar{\rho}_0}{M}f(\nu)d\nu\nonumber\\
&=&\frac{\bar{\rho}_0}{M}A[1+(a\nu)^{-p}]\sqrt{a\nu}\exp\left(-\frac{a\nu}{2}
\right)\frac{d\nu}{\nu},
\label{eqn:massfun}
\end{eqnarray}
where $\nu$ is the peak height defined by
\begin{equation}
\nu=\left[\frac{\delta_c(z)}{D(z)\sigma(M)}\right]^2.
\end{equation}
Here $\sigma(M)$ is the present-day rms fluctuation in the mass
density, smoothed with a top-hat filter of radius $R_M\equiv
(3M/4\pi\bar{\rho}_0)^{1/3}$, $\delta_c$ is the threshold overdensity
for the spherical collapse model (see Nakamura \& Suto 1997 and Henry 2000
for a useful fitting function) and $D(z)$ is the growth factor (e.g.,
Peebles 1980).  The numerical coefficients $a$ and $p$ are
empirically fitted from $N$-body simulations
as $a=0.75$ and $p=0.3$. Note that the value of $a$ is modified from the
$a=0.707$ in Sheth \& Tormen (1999) to better fit
the mass function at cluster mass scales in the Hubble volume simulations (R. Sheth; 
private communication).  The coefficient $A$ is set by the normalization
condition $\int_0^\infty\!d\nu f(\nu)=1$, leading to $A = 0.129$. 
This condition reflects the assumption that all the matter
 is in halos.  Note that the
peak hight $\nu$ is specified as a function of $M$ at all redshifts once
the cosmological model is fixed.

Recently, various authors (e.g., Jenkins et al. 2001; White 2002; Hu \&
Kravtsov 2003) have addressed the non-trivial problem of how the mass
function seen in the simulations depends on the halo identification scheme and
the mass estimator.  There is no clear
boundary between a halo and the surrounding large-scale structure, 
therefore the halo mass does depend on the algorithm used 
(e.g., the friend-of-friend method or the spherical
overdensity method).  Jenkins et al. (2001) showed that
if one employs the halo mass estimator, $M_{180}$, enclosed within
a sphere of radius $r_{180}$ (interior to which the mean density is $180$
times the background density), the mass function measured from the
simulation can be well fitted by the universal form of equation (\ref{eqn:massfun}).
This conclusion was also verified by White (2002; also see Hu \&
Kravtsov 2003).  Despite these facts, in this paper we employ the virial
halo mass to describe the mass function for simplicity, since it is
based on the more physically-motivated spherical collapse
model that can be applied to any cosmology, irrespective of the
halo density profile:
\begin{equation}
M=\frac{4\pi}{3}\bar{\rho}_0 \Delta_v(z)\rvir^3,
\label{eqn:mv2}
\end{equation}
where $\bar{\rho}_0$ is the present-day mean density of matter, $\rvir$
is the virial radius, and $\Delta_v(z)$ is the overdensity of collapse
given as a function of redshift (e.g., see Nakamura \& Suto 1997 and
Henry 2000 for a useful fitting formula).  One justification of our
treatment is the result of Figure 5 in White (2002), which showed that
the form of equation (\ref{eqn:massfun}) fits the 
simulations when the virial mass estimator is employed
(although the agreement is not as good as the case for $M_{180}$).
The difference in the halo model
predictions for lensing statistics due to the two definitions
is small, as will be shown in Figure \ref{fig:2ptbound}.  
Finally, it is worth pointing out the advantage of the mass function
of equation (\ref{eqn:massfun}) over that of Jenkins et al. (2001): it is well
behaved over the full range of mass and satisfies the normalization
condition $\int_0^\infty d\nu f(\nu)=1$, while the mass function of
Jenkins et al. (2001) cannot be safely extrapolated outside of the range
of their fit and does not satisfy the normalization condition. 

Mo \& White (1996) developed a useful formula to describe the bias
relation between the halo distribution and the underlying mass
distribution.  This idea has been improved by several authors using
$N$-body simulations \cite{Mo97,Jingbias98,Sheth99,ST99}; we will use
the fitting formula of Sheth \& Tormen (1999) for consistency with the
mass function (\ref{eqn:massfun}):
\begin{equation}
b(\nu)=1+\frac{a\nu-1}{\delta_c}+\frac{2p}{\delta_c(1+(a\nu)^p)},
\label{eqn:bias}
\end{equation}
where we have assumed scale-independent bias and neglected the higher
order bias functions $(b_2, b_3,\cdots)$.  This bias model is used for
calculations of the 2-halo term in the 2PCF 
and the 2- and 3-halo terms of the 3PCF. It is not
important at the small, non-linear scales
where the 1-halo term arising from correlations within a single halo provides 
the dominant contribution.

\subsection{Convergence and shear profiles for an NFW halo}

In this subsection, we derive useful, analytical expressions for the
convergence and shear profiles around an NFW halo, which is the most
essential ingredient for small scales.

We consider halo density profiles given by the form
\begin{equation}
\rho_h(r)=\frac{\rho_s}{(cr/\rvir)^\alpha (1+cr/\rvir)^{3-\alpha}},
\label{eqn:nfw}
\end{equation}
where $\rho_s$ is the central density parameter and $c$ is the
concentration parameter. It is used instead of the scale radius $r_s=\rvir/ c$
which is the transition radius between $\rho_h\propto r^{-\alpha}$
and $r^{-3}$.  For most of the paper, we will use the NFW profile
with $\alpha=1$. However, since simulations with higher spatial
resolution have indicated $\alpha\approx -1.5$
\cite{Fukushige97,Moore98,JS00,Ghigna00}, we will also consider the effect of
variations in $\alpha$ for lensing statistics.  Given the halo profile, the parameter 
$\rho_s$ can be eliminated from the definition of the virial mass:
\begin{equation}
M=\int_0^{\rvir}\!\!4\pi r^2dr \rho_h(r)
=\frac{4\pi \rho_s \rvir^3}{c^3}
\left\{
\begin{array}{ll}
\displaystyle f^{-1}, & (\mbox{NFW}; \alpha=1),\\
\displaystyle \frac{c^{3-\alpha}}{3-\alpha}
~ {}_2F_1(3-\alpha,3-\alpha,4-\alpha,-c), & (\mbox{otherwise}), 
\end{array}
\right.
\label{eqn:mv}
\end{equation}
where $f = 1/[\ln(1+c)-c/(1+c)]$ and ${}_2F_1$ denotes the
hypergeometric function.  Note that equation (\ref{eqn:mv2}) gives the
virial radius is given in terms of the halo mass $M$ and redshift $z$.

To express the halo profile in terms of $M$ and $z$, we further need the
dependence of the concentration parameter $c$ on $M$ and $z$;
however, this still remains somewhat uncertain.  Following TJ02 and TJ03b 
we use
\begin{equation}
c(M,z)=c_0(1+z)^{-1}\left(\frac{M}{M_\ast(z=0)}\right)^{-\beta},
\label{eqn:conc}
\end{equation}
where $M_\ast(z=0)$ is the nonlinear mass scale at present defined by
$\delta_c(z=0)=\sigma(M_\ast)$.  The redshift dependence $(1+z)^{-1}$
and our fiducial choice of $(c_0,\beta)=(9,0.13)$ are based on the
numerical simulation results in Bullock et al. (2001).

For an NFW profile stated above, we can derive an analytical expression
for the the convergence field. 
As we discussed in TJ03b, the halo profile is taken to be 
truncated at the virial radius in order to
maintain mass conservation given by the normalization  of the mass function. 
Hence, the convergence field for a halo of mass $M$ can be defined as
\begin{equation}
\kappa_M(\theta)=4\pi Ga^{-1}\frac{d_A(\chi)d_A(\chi_s-\chi)}{d_A(\chi_s)}
\Sigma_M(\theta)
\label{eqn:convm}
\end{equation}
where $\Sigma_M$ is the projected density field defined by:
\begin{equation}
\Sigma_M(\theta)\equiv\int_{-r_{\rm vir}}^{r_{\rm vir}}
\!\!dr_{\parallel}\rho_M(r_\parallel, d_A\bm{\theta})
=\frac{M f c^2}{2\pi \rvir^2} F(c\theta/\theta_{\rm vir}).
\end{equation}
The explicit form for $F(x)$ is given by equation (27) in TJ03b.  

%%%%%%%%%%%%%%%%%%%%%%%%%%%%%%%%%%%%%%%%%%%%%%%%%%%%%%%%%%%%%%%%%%%%%%
\begin{figure}
  \begin{center}
    \leavevmode\epsfxsize=6.cm \epsfbox{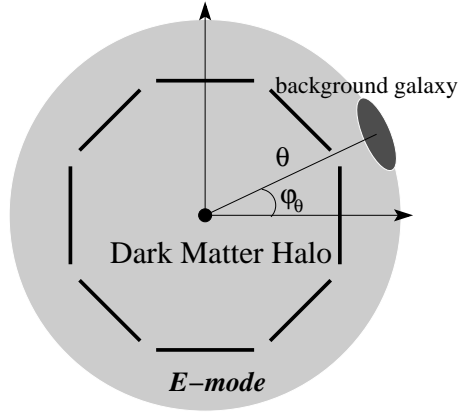}
  \end{center}
\caption{Illustration of the shear pattern around an axially-symmetric halo. 
 A background galaxy image is tangentially deformed with
 respect to the halo center, corresponding to 
 a pure $E$ mode in the shear field. } \label{fig:nfwillust}
\end{figure}
%%%%%%%%%%%%%%%%%%%%%%%%%%%%%%%%%%%%%%%%%%%%%%%%%%%%%%%%%%%%%%%%%%%%%%
For an axially-symmetric mass distribution, the shear amplitude can be
expressed in terms of the convergence field as a function of the radius
from the halo center (e.g., Miralda-Escude 1991):
\begin{equation}
\gamma_M(\theta)=\bar{\kappa}_M(\theta)-\kappa_M(\theta),
\end{equation}
where $\bar{\kappa}_M$ is the mean surface mass density inside a
circle of radius $\theta$: $\bar{\kappa}_M=(1/\pi
\theta^2)\int^\theta_0~ 2\pi \theta'd\theta'~ \kappa_M(\theta')$.  The
equation above reflects the non-local nature of the shear field, since
it is induced by non-local tidal forces.  The
shear field does not vanish outside the projected virial region because
$\bar{\kappa}_M$ is non-zero, even if $\kappa_M(\theta)=0$.  
From equation (\ref{eqn:convm}), the shear profile for an NFW
halo of mass $M$ can be analytically expressed as
\begin{equation}
\gamma_M(\theta)= 4\pi Ga^{-1}\frac{d_A(\chi)d_A(\chi_s-\chi)}{d_A(\chi_s)}
\frac{Mf c^2}{2\pi \rvir^2}G(c\theta/\theta_{\rm vir}),
\label{eqn:gammam}
\end{equation}
with 
\begin{eqnarray}
G(x)=
\left\{
\begin{array}{ll}
\displaystyle \frac{1}{x^2(1+c)}\left[
\frac{(2-x^2)\sqrt{c^{2}-x^2}}{1-x^2}-2c\right]+\frac{2}{x^2}
\ln\frac{x(1+c)}{c+\sqrt{c^{2}-x^2}}+\frac{2-3x^2}{x^2(1-x^2)^{3/2}}
{\rm arccosh}\frac{x^2+c}{x(1+c)},
& (x<1)\\
\displaystyle
\frac{1}{3(1+c)}\left[\frac{(11c+10)\sqrt{c^{2}-1}}{1+c}-6c
\right]+2\ln\frac{1+c}{c+\sqrt{c^{2}-1}},& (x=1)\\
\displaystyle \frac{1}{x^2(1+c)}\left[
\frac{(2-x^2)\sqrt{c^{2}-x^2}}{1-x^2}-2c\right]+\frac{2}{x^2}
\ln\frac{x(1+c)}{c+\sqrt{c^{2}-x^2}}
-\frac{2-3x^2}{x^2(x^2-1)^{3/2}}{\rm arccos}\frac{x^2+c}{x(1+c)},&
(1<x \le c)\\
\displaystyle \frac{2 f^{ -1}}{x^2},& (x>c).
\end{array}
\right.
%\nonumber \\
\end{eqnarray}
where $f^{}=1/[\ln(1+c)-c/(1+c)]$.  Note the asymptotic behavior
$G(x)\rightarrow 1/2$ for $x\rightarrow 0$, while the convergence 
$\kappa_M\rightarrow \infty$ in this limit (following from the NFW inner slope
$\rho_h\propto r^{-1}$).  This feature is in contrast to the case of a
power law density profile $\rho_h\propto r^{-n}$ ($1< n\le
3$), which leads to $\kappa_M,\gamma_M\propto \theta^{1-n}$ and the
asymptotic behavior $\kappa_M,\gamma_M\rightarrow \infty$ for
$\theta\rightarrow 0$.  
Outside the virial region the shear amplitude behaves like the field around a
point mass, $\gamma_M\propto 1/\theta^2$, due to the sharp cutoff of the
mass distribution.  Taking $c\rightarrow \infty$ in the equation
above leads to the expression for $\gamma_M$ in Bartelmann (1996; also
see Wright \& Brainerd 2000), which is derived from a
line-of-sight projection of the NFW profile under the assumption that
the profile is valid for an infinite range beyond the virial region.  As
shown in Figure \ref{fig:2ptbound}, if one adopts the expression
for $\gamma_M$ in Bartelmann (1996) the amplitude of shear correlations
is significantly overestimated. Therefore, the halo profile boundary
should be carefully considered to obtain accurate results as well as
a consistent formulation. 

The sketch in Figure \ref{fig:nfwillust} illustrates how a background
galaxy image is {\em tangentially} deformed around a foreground lens
with an axially-symmetric profile. In the weak lensing regime, the shear
pattern is described by a pure $E$-mode.  In  Cartesian
coordinates with the center taken as the halo center, the two
shear components can be expressed as
\begin{equation}
\gamma_{M,1}(\bm{\theta})=-\gamma_M(\theta)\cos2\varphi_{\bm{\theta}},
\hspace{2em}
\gamma_{M,2}(\bm{\theta})=-\gamma_M(\theta)\sin2\varphi_{\bm{\theta}}, 
\end{equation}
where $\varphi_{\bm{\theta}}$ is the angle between the first x-axis and
the line connecting the halo center and the source galaxy position (see
Figure \ref{fig:nfwillust}).

%%%%%%%%%%%%%%%%%%%%%%%%%%%%%%%%%%%%%%%%%%%%%%%%%%%%%%%%%%%%%%%%%%%%%%
\begin{figure}
  \begin{center}
    \leavevmode\epsfxsize=17.cm \epsfbox{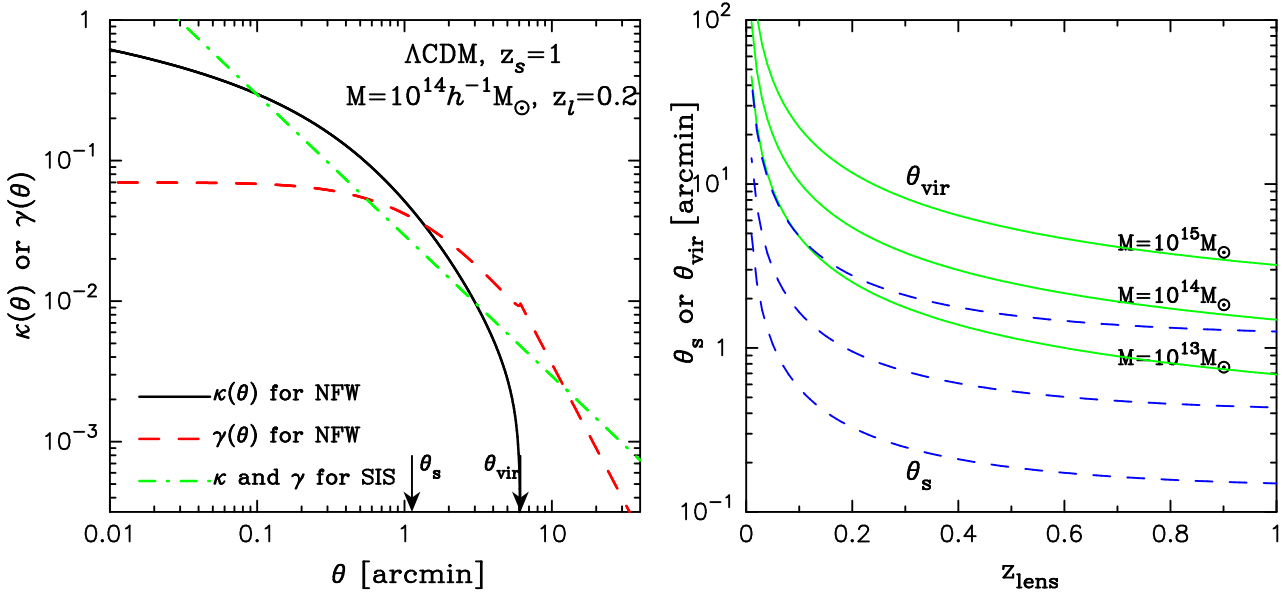}
  \end{center}
\caption{{\em Left panel:}
Radial profiles of the convergence (solid curve) and shear
(dashed curve) fields for an NFW halo. The halo mass is 
$M=10^{14}h^{-1}M_\odot$, lens redshift $z_l=0.2$ and source redshift 
$z_s=1$.  For comparison, the dot-dashed curve shows
a singular isothermal sphere $(\rho_h\propto r^{-2})$ with the same
virial mass. The two arrows show 
the scale radius and the virial radius of the NFW profile. 
{\em Right panel:} The angular scales of the scale and
virial radii for a lens halo are plotted against its redshift.
The three curves show the results for halo masses $M/M_\odot=10^{13}$, 
$10^{14}$ and $10^{15}$, from bottom to top. 
Halos relevant for lensing statistics have
$\theta_{\rm s}\simlt 2'$ for their scale radii at redshift
$z=0.4$. Thus the effect of the inner structures of the halo profile
on lensing statistics should appear at scales $\simlt 2'$.  
} \label{fig:nfwshear}
\end{figure}
%%%%%%%%%%%%%%%%%%%%%%%%%%%%%%%%%%%%%%%%%%%%%%%%%%%%%%%%%%%%%%%%%%%%%%
The left plot in Figure \ref{fig:nfwshear} shows the radial profiles of 
the convergence (solid curve) and shear (dashed curve) for an NFW halo of
mass $10^{14}h^{-1}M_\odot$, computed using equations
(\ref{eqn:convm}) and (\ref{eqn:gammam}). We use the
$\Lambda$CDM model, lens redshift  $z_l=0.2$ and  source
redshift $z_s=1$. The two arrows 
denote the angular scales $\theta_s$ and $\theta_{\rm vir}$ corresponding
to the scale radius and the virial radius, respectively. 
In contrast to the convergence, the shear is almost constant in the inner
region and does not vanish for $\theta\ge \theta_{\rm vir}$, rather it 
follows the point mass profile $\propto \theta^{-2}$.  
It should be noted that the shear profile appears
to be not a smoothly varying function at $\theta=\theta_{\rm vir}$
because of the artificial sharp halo boundary. 
For comparison, the dot-dashed curve shows the
shear amplitude for a singular isothermal sphere (SIS) with the profile
$\rho_h\propto r^{-2}$ (for this case the convergence and shear fields
coincide).  
The comparison manifests the characteristic feature for the NFW shear
field. For example, the plateau profile, $\gamma_M\approx {\rm
constant}$, at $\theta\simlt \theta_s$ could be a direct test of the
inner slope $\rho_h\propto r^{-1}$. 

The right panel in Figure \ref{fig:nfwshear} explicitly shows the
angular scales of the scale and virial radii for lens halos as a function 
of redshift. We consider the mass scales, $M/M_\odot=10^{13}$,
$10^{14}$ and $10^{15}$. 
Halos with $M>10^{13}M_\odot$, which dominate the
contribution to lensing statistics on small, non-linear scales, 
have $\theta_{\rm s}\simlt 2'$ for their
scale radii at redshift $z=0.4$. Thus, the figure implies that the 
the inner region likely affects the lensing statistics on angular scales
$\simlt 2'$, as we will show explicitly in Figure \ref{fig:3ptalpha}.

\subsection{Real-space halo approach}
\label{halomodel}

In the halo model, the 2PCF of
the convergence field, $\xi_{\kappa}(\theta)$, can be expressed as the
sum of correlations within a single halo (1-halo term) and between
two halos (2-halo term). In TJ03b, we obtained the real-space
relation for the 1-halo term:
\begin{equation}
\xi^{1h}_\kappa(\theta)\equiv \skaco{\kappa(\bm{\theta}_1)
\kappa(\bm{\theta}_2)}^{1h}
= \int^{\chi_s}_{0}\!\!d\chi~ 
\frac{d^2V}{d\chi d\Omega}\int\!\!dM~ n(M; \chi)
\int_0^{\theta_{\rm vir}}\!\!ds \int^{2\pi}_0\!\!d\varphi
~ s~ \kappa_M(s; \chi)
\kappa_M(|\bm{s}+\bm{\theta}|; \chi), 
\label{eqn:conv2pt}
\end{equation}
where $d^2V/d\chi d\Omega=d_A^2(\chi)$ for a flat universe, and we have
used polar coordinate $\bm{s}=s(\cos\varphi,\sin\varphi)$ to write
$d^2\bm{s}=s ds d\varphi$.  $\skaco{\cdots}$ denotes the ensemble
average.  From statistical symmetry, we can set the separation vector
$\bm{\theta}$ to be along the first axis, so that
$|\bm{s}+\bm{\theta}|=(s^2+\theta^2+2s\theta\cos\varphi)^{1/2}$. 
 To obtain $\xi^{1h}_{\kappa}$ for a given cosmological
model, we perform a 4-dimensional numerical integration.  
Equation
(\ref{eqn:conv2pt}) shows that $\xi_\kappa^{1h}$ 
is given by the sum of lensing contributions due to halos along the
line of sight weighted with the halo number density on the light cone. Hence, 
this form correctly accounts for multiple lensing due to
halos at different redshifts.  

\begin{figure}
  \begin{center}
    \leavevmode\epsfxsize=7.4cm \epsfbox{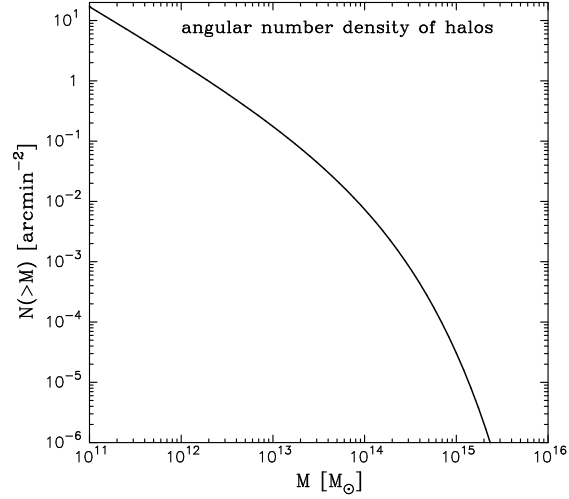}
  \end{center}
\caption{Projected number density of halos more massive than a given mass
 scale $M$, $N(>M)$, between $z=0$ and $1$ for the $\Lambda$CDM
 model. } \label{fig:angnum}
\end{figure}

Figure \ref{fig:angnum}
plots the angular number density of halos more massive than a given mass
scale $M$ between the source redshift $z_s=1$ and the present:
$N(>M)=\int^{z_s}_0\!\!dz~ d^2V/(dzd\Omega) \int_M^\infty\!\!  dM'~
n(M'; z)$. One can see $N(>M)\simlt 10^{-1}$ arcmin$^{-2}$ for 
halos with $M>10^{13}M_\odot$, which provide dominant contribution to the
lensing statistics on small angular scales (see Figure
\ref{fig:3ptMmax}). This clarifies that the 1-halo contribution is
primarily due to a single lens halo, and that there is only a small probability for
multiple lensing due to such massive halos at different redshifts.
However, one should keep mind the importance of multiple
lensing due to such cluster-scale halos and less massive halos, as
shown by White, Van Waerbeke \& Mackey (2002) and Padmanabhan, Seljak \& Pen (2003). 

%%%%%%%%%%%%%%%%%%%%%%%%%%%%%%%%%%%%%%%%%%%%%%%%%%%%%%%%%%%%%%%%%%%%%%
%%%%%%%%%%%%%%%%%%%%%%%%%%%%%%%%%%%%%%%%%%%%%%%%%%%%%%%%%%%%%%%%%%%%%% 
 
The form of equation (\ref{eqn:conv2pt}) allows us to straightforwardly
extend it to compute the 2PCF of shear fields by replacing the
convergence profile, $\kappa_M$, with the shear profile, $\gamma_M$, for a
given halo.  The 1-halo term in Cartesian coordinates is 
\begin{equation}
\skaco{\gamma_\mu(\bm{\theta}_1)\gamma_\nu(\bm{\theta}_2)}^{1h}
=\int^{\chi_s}_{0}\!\!d\chi~ 
d_A^2(\chi)\int\!\!dM~ n(M; \chi)
\int_0^{\infty}\!\!ds\int^{2\pi}_0d\varphi~ 
s~ \gamma_M(s; \chi)
\gamma_{M}(|\bm{s}+\bm{\theta}|; \chi) 
\epsilon_\mu(\bm{s})\epsilon_\nu(\bm{s}+\bm{\theta}),  
\label{eqn:shear2pt}
\end{equation}
where $\epsilon_\mu$ is the phase factor of the shear
field as given by equation (\ref{eqn:spattern});
$\epsilon_\mu(\bm{s})=-(\cos 2\varphi,\sin2\varphi)$ and
$\epsilon_\mu(\bm{s}+\bm{\theta})\equiv
-(\cos2\varphi_{\bm{s}+\bm{\theta}},\sin2\varphi_{\bm{s}+\bm{\theta}})$
with 
$(\cos\varphi_{\bm{s}+\bm{\theta}},\sin\varphi_{\bm{s}+\bm{\theta}})
=-|\bm{s}+\bm{\theta}|^{-1} (s\cos\varphi+\theta,s\sin\varphi)$.  In
contrast to equation (\ref{eqn:conv2pt}), the equation above employs an
infinite integration range for $d^2\bm{s}$ in order to account for
the non-local property of the shear fields.  In practice, 
setting the upper bound of $\int\!\!ds$ to be three
times the projected virial radius gives the same result, to within
a few percent.  We investigate the accuracy
 of equation (\ref{eqn:shear2pt}) as well as its self-consistency 
with other methods
in \S \ref{valid}. 

Since the two shear components are not invariant under coordinate
rotation, the shear 2PCFs
generally depend on the relative orientation of the two points (e.g., Kaiser
1992).  This issue has been well studied in the literature (Kamionkowski
et al. 1998; Crittenden et al. 2002; Schneider et al. 2002a). One way to
avoid the coordinate dependence is to use the $+/\times$ decompositions,
where the $+$ component is defined as the shear field in parallel or
perpendicular direction relative to the line connecting the two points
taken, while the $\times$ component is the $45^\circ$ rotated shear
field.  We can thus define the rotationally invariant 2PCFs of the shear
field:
\begin{eqnarray}
\xi_{\gamma,+}(\theta)&=&\skaco{\gamma_+(\bm{\theta}_1)
\gamma_+(\bm{\theta}_2)},\nonumber\\
\xi_{\gamma,\times}(\theta)&=&\skaco{\gamma_\times(\bm{\theta}_1)
\gamma_\times(\bm{\theta}_2)}.
\label{eqn:deftshear2pt}
\end{eqnarray}
It should be noted that
 $\skaco{\gamma_+(\bm{\theta}_1)\gamma_\times(\bm{\theta}_2)}$
 and
$\skaco{\gamma_\times(\bm{\theta}_1)\gamma_+(\bm{\theta}_2)}$ vanish
because of invariance under parity transformation.  We will also
consider the following 2PCF:
\begin{equation}
\xi_{\gamma}(\theta)=\skaco{\bm{\gamma}(\bm{\theta}_1)
\cdot\bm{\gamma}^\ast(\bm{\theta}_2)}.
\end{equation}
Note that $\xi_\kappa(\theta)=\xi_{\gamma}(\theta)$. 
The 1-halo term contributions to these shear 2PCFs can be calculated by
replacing the phase factors $\epsilon_\mu\epsilon_\nu$ in equation
(\ref{eqn:shear2pt}) with
\begin{eqnarray}
\epsilon_\mu(\bm{s})\epsilon_\nu(\bm{s}+\bm{\theta})
\rightarrow
\left\{
\begin{array}{ll}
\cos2\varphi\cos2\varphi_{\bm{s}+\bm{\theta}}
+\sin2\varphi\sin2\varphi_{\bm{s}+\bm{\theta}},& \mbox{for }~ \xi_\gamma,\\
\cos2\varphi\cos2\varphi_{\bm{s}+\bm{\theta}}, & \mbox{for }~ \xi_{\gamma,+},\\
\sin2\varphi\sin2\varphi_{\bm{s}+\bm{\theta}}, & \mbox{for }~ \xi_{\gamma,\times},
\end{array} 
\right.
\label{eqn:2ptphase}
\end{eqnarray}

An alternative approach to the lensing 2PCF, which is
conventionally used in the literature, is based on a model of the
non-linear 3D power spectrum of the mass, $P(k)$. Combining $P(k)$ with 
Limber's approximation (Limber 1954; Kaiser 1992) allows us to compute
the shear 2PCFs:
\begin{equation}
\skaco{\gamma_\mu\gamma_\nu}(\theta)=\int^{\chi_s}_{0}\!\!d\chi~
W^2(\chi,\chi_s)d_A^{-2}(\chi)\int\!\!\frac{ldl}{2\pi}~ 
P\!\left(k=\frac{l}{d_A(\chi)}\right)F(l\theta). 
\label{eqn:fshear2pt}
\end{equation}
Setting the window function to $F(x)=J_0(x)$, $[J_0(x)+J_4(x)]/2$ and
$[J_0(x)-J_4(x)]/2$ yields $\xi_\gamma$, $\xi_{\gamma,+}$ and
$\xi_{\gamma,\times}$, respectively.  So far, there are two well 
studied models for the nonlinear 
$P(k)$.  One is the fitting formula calibrated from
$N$-body simulations (e.g., Jain, Mo \& White 1995, Peacock \& Dodds
1996, and Smith et al. 2002). This method has been
extensively used for the interpretation of cosmic shear
measurements in terms of cosmological parameters (e.g., Van Waerbeke
2001b). The other is the recently developed Fourier-space halo approach
(Seljak 2000; Ma \& Fry 2000b,c; Scoccimarro et al. 2001). In the halo
model, $P(k)$ is similarly expressed as sum of the 1- and 2-halo terms;
$P(k)=P_{1h}(k)+P_{2h}(k)$ (see \S 2.2 in TJ0b for details).
Inserting $P_{1h}(k)$ into equation (\ref{eqn:fshear2pt}) leads to the
1-halo term of the shear 2PCFs in the Fourier-space halo model. 
Note that the condition $\xi_\kappa=\xi_\gamma$ holds,
since they both have the same window function $J_0$ in equation
(\ref{eqn:fshear2pt}).
  
%%%%%%%%%%%%%%%%%%%%%%%%%%%%%%%%%%%%%%%%%%%%%%%%%%%%%%%%%%%%%%%%%%%%%%
\begin{figure}
  \begin{center}
    \leavevmode\epsfxsize=8.cm \epsfbox{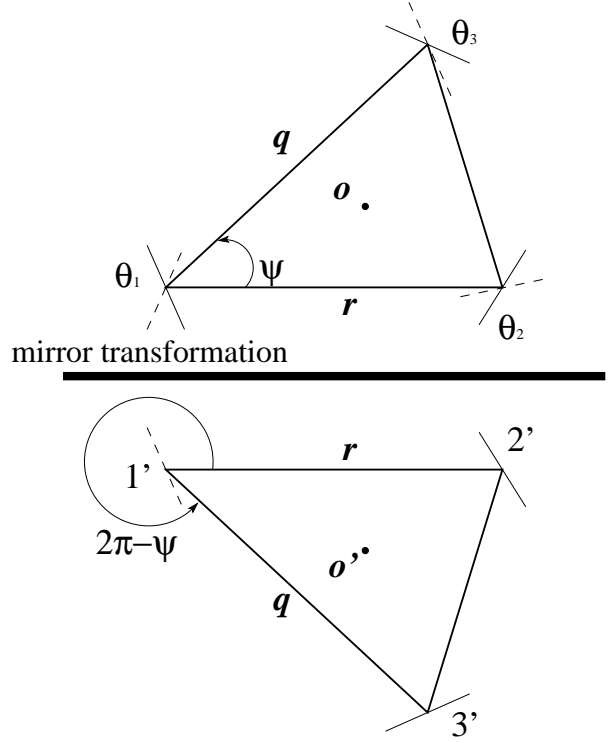}
  \end{center}
\caption{{\em Upper}: A sketch of the triangle configuration, parameterized
by $r$, $q$ and $\psi$, used to describe the 3PCF. 
The solid and dashed lines at each vertex show the positive
directions of the $+$ and $\times$ components of the shear field,
defined with respect to the triangle center denoted by $\bm{o}$.
{\em Lower}: The mirror transformation of the triangle with respect to
the side vector $\bm{r}$ for $\xi_{\times ++}$ is shown as one
example. The $\gamma_{\times}$ component at each vertex changes sign
under the mirror transformation (at vertices $1$ and $1'$ in this
case). } \label{fig:triang}
\end{figure}
%%%%%%%%%%%%%%%%%%%%%%%%%%%%%%%%%%%%%%%%%%%%%%%%%%%%%%%%%%%%%%%%%%%%%%
We turn to the 3PCF of lensing fields, which is main
focus of this paper.  The 1-halo term of the 3PCF of the convergence
field is given by equation (52) in TJ03b:
\begin{eqnarray}
\zeta^{1h}_{\kappa}(r,q,\psi)\equiv 
\skaco{\kappa(\bm{\theta}_1)\kappa(\bm{\theta}_2)\kappa(\bm{\theta}_3)}^{1h}
=\int^{\chi_s}_0\!\!d\chi~ d_A^2(\chi)\int\!\!dM~ n(M; \chi)
\int^{\theta_{\rm vir}}_0\!\!ds\int_0^{2\pi}\!\!d\varphi~
s~ \kappa_M(s)
\kappa_M(|\bm{s}+\bm{r}|)\kappa_M(|\bm{s}+\bm{q}|), 
\label{eqn:conv3pt}
\end{eqnarray}
where we have set $|\bm{s}+\bm{r}|=(s^2+r^2+2rs\cos\varphi)^{1/2}$ and
$|\bm{s}+\bm{q}|=[s^2+q^2+2sq\cos(\varphi-\psi)]^{1/2}$.  From 
statistical symmetry, the convergence 3PCF can be expressed as a function of
three parameters $r$, $q$ and $\psi$ characterizing the triangle
configuration, as shown in Figure
\ref{fig:triang}.  Note that we often omit $\chi$ in the
argument of $\kappa_M$ or $\gamma_M$ for simplicity.  This equation
means that we can compute $\zeta^{1h}_\kappa$ by a 4-dimensional
integration for any triangle configuration, which is the same level of 
computation as the 2PCF.  This holds for
higher-order moments as well (TJ03b), which is a great advantage of the
real-space halo model.  For comparison, the Fourier-space halo model
requires a 6-dimensional integration  to get $\zeta^{1h}_\kappa$.
In TJ03b, we have also developed  approximations for computing the 2-
and 3-halo terms of $\zeta_{\kappa}$. The 2-halo
term is relevant for a range of transition scales between the
non-linear and quasi-nonlinear regimes, while the 3-halo term dominates
in the large scale, quasi-linear regime where perturbation theory (PT) is valid. 

We can extend equation (\ref{eqn:conv3pt}) to 
the shear 3PCFs.  Since the shear field has two components at
each vertex of a given triangle configuration, we can formally construct
eight ($2^3=8$) components of the shear 3PCFs.  In Cartesian coordinates 
the functions are
\begin{equation}
\zeta_{\gamma,abc}(\bm{\theta}_1,\bm{\theta}_2,\bm{\theta}_3)
\equiv\skaco{\gamma_a(\bm{\theta}_1)\gamma_b(\bm{\theta}_2)
\gamma_c(\bm{\theta}_3)},  
\end{equation}
where $a,b,c=1, 2$.  Note that the subscripts $a,b,c$ correspond to
the shear components at vertices $\bm{\theta}_1$, $\bm{\theta}_2$ and
$\bm{\theta}_3$, respectively, in our convention (see Figure
\ref{fig:triang}).  As for
the shear 2PCFs, the  3PCFs defined above are not invariant under
 coordinate rotation. Therefore, in contrast to the 3PCF of the
convergence or any scalar quantity, they
are not specified by three parameters. Instead four parameters are needed, 
e.g., $r$, $q$, $\psi$ and the orientation angle of the side vector $\bm{r}$
relative to the first coordinate axis. The real-space halo model
yields the following expressions for the 1-halo terms of the shear 3PCFs:
\begin{eqnarray}
\zeta^{1h}_{\gamma,abc}(\bm{r},\bm{q})
&=&\int^{\chi_s}_{0}\!\!d\chi~ 
d_A^2(\chi) \int\!\!dM~ n(M; \chi)
\int^{\infty}_0\!\!ds\int_0^{2\pi}\!\!d\varphi~
s ~ \gamma_M(s)
\gamma_M(|\bm{s}+\bm{r}|)
\gamma_M(|\bm{s}+\bm{q}|)\epsilon_a(\bm{s})
\epsilon_b(\bm{s}+\bm{r})\epsilon_c(\bm{s}+\bm{q}), 
\label{eqn:shear3pt}
\end{eqnarray}
with 
\begin{eqnarray}
\epsilon_a({\bm{s}})&=&-(\cos2\varphi,\sin2\varphi),\nonumber\\
\epsilon_b(\bm{s}+\bm{r})&=&-(\cos2\varphi_{\bm{s}+\bm{r}},
\sin2\varphi_{\bm{s}+\bm{r}}),\nonumber\\ 
\epsilon_c(\bm{s}+\bm{q})&=&-(\cos2\varphi_{\bm{s}+\bm{q}},
\sin2\varphi_{\bm{s}+\bm{q}})
\end{eqnarray}
where $(\cos\varphi_{\bm{s}+\bm{r}},\sin\varphi_{\bm{s}+\bm{r}})=
(r+s\cos\varphi,s\sin\varphi) /|\bm{s}+\bm{r}|$,
$(\cos\varphi_{\bm{s}+\bm{q}},\sin\varphi_{\bm{s}+\bm{q}})=
(q\cos\psi+s\cos\varphi,q\sin\psi +s\sin\varphi)/|\bm{s}+\bm{q}|$,
$|\bm{s}+\bm{r}|$ and $|\bm{s}+\bm{q}|$ are given below equation
(\ref{eqn:conv3pt}), and we again take the halo center as the coordinate
center, using statistical symmetry.

To avoid the coordinate dependence for the shear 3PCFs, we again
use the $+/\times$ decomposition of the shear fields, as stressed in
SL03. For the three-point case, however, there is no unique choice of
the reference direction to define the $+/\times$ components.  
Following ZS03 and TJ03a, we take the
`center of mass', $\bm{o}$, of the triangle, defined by
\begin{equation}
\bm{o}=\frac{1}{3}\sum_{i=1}^3 \bm{\theta}_i=\bm{s}
+\frac{1}{3}(r+q\cos\psi,q\sin\psi). 
\label{eqn:center}
\end{equation} 
The projection operators, which transform $\gamma_1$ and $\gamma_2$ at
each vertex into the $+/\times$ components, with respect to $\bm{o}$, are
\begin{eqnarray}
\bm{P}_+(\bm{\theta}_i)
=-(\tilde{\theta}_{i1}^2-
\tilde{\theta}_{i2}^2
,2\tilde{\theta}_{i1}\tilde{\theta}_{i2})/\tilde{\theta}_i^2,
\hspace{2em}
\bm{P}_\times(\bm{\theta}_i)=
-(-2\tilde{\theta}_{i1}\tilde{\theta}_{i2},
\tilde{\theta}_{i1}^2-\tilde{\theta}_{i2}^2)/\tilde{\theta}_i^2,
\label{eqn:project}
\end{eqnarray}
where $\tilde{\bm{\theta}}_i=\bm{\theta}_i-\bm{o}$ ($i=1,2,3$).  The
geometry of the triangle we consider is illustrated in Figure
\ref{fig:triang}, where the solid and dashed lines at each vertex denote
positive directions of the $+$ and $\times$ components. Using these
projections, we can thus define the shear 3PCFs from combinations of the
$+/\times$ components so that the resulting 3PCFs are invariant under
rotations of the triangle with respect to $\bm{o}$:
\begin{equation}
\zeta_{\gamma,\nu\mu\tau}
(r,q,\psi)
=P_\mu^a(\bm{\theta}_1)P_{\nu}^b(\bm{\theta}_2)
P_\tau^c(\bm{\theta}_3)\skaco{\gamma_a(\bm{\theta}_1)\gamma_b(\bm{\theta}_2)
\gamma_c(\bm{\theta}_3)},
\label{eqn:pshear3pt}
\end{equation}
where $\mu,\nu,\tau=+$ or $\times$. Inserting equation
(\ref{eqn:shear3pt}) into the r.h.s above yields the 1-halo term for
$\zeta_{\gamma,\nu\mu\tau}$.  The projection operators and
$\skaco{\cdots}$ commute, since the projection
operators do not depend on the integration variable $\bm{s}$, as seen
from equation (\ref{eqn:project}).  The shear 3PCFs defined in this way
are functions of the three parameters $r$, $q$ and $\psi$.
Although we adopt the center of mass throughout this paper, SL03 proved
that the eight shear 3PCFs with respect to {\em any} center can be
expressed as linear combinations of the eight 3PCFs above.
To complete the halo model predictions, it is necessary to develop the
perturbation theory predictions for the shear 3PCFs, which are
relevant on large scales.  For the convergence we have obtained these, but 
the complex spin-2 properties
of the shear fields make it non-trivial to obtain the PT prediction.
We will consider only the 1-halo term for predictions of the
shear 3PCFs. Nevertheless, the 1-halo term agrees with the simulation
results on scales of interest, as shown below.

\section{Triangle configuration dependences of the shear 3PCF}
\label{parity}

From the observation that there are eight shear 3PCFs, SL03 and ZS03
addressed the  questions: how many functions are non-zero?
How does each function carry information about the $E/B$-modes?  In
TJ03a, using ray-tracing simulations, we qualitatively verified the
conclusions of SL03 and ZS03, which were based on analytical studies. 
We found that:
(1) The eight 3PCFs are generally non-zero. (2) For a pure
$E$-mode,  two or four components vanish for isosceles or equilateral
triangles, respectively. (3) We also studied the triangle configuration
dependence of the shear 3PCF 
given by equation (\ref{eqn:este}), and pointed out that they offer
a promising way to disentangle the $E/B$-modes for measured shear 3PCFs.  
Here we summarize these characteristics of the shear 3PCFs.

We restrict ourselves to a pure $E$ field, expected in the weak
lensing regime.  We consider a
mirror transformation as shown in Figure \ref{fig:triang}, which
illustrates the transformation with respect to the side vector $\bm{r}$
for $\zeta_{\times++}$.  It corresponds to $\psi \rightarrow 2\pi-\psi$
in our parameterization. From statistical homogeneity and symmetry, the
amplitude of the shear 3PCF depends only on the distances between the
center and each vertex. Hence, the absolute amplitudes of
$\zeta_{\times++}$ for the two triangles shown should be same. But the
sign of $\gamma_\times$ at the vertex $1'$ changes under this mirror
transformation.  
According to this property, we can divide the eight
3PCFs into two groups:
\begin{eqnarray}
\mbox{Parity-even functions:}&& 
\zeta_{\mu\nu\tau}(r,q,\psi)=\zeta_{\mu\nu\tau}(r,q,2\pi-\psi),
\mbox{ for } \ \ (\mu,\nu,\tau)=(+,+,+),(+,\times,\times),(\times,+,\times),
(\times,\times,+),
\nonumber\\
\mbox{Parity-odd functions:}&& 
\zeta_{\mu\nu\tau}(r,q,\psi)=-\zeta_{\mu\nu\tau}(r,q,2\pi-\psi),
\mbox{ for } \ \ (\mu,\nu,\tau)=(\times,\times,\times),(\times,+,+)
,(+,\times,+),(+,+,\times).
\label{eqn:mirror}
\end{eqnarray}
Note that the 3PCF of a scalar quantity transforms 
as $\zeta(r,q,\psi)=\zeta(r,q,2\pi-\psi)$.

Next we consider special triangle configurations: the
first is an isosceles triangle with $r=q$.  In this case, the
$\gamma_\times$ components at vertices $1$ and $1'$ in Figure
\ref{fig:triang} are statistically identical (viewed from the center of
the triangle, they should have equal contributions when averaged over the
matter distribution)\footnote{This argument is true only for a pure $E$
field, since it relies on the invariance of the
$E$-mode under parity transformation.}.  We thus have additional
symmetries for the two 3PCFs:
$\zeta_{\times++}(r,q,\psi)=\zeta_{\times++}(r,q,2\pi-\psi)$ and $
\zeta_{\times\times\times}(x_1,x_1,\psi)
=\zeta_{\times\times\times}(x_1,x_1,2\pi-\psi)$.  These relations and
equation (\ref{eqn:mirror}) yield
\begin{equation}
\mbox{Isosceles Triangles: } \  \ 
\zeta_{\times++}=\zeta_{\times\times\times}=0.
\label{eqn:parityinv}
\end{equation}
Note that the other two parity-odd functions, $\zeta_{+\times+}$ and
$\zeta_{++\times}$, do not vanish in general, since the component
$\gamma_\times$ is at a vertex bounded by unequal sides.  For
equilateral triangles, however, all four parity-odd functions vanish:
\begin{equation}
{\rm Equilateral \ Triangles: } \ \ 
\zeta_{\times\times\times}=\zeta_{\times ++}
=\zeta_{+\times+}=\zeta_{++\times}=0. 
\label{eqn:equil}
\end{equation}
The 3PCFs discussed above do not vanishing for
a $B$-mode shear field, as explained below. Hence these 3PCFs provide a direct,
simple test of the $B$-mode contribution to the measured signal.

Let us consider again generic triangle configurations, but now for 
a pure $B$ field. As discussed by TJ03a, the analog of equation 
(\ref{eqn:mirror}) for a B-mode spin-2 field is:
\begin{eqnarray}
&& \zeta_{\mu\nu\tau}(r,q,\psi)=-\zeta_{\mu\nu\tau}(r,q,2\pi-\psi),
\mbox{ for } \ \ (\mu,\nu,\tau)=(+,+,+),(+,\times,\times),(\times,+,\times)
,(\times,\times,+), \nonumber\\
&& \zeta_{\mu\nu\tau}(r,q,\psi)=\zeta_{\mu\nu\tau}(r,q,2\pi-\psi),
\mbox{ for } \ \ (\mu,\nu,\tau)=(\times,\times,\times),(\times,+,+)
,(+,\times,+),(+,+,\times).
\label{eqn:bmirror}
\end{eqnarray}
Thus the symmetric and anti-symmetric functions are reversed compared to 
the $E$-mode. This follows from the fact that given a pure
$E$ field, we can generate a pure $B$ field by
rotating the $E$ field at each point by $45$ degrees (Kaiser 1992). We
can then consider the eight 3PCFs of a pure $B$ mode similarly as done
for the shear 3PCFs. Since this procedure transforms the original
$E$-mode components $\gamma^{E}_{+}$ and $\gamma^{E}_{\times}$ at each
point into $\gamma^{B}_{\times}$ and $\gamma^B_{+}$ in the transformed
$B$ field, respectively, we get $\zeta^E_{+++}\rightarrow
\zeta^B_{\times\times\times}$ and so on. For a general spin-2
field that contains both $E/B$ modes, 
the configuration dependences of
equations (\ref{eqn:mirror}) and (\ref{eqn:bmirror}) do not hold. The eight
3PCFs therefore have to be measured over the full range $\psi=[0,2\pi]$, unlike
the 3PCF of a scalar quantity.  

From the symmetry properties discussed above,  we
propose simple estimators for the $E/B$-mode contributions to 
measured shear 3PCFs:
\begin{eqnarray}
\mbox{Estimator of $E$-mode}: & {\displaystyle 
\zeta^E_{\mu\nu\tau}(r,q,\psi)
=\frac{1}{2}\left[\zeta_{\gamma,\mu\nu\tau}(r,q,\psi)
\pm \zeta_{\gamma,\mu\nu\tau}(r,q,2\pi-\psi)\right]},\nonumber\\ 
\mbox{Estimator of $B$-mode}: & {\displaystyle 
\zeta^B_{\mu\nu\tau}(r,q,\psi)
=\frac{1}{2}\left[\zeta_{\gamma,\mu\nu\tau}(r,q,\psi)
\mp \zeta_{\gamma,\mu\nu\tau}(r,q,2\pi-\psi)\right]},
\label{eqn:este}
\end{eqnarray}
where the upper and lower signs in $\pm$ or $\mp$ 
are meant  for
$(\mu,\nu,\tau)=(+,+,+),(+,\times,\times),(\times,+,\times),(\times,\times,+)$
and $(\times,\times,\times),(\times,+,+),(+,\times,+) ,(+,+,\times)$,
respectively. Note that this argument holds if the $E$ and $B$ modes are
statistically uncorrelated.  These estimators
allow one to separate the lensing $E$-mode contribution from the
measured 3PCFs that are in general contaminated by the $B$-modes 
contribution of intrinsic alignments, source galaxy clustering and 
observational systematics.  In comparison, for the case of the shear
2PCFs, a non-local integration is required to discriminate
the lensing $E$-mode (Schneider et al. 1998; Crittenden et al. 2002;
Schneider et al. 2002a).

After the submission of this paper, Schneider (2003) pointed out that
if a $B$ field is parity invariant in a statistical sense, any
correlation function that contains an odd number of $B$-mode shear
components vanishes.  This is likely to be true for a cosmological $B$
field such as intrinsic alignments if we have a sufficient survey area.
Therefore, the argument we have made above is valid only for a parity 
non-invariant $B$ field. This can be seen in that the 45 degree rotation
procedure described above to generate a $B$ field from simulations
would produce shear fields around clusters with a clockwise
curl direction, which violates statistical parity invariance. 
Nevertheless, we believe our discussion
is useful, because generic observational systematics are likely lead to the
violation of parity invariance. Hence, these arguments
strength practical usefulness of the shear 3-point functions to
disentangle $E/B$ modes from the measurement.
%%%%%

\section{Comparison with ray-tracing simulations} \label{results}

In this section, we address the accuracy of the halo model for
predicting lensing statistics, in particular the 3PCFs, by comparing
model predictions with ray-tracing simulation results.  The
validity of the real-space halo model for shear
correlations, developed in \S \ref{model}, is carefully investigated in
\S \ref{valid}.

\subsection{Cosmological models}
\label{cosmo}

In this paper, we mainly consider two CDM models whose cosmological
parameters are chosen to facilitate comparison with the ray-tracing
simulations used below (Jain et al. 2000; M\'enard et al. 2003; Hamana
et al. 2003).  One is the SCDM model ($\Omega_{m0}=1$, $h=0.5$ and
$\sigma_8=0.6$), and the other is the $\Lambda$CDM model
($\Omega_{m0}=0.3$, $\Omega_{\lambda0}=0.7$, $\Omega_{b0}=0.04$, $h=0.7$
and $\sigma_8=0.9$).  For the \LCDM model, we need to care about the
baryon contribution to the input primordial power spectrum.  Although we
assume a scale-invariant power spectrum for the primordial fluctuations,
we employ different CDM transfer functions for the SCDM and \LCDM
models.
For the SCDM model, we use the transfer function in Bond \& Efstathiou
(1984) with the shape parameter $\Gamma=\Omega_{\rm m0}h=0.5$. On the
other hand, for the \LCDM model we employ the BBKS transfer function
(Bardeen et al. 1986) with the shape parameter in Sugiyama (1995), since
the shape parameter approximately describes the baryon contribution.

\subsection{Ray-tracing simulations}
\label{simul}

We use ray-tracing simulations of the lensing convergence and
shear fields. We will mainly consider two cosmological model
simulations, the $\Lambda$CDM model (kindly made available to us
by T. Hamana; see M\'enard et al. 2003; Hamana et al. 2003 for details)
and the SCDM model (Jain et al. 2000), as described in \S \ref{cosmo}.  
The $N$-body simulations on which these are based
were carried out by the Virgo Consortium \footnote{ see {\tt
http://star-www.dur.ac.uk/\~{}frazerp/virgo/virgo.html} for the details}
(also see Yoshida, Sheth \& Diaferio 2001), and were run using the
particle-particle/particle-mesh (P$^3$M) code with a force softening
length of $l_{\rm soft}\sim 30h^{-1}$kpc.  The linear power spectra of
the initial conditions of the simulations were set up using the transfer
function from CMBFast (Seljak \& Zaldarriaga 1996) 
for the \LCDM model and the fitting function in
Bond \& Efstathiou (1984) for the SCDM model, respectively. 
For the halo model predictions for \LCDM, we employ the BBKS plus Sugiyama 
transfer function as described above. We verify in Figure
\ref{fig:smithpd} that the transfer function is sufficiently accurate 
(see also Eisenstein \& Hu 1999).

The $\Lambda$CDM simulation employs $512^3$ CDM particles in a cubic box 
$479h^{-1}$Mpc on a side, while the SCDM simulation uses $256^3$
particles in a $84.5h^{-1}$Mpc side-length box. The particle mass 
is $m_{\rm part}=6.8\times 10^{10}h^{-1}M_\odot$ and
$1.0\times 10^{10}h^{-1}M_\odot$,
respectively.  The resolution scale, which is not affected by the
discreteness of the $N$-body simulations, is estimated roughly as being
ten times the grid size, leading to $\lambda \simgt 94h^{-1}$kpc and
$33h^{-1}$kpc for the \LCDM and SCDM models, respectively. Therefore,
the SCDM model simulation has better spatial resolution than the \LCDM model. 

The multiple-lens plane algorithm to simulate the lensing maps from the
$N$-body simulations is detailed in
Jain et al. (2000) and Hamana \& Mellier (2001). Throughout this paper,
we use a single source redshift of $z_s=1$. The simulated areas
of the lensing maps are $\Omega_s=11.7$ and $7.69$ degree$^2$ for the
\LCDM and SCDM models, respectively.  The map is given on $1024^2$ grids
with grid spacing $\theta_{\rm grid}=0.\! \! '2$ and $0.\! \!  '16$ for the
two models.  To analyze the correlation functions 
we should be careful about the possibility that the finite angular
resolution could affect computations of the $n$-point correlation
functions from the simulated maps on small scales. 
It is not easy to infer the effective angular resolution from the
spatial resolution of the $N$-body simulations due to the broad lensing
projection kernel. Further, the projected density field 
was smoothed to suppress the discreteness effect of the $N$-body
simulations.  The smoothing scale is likely to determine the
angular resolution rather than the spatial resolution of the $N$-body
simulation.  Note that the \LCDM simulation we employ below
is the one labelled with {\em small-scale smoothing} in M\'enard et
al. (2003).  These resolution issues were carefully discussed in
M\'enard et al. (2003) and Jain et al. (2000), as can be seen from
Figure 3 in M\'enard et al (2003) and Figure 2 in Jain et al. (2000).
The former shows the projected angular scale of the smoothing as a
function of redshift, while the latter shows the angular scale of the
force-softening length of the $N$-body simulation. These scales are
$\theta_{\rm res}\sim 0.\!  \!  '3$ and $0.\!\!'2$ at $z=0.4$ for the
\LCDM and SCDM models, respectively, where $z=0.4$ is approximately
the peak redshift of the lensing efficiency for source redshift
$z_s=1$. Therefore, angular scales that are not affected by
finite resolution are likely to be $\simgt 1'$, although the
resolution of the \LCDM model simulation might be slightly worse than
that of the SCDM model, as stated above.
We will keep in mind these resolution issues in the following analysis.

To compute the sample variance of the lensing statistics from the
simulations, we use 36 and 9 realizations of the simulated lensing maps
for the \LCDM and SCDM models, respectively. The realizations are
generated by randomly rotating and translating the N-body simulation boxes
(using the periodic boundary conditions of the $N$-body boxes) when the 
ray-tracing simulations are
performed.  However, they were built from one realization of the
$N$-body simulation, which is a sequence of the redshift-space evolution
of large-scale structure.  Therefore the
different realizations of the simulated lensing maps are not fully independent
of each other.  In particular, this could matter when
we compute the covariance for the $n$-point correlations in the
different bins from the simulations.  This issue is still an open
question, to be addressed in the future using a sufficient number of truly
independent ray-tracing simulations.  

\subsection{Algorithm for computing the 3PCF from simulated maps} 
\label{3ptalg}

To compute the 3PCFs of the lensing fields from the
ray-tracing simulations, we implement the method described in \S 3.2.2
in Barriga \& Gazta\~naga (2002). 
The 3PCF is given as a function of three parameters
($r$, $q$ and $\psi$) specifying the triangle configuration (see Figure
\ref{fig:triang}).  The question is how  we can efficiently find
triplets from the simulated lensing map with $N_{\rm grid}$ grid points,
subject to the constraint that the triplet forms a given triangle 
configuration within the bin widths.  We first select vertex 1 on the
grid, and then search vertex 2 in the upper half plane in an annulus of
radius $r$ with given bin width, centered on vertex 1. 
For given pair 1-2, we look for vertex 3 in a semi-annulus
of radius $q$ in the upper plane above the line connecting vertices $1$
and 2 (in the anticlockwise direction from vertex 1 as our
convention), imposing the condition that the three vertices form the
required triangle configuration within the bin widths. This 
results in all triangles being counted once, if we impose the
conditions $\theta_{12}\le \theta_{23}\le \theta_{31}$.  We can  use
the same list of neighbors to find vertices 2 and 3 for each
vertex 1.  This process is illustrated in Figure 3 in Barriga \&
Gazta\~naga (2002).  Further, to compute the 3PCFs of the shear fields,
we need to compute the $+/\times$ components
of the shear fields at each vertex of the triangle.  The projection
operators to compute these components are given as a function of 
the list of neighbors of
vertices 2 and 3, independent of the position of vertex 1, as can be
seen from the definition of equation (\ref{eqn:project}).  In summary, the 
3PCF computation from the simulation map requires roughly $O(N_{\rm grid})$
operations on sufficiently small scales. This is significantly faster
than a naive, direct implementation which requires 
$O(N_{\rm grid}^3)$ operations.

\subsection{The two-point correlation function of the shear fields}
%%%%%%%%%%%%%%%%%%%%%%%%%%%%%%%%%%%%%%%%%%%%%%%%%%%%%%%%%%%%%%%%%%%%%%
\begin{figure}
  \begin{center}
    \leavevmode\epsfxsize=18.cm \epsfbox{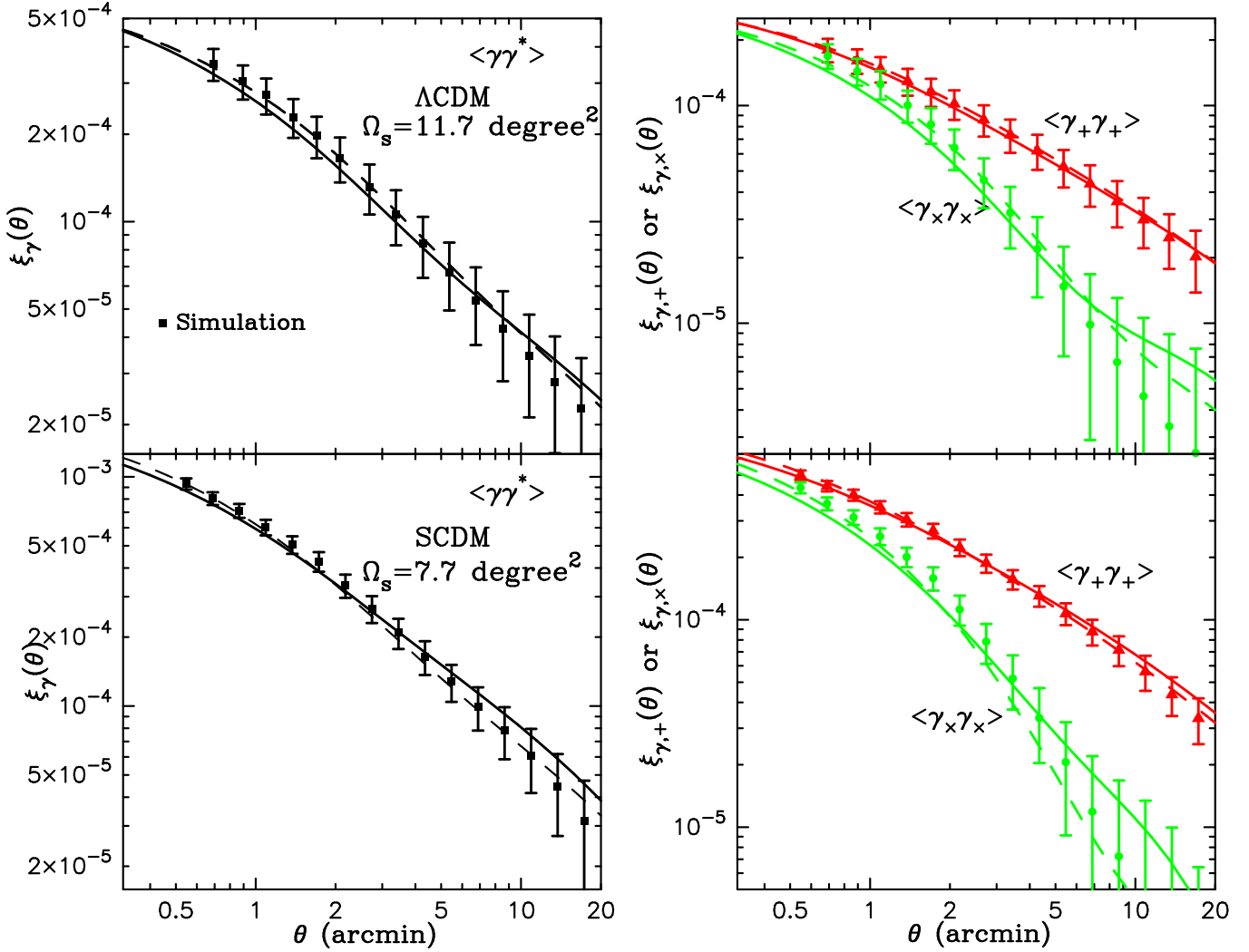}
  \end{center}
\caption{Comparison of model predictions for the shear 2PCFs
 with measurements from the ray-tracing simulations.  The left panel
 shows the shear 2PCF,
 $\xi_{\gamma}=\skaco{\bm{\gamma}\cdot\bm{\gamma}^\ast}$, for the \LCDM
 (upper panel) and SCDM (lower panel) models.  The solid and dashed
 curves are the predictions from the halo model and the Smith02 formula,
 respectively, while the square symbols with error bars denote the
 simulation result.  
 Note that errors in different bins are highly
 correlated with each other.  Similarly, the right panel shows the
 comparison for $\xi_{\gamma,+}$ and $\xi_{\gamma,\times}$. 
  } \label{fig:2ptlcdm}
\end{figure}
%%%%%%%%%%%%%%%%%%%%%%%%%%%%%%%%%%%%%%%%%%%%%%%%%%%%%%%%%%%%%%%%%%%%%%
In Figure \ref{fig:2ptlcdm} we present a comparison of 
the halo model predictions for the
shear 2PCFs with those measured from the ray-tracing simulations for the
$\Lambda$CDM (upper panel) and SCDM (lower panel) models.  The left
panel is the comparison for $\xi_\gamma$. The solid curve is the halo
model prediction, while the square symbol denotes the simulation result.
The dashed curve denotes the result computed from the
Smith02 formula. The theoretical predictions agree very well with the
simulation results for the two CDM models.  The smallest scale data from the
simulations corresponds to three times the grid size, and the
error bar in each bin is the sample variance for area
$\Omega=11.7$ and $7.7$ degree$^2$ for the \LCDM and SCDM model 
simulations,
respectively, as estimated from the scatter among their 36 and 9
realizations (see \S \ref{simul}).  We note that the errors in different
bins are highly correlated with each other. Even if one combines two
neighboring bins, the error amplitude remains almost
unchanged.  We have confirmed that this is true for all the statistical
quantities we consider below, over angular scales of interest.

Similarly, the right panel in Figure \ref{fig:2ptlcdm} shows the
comparison for $\xi_+$ and $\xi_\times$. The theoretical
predictions for $\xi_{\gamma,+}$ are again in agreement with the
simulation results.  On the other hand,
for $\xi_{\gamma,\times}$, the halo model prediction lies slightly below
the simulation results at $\theta\simlt 5'$, while the Smith02
prediction agrees better.  
If we employ the halo boundary of
$r_{180}$ as discussed in Figure \ref{fig:2ptbound}, the 
model predictions agree better with the simulations. 
The relation between the amplitudes of $\xi_{\gamma,+}$ and
$\xi_{\gamma,\times}$ is physically determined by the $k$-slope of the
underlying mass power spectrum, as can be seen from equation
(\ref{eqn:fshear2pt}).  Within the framework of the halo model, the
slope is determined by the combined effects of the halo profile, 
the mass function and the slope of the primordial power spectrum 
(Seljak 2000; Ma \& Fry 2000a,b,c; Scoccimarro et al. 2001; TJ02;
TJ03b).  Thus, separate measurements  of $\xi_+$ and $\xi_\times$
could constrain these physical ingredients. 

\subsection{The three-point correlation function of the convergence field}

In the following, we address the accuracy of halo model predictions
for the 3PCFs of lensing fields by comparison with ray-tracing
simulations.  Until our recent work (TJ03a,b), there had been no
analytical model for the lensing 3PCFs on small 
angular scales, except for investigations of the skewness and bispectrum
of the convergence field based on extended perturbation theory
(Scoccimarro \& Frieman 1999; Hui 1999; Van Waerbeke et al. 2001a) or
the Fourier-space halo model (Cooray \& Hu 2001a; TJ02). Hence we
present a detailed analysis of the accuracy of halo model predictions
for the lensing 3PCFs.

First, we consider the 3PCF of the convergence field.  As in the
literature, we define the reduced 3PCF as
\begin{equation}
Q_\kappa(r,q,\psi)=\frac{\zeta_\kappa(r,q,\psi)}
{\xi_\kappa(r)\xi_\kappa(q)+\xi_\kappa(r)
\xi_\kappa(|\bm{r}-\bm{q}|)
+\xi_\kappa(q)\xi_\kappa(|\bm{r}-\bm{q}|)},
\label{eqn:convq}
\end{equation}
where we have used the parameters $r$, $q$ and $\psi$ to describe the
triangle configuration and $|\bm{r}-\bm{q}|=(r^2+q^2-2rq\cos\psi)^{1/2}$
(see Figure \ref{fig:triang}).  The reduced 3PCF is sensitive to
$\Omega_{\rm m0}$, but insensitive to the power spectrum normalization
$\sigma_8$, scaling roughly as $Q_\kappa\propto \Omega_{\rm
m0}^{-1}$ (Bernardeau et al. 1997; Jain \& Seljak 1997; TJ02).
Therefore, measuring the 3PCF is expected to break degeneracies in the
determination of $\sigma_8$ and $\Omega_{\rm m0}$ from
measurements of the shear 2PCFs.

%%%%%%%%%%%%%%%%%%%%%%%%%%%%%%%%%%%%%%%%%%%%%%%%%%%%%%%%%%%%%%%%%%%%%%
\begin{figure}
  \begin{center}
    \leavevmode\epsfxsize=13.cm \epsfbox{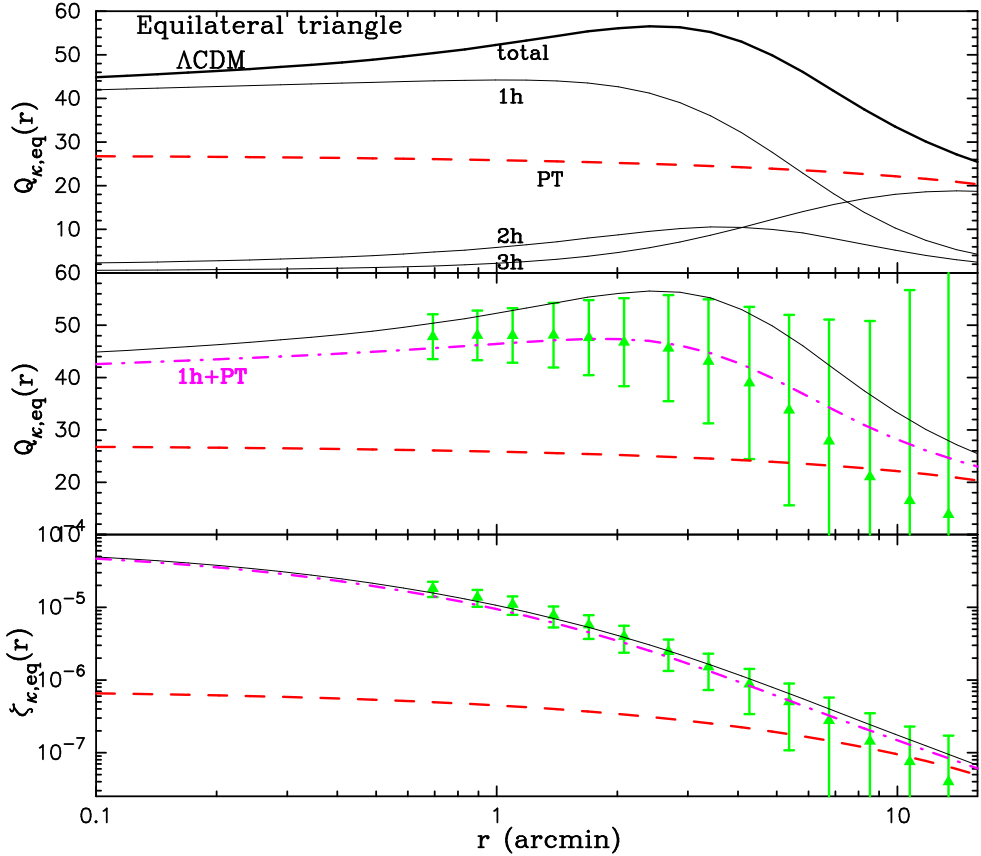}
  \end{center}
\caption{The 3PCF of the convergence field
for equilateral triangles against side length $r$ in arcminutes.
{\em Top panel}: The thick solid curve shows the halo model prediction
for the reduced 3PCF, $Q_\kappa$, defined by equation
(\ref{eqn:convq}). The thin solid curves denote the 1-, 2- and 3-halo
term contributions separately. 
The dashed curve is the perturbation theory (PT) prediction.  {\em
Middle panel}: The comparison with simulations for $Q_\kappa$,
as in Figure \ref{fig:2ptlcdm}. The simulation result does not display
the bump in $Q_\kappa$ seen in the halo model prediction.  This
is likely to be due to halo boundary effects in the standard
implementation of the halo model (see text for details).  The dot-dashed
curve denotes the result if we replace the 2- and 3-halo terms with the
PT prediction.  This modified halo model 
agrees well with the simulation result.  {\em Bottom panel}: The 
comparison for the 3PCF itself, as in the middle panel. 
} \label{fig:c3ptlcdmeq}
\end{figure}
%%%%%%%%%%%%%%%%%%%%%%%%%%%%%%%%%%%%%%%%%%%%%%%%%%%%%%%%%%%%%%%%%%%%%%
Figure \ref{fig:c3ptlcdmeq} shows the convergence 3PCF for equilateral
triangles against side length for the \LCDM model. The thick solid
curve in the top panel shows the halo model prediction for
$Q_\kappa$. A bump feature is evident over the range $1\simlt
r\simlt5'$.  As discussed below, this feature is unlikely
to be real. The three
thin curves show the 1-, 2- and 3-halo terms separately. 
These contributions to the convergence 3PCF are present in the
numerator of $Q_{\kappa}$, but the 2PCF in the denominator includes the
full contribution (1- plus 2-halo terms).  The 1-halo term provides the
dominant contribution to $Q_{\kappa}$ at small scales, $r\simlt 3'$. The
2-halo term is relevant over the transition scales between the
non-linear and linear regimes. The bump feature in
$Q_\kappa$ is mainly due to the 2-halo term contribution.  The 3-halo
term eventually dominates on larger scales, and gives the 
perturbation theory (PT) result for $Q_\kappa$ at $r\simgt 10'$.  
The 2- and 3-halo terms appear to be relevant over a wide range of 
angular scales
compared to the 3D mass 3PCF, $Q$, shown in Figure 7 in TJ03b, since
the lensing projection causes various length scales
at different redshifts to contribute to the lensing statistics. 
Even at the non-linear scale of $r=1'$, the 2- and
3-halo terms make $11\%$ and $4\%$ contributions to $Q_{\kappa}$. 

The middle panel shows a comparison of the halo model prediction with
the simulation result for $Q_\kappa$, as in Figure \ref{fig:2ptlcdm}.  
The bump feature in the halo model prediction cannot be seen in the
simulation result and, as a result, the halo model overestimates the
simulation result at more than $1\sigma$.  We have confirmed
that this is also true for the SCDM model (also see Figures 8 and 10 in
TJ03b for the discrepancy between the halo model prediction and the
simulation result for
the 3D mass 3PCF).  There are two effects to be considered in finding the
origin of the bump feature. First, the standard halo model does not take into
account halo exclusion effects: the 2- and 3-halo terms should 
require that {\em different} halos be separated by
at least the sum of their virial radii. Second, so far we have
used a sharp cutoff for the halo profile, which could lead to
inaccuracy in the prediction as discussed for Figure \ref{fig:2ptbound}
\footnote{Note that, 
even if we employ the halo boundary of $r_{180}$ as discussed
in Figure \ref{fig:2ptbound}, the bump feature for $Q_\kappa$
remains at the same level}.
In fact, Figure 8 in TJ03b clarified that modifications aimed at
resolving these two effects do suppress the bump feature
in the 3D mass $Q$ (see also Somerville et al. 2001, Bullock et al. 2002
and Zehavi et al. 2003 for discussions on the halo exclusion
effect). However, resolving these problems requires a more careful and
systematic study, in combination with $N$-body simulations.  This is
beyond the scope of this paper. Therefore, we instead employ a simple
prescription to avoid the overestimation of $Q_\kappa$ -- we replace the
contribution from the 2- plus 3-halo terms in the 3PCF with the PT
prediction.  This treatment preserves two merits of the halo
model: the quasi-linear 3PCF at large scales is
reproduced by the PT result alone, and the non-linear 3PCF
on small scales is described primarily by the 1-halo term.  The
dot-dashed curve shows the modified halo model prediction for
$Q_\kappa$.  It displays excellent agreement with the simulation
results over the scales we have considered.  An alternative method is to
simply ignore the 2-halo term contribution to the 3PCF, leading to 
similar agreement.  Hereafter, we
will use the 1-halo term plus the PT result as the halo model
prediction for the convergence 3PCF.

\begin{figure}
  \begin{center}
    \leavevmode\epsfxsize=13.cm
    \epsfbox{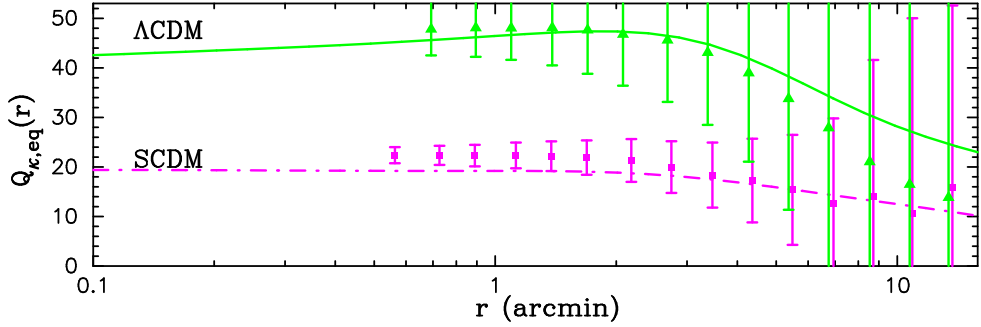}
  \end{center}
\caption{Plots for the SCDM model, as the middle panel in Figure
\ref{fig:c3ptlcdmeq}.  The square symbols show the simulation result for
$Q_\kappa$ for the SCDM model.  
The triangles denote the
simulation results for the $\Lambda$CDM model, with the error bars
scaled to the same area.
The comparison shows the sensitivity of $Q_\kappa$ to the cosmological 
model, in particular to $\Omega_{m0}$.  } \label{fig:c3ptscdmeq}
\end{figure}

The halo model and the simulations show a flattening of $Q_\kappa$
at small scales, $r\simlt 3'$ (the halo model predicts a slight 
decrease of $Q_\kappa$ with decreasing $r$).  The same feature is 
found for the SCDM model, as shown in Figure \ref{fig:c3ptscdmeq}.  
This is consistent with the hierarchical ansatz for relating higher
order correlations with the 2PCF (e.g. Peebles 1980). For the halo
model this behavior results mainly from the NFW profile and the halo
concentration used (Bullock et al. 2001). 
It will be of great interest to address these issues in detail
using ray-tracing simulations with higher angular resolution that
probes sub-arcminute scales. 

The bottom panel compares the halo model prediction for the 3PCF 
(1-halo term plus PT) with the simulation result.  The 
agreement is excellent, while it is clear that PT substantially
underestimates the 3PCF at scales $\theta\simlt
5'$.  However, in contrast to the middle panel, the standard halo
model prediction (1h+2h+3h terms) denoted by the solid curve displays
equally good agreement.  This explains why the $Q_\kappa$ parameter
is a more sensitive indicator of gravitational clustering
(see Bernardeau et al. 2002b for
an extensive review). For example, the amplitude and configuration
dependence of the $Q$ parameter have been widely
used in the literature in connection with questions of stable
clustering and the hierarchical ansatz (e.g., Peebles 1980; also see
Jain 1997; Ma \& Fry 2000b; TJ03b).  In fact, from a comparison between
the middle and bottom panels, one can see that the $Q_\kappa$ parameter
displays a pronounced 
transition between the non-linear and quasi-linear regimes --
$Q_\kappa\approx 45$ and $\simlt 30$ at $\theta \simlt 3'$ and $\simgt
5'$, respectively.  This feature originates from the
transitions in the 2PCF and 3PCF of the 3D mass distribution 
predicted for CDM structure formation (e.g., Figures 1 and 11 in TJ03b). 

%%%%%%%%%%%%%%%%%%%%%%%%%%%%%%%%%%%%%%%%%%%%%%%%%%%%%%%%%%%%%%%%%%%%%%
%%%%%%%%%%%%%%%%%%%%%%%%%%%%%%%%%%%%%%%%%%%%%%%%%%%%%%%%%%%%%%%%%%%%%%
Figure \ref{fig:c3ptscdmeq} shows a comparison of the halo
model prediction for $Q_\kappa$ (dot-dashed curve) with simulation
results (square symbol) for the SCDM model, as in the middle panel of the
previous figure. The halo model prediction matches the simulation
results.  
For comparison, the triangle symbols denote the simulation results for
the \LCDM model, which shows the strong sensitivity of $Q_\kappa$ to
the cosmological model, especially to $\Omega_{\rm m0}$.

%%%%%%%%%%%%%%%%%%%%%%%%%%%%%%%%%%%%%%%%%%%%%%%%%%%%%%%%%%%%%%%%%%%%%%
\begin{figure}
  \begin{center}
    \leavevmode\epsfxsize=13.cm \epsfbox{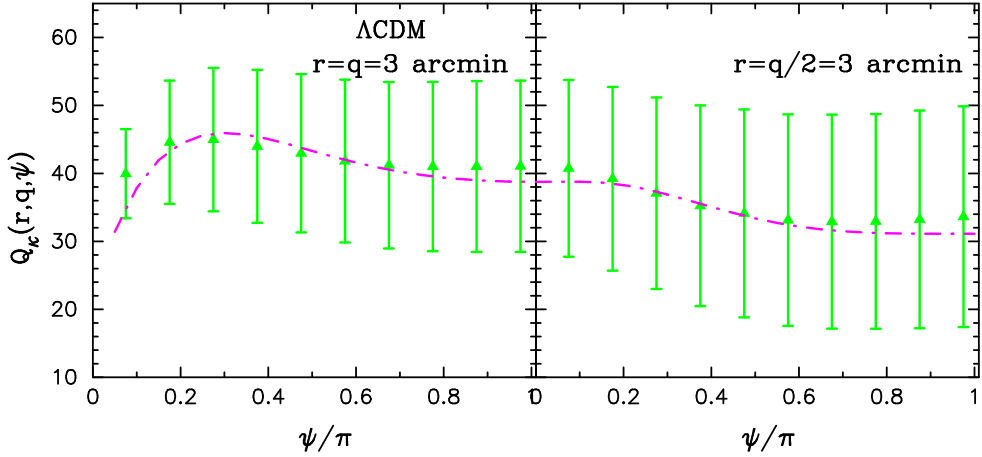}
  \end{center}
\caption{The reduced 3PCF of the convergence field, $Q_\kappa$, 
for the \LCDM model against
triangle configurations parameterized by $r$, $q$ and $\psi$ (see Figure
\ref{fig:triang}).  The left panel shows the result for isosceles
triangles with $r=q=3'$ fixed and varying $\psi$, while the right panel
for more elongated triangles with $r=q/2=3'$. 
} \label{fig:c3ptlcdm}
\end{figure}
%%%%%%%%%%%%%%%%%%%%%%%%%%%%%%%%%%%%%%%%%%%%%%%%%%%%%%%%%%%%%%%%%%%%%%
The accuracy of the halo model is further explored in Figure
\ref{fig:c3ptlcdm}, 
which compares the prediction for the reduced convergence
3PCF, $Q_\kappa$,  with simulation results against triangle 
configurations for the $\Lambda$CDM model.  
Since the halo model employs the spherically
symmetric NFW profile, it is interesting to examine whether or 
not the halo model
can properly describe the configuration dependence of the 3PCF seen in
simulations which include contributions from realistic
aspherical halos with substructure.  The left panel 
shows the result for isosceles triangles with $r=q=3'$ 
against varying the interior angle $\psi$ (see
Figure \ref{fig:triang} for the triangle geometry). The right
panel is for more elongated triangles with $r=2q\approx 3'$. 
Here, the scale $r\approx 3'$ is chosen based on the fact that it 
is in the non-linear regime and unlikely to be affected by the resolution
of the simulations. 
All the plots display excellent agreement between the halo model
predictions and the simulations.

\subsection{The three-point correlation functions of the shear fields}
%%%%%%%%%%%%%%%%%%%%%%%%%%%%%%%%%%%%%%%%%%%%%%%%%%%%%%%%%%%%%%%%%%%%%%
\begin{figure}
  \begin{center}
    \leavevmode\epsfxsize=13.cm \epsfbox{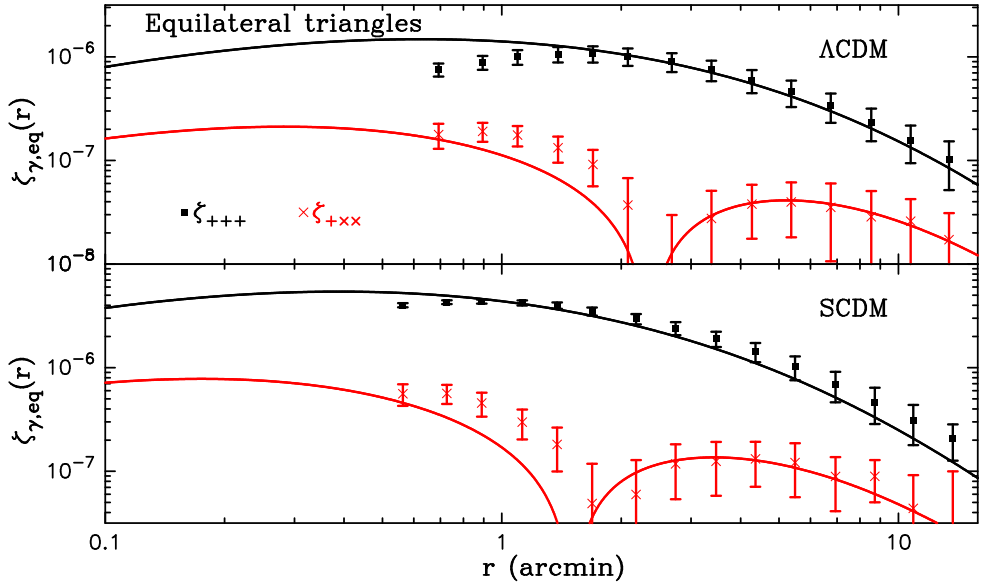}
  \end{center}
\caption{The shear 3PCFs for equilateral triangles against side
length $r$ for the $\Lambda$CDM (upper panel) and SCDM (lower panel)
models, as in the bottom panel of Figure \ref{fig:c3ptlcdmeq}.  From
symmetry considerations, the four parity-odd functions defined by
equation (\ref{eqn:mirror}) vanish for an $E$-mode, 
and three of the parity even functions are equal.
Hence, only the results for $\zeta_{+++}$ and
$\zeta_{+\times\times}$ are shown. Note that the absolute value of
$\zeta_{+\times\times}$ is plotted, since it becomes negative at large
scales. The two solid curves are the halo model predictions, using only 
the 1-halo term. 
} \label{fig:s3pt}
\end{figure}
%%%%%%%%%%%%%%%%%%%%%%%%%%%%%%%%%%%%%%%%%%%%%%%%%%%%%%%%%%%%%%%%%%%%%%

We turn to the shear 3PCFs, which are more complex but are 
important because they are easier to measure from survey 
data than the convergence 3PCF. We will focus on the shear 3PCFs 
rather than the reduced 3PCF, since there is an ambiguity in defining
them for the shear. The ambiguity arises from the fact that the
$+/\times$ components are defined with respect to
the triangle center (see equation (\ref{eqn:pshear3pt}) to form the 3PCF, 
hence there is no clear choice for defining the 2PCFs that enter in the 
denominator of $Q$. One way to do this is to use $\xi_{\gamma,+}$ or
$\xi_{\gamma,\times}$ defined with respect to the side of the triangle
connecting the two vertices, but we will not pursue this. 
Figure \ref{fig:s3pt} compares the halo model predictions with simulation
results for the shear 3PCFs for equilateral triangles, as in the bottom
panel of Figure \ref{fig:c3ptlcdmeq}.  For equilateral
triangles there are only two independent, non-zero 3PCFs: $\zeta_{+++}$
and $\zeta_{+\times\times}=\zeta_{\times+\times}=\zeta_{\times\times+}$, 
while $\zeta_{\times\times\times}=\zeta_{\times++}=
\zeta_{+\times+}=\zeta_{++\times}=0$ (see \S \ref{parity} and also SL03
and TJ03a).  
The figure thus shows only the results for $\zeta_{+++}$
and $\zeta_{+\times\times}$. The upper and lower panels show the
results for the \LCDM and SCDM models, as indicated. Note that the plots
are on a logarithmic scale and the absolute value
of $\zeta_{+\times\times}$ is plotted, since it becomes negative on
large scales.  The halo model predictions include only the 1-halo 
contribution.
The $\zeta_{+++}$ component carries most of the information of the 
lensing signal for equilateral triangles (ZS03; TJ03a). 

Figure \ref{fig:s3pt} shows that the halo model agrees with the simulation 
results for angular scales $r \simgt 2'$ and $\simgt 1'$ for the \LCDM and
SCDM models, respectively. Whether or not the discrepancy
at the smaller scales is genuine is unclear due to the resolution limit
of the simulations (see \S \ref{simul}).  We found that the shear
correlations are more sensitive to the angular resolution of the simulations
than the convergence field. The agreement extends
to scales $\simgt 10'$, though the PT contribution
to the shear 3PCFs is not included,  
in contrast to the convergence 3PCF where the PT
contribution is dominant at $\simgt 5'$.  This agreement is somewhat
surprising, but it might be explained as follows.  The spin-2 field
properties of the shear fields cause cancellations between the shear
3PCFs to some extent, which explains why the shear 3PCF amplitude is
smaller than the convergence 3PCF by an order of magnitude (see Figure
\ref{fig:c3ptlcdm}).  The shear 3PCFs appear to be dominated by the
coherent shear pattern around a halo rather than
filamentary structures in the quasi non-linear regime, which are
described by PT (at least for triangles that are close to equilateral). 

Both the halo model and the simulations show
a flattening and decline in  $\zeta_{+++}$ at small scales. 
The halo model predicts the decline at slightly smaller scales $r\simlt 
0.\!  \! '5$ compared to the simulations.  
The origin of this feature may be explained as
follows.  The shear 3PCFs vanish as one goes to zero triangle size 
(as the three vertices approach the same point), since in this limit
it is equivalent to the shear skewness which vanishes from statistical 
symmetry.  This limiting behavior is another way to see why the amplitude
of the shear 3PCF is smaller than that of the convergence. 
Nevertheless, it is possible that the 
flattening scale of the shear 3PCF reflects a scale related to the
size of typical halos that provide the dominant
contribution. It will be interesting to study this 
feature more precisely with higher resolution simulations.

For $\zeta_{+\times\times}$, the agreement between the
halo model prediction and the simulation result is not good as for
$\zeta_{+++}$. However, the accuracy of the simulation results is likely
to be worse because of the smaller amplitude of $\zeta_{+\times\times}$.  

%%%%%%%%%%%%%%%%%%%%%%%%%%%%%%%%%%%%%%%%%%%%%%%%%%%%%%%%%%%%%%%%%%%%%%
\begin{figure}
  \begin{center}
    \leavevmode\epsfxsize=15.cm \epsfbox{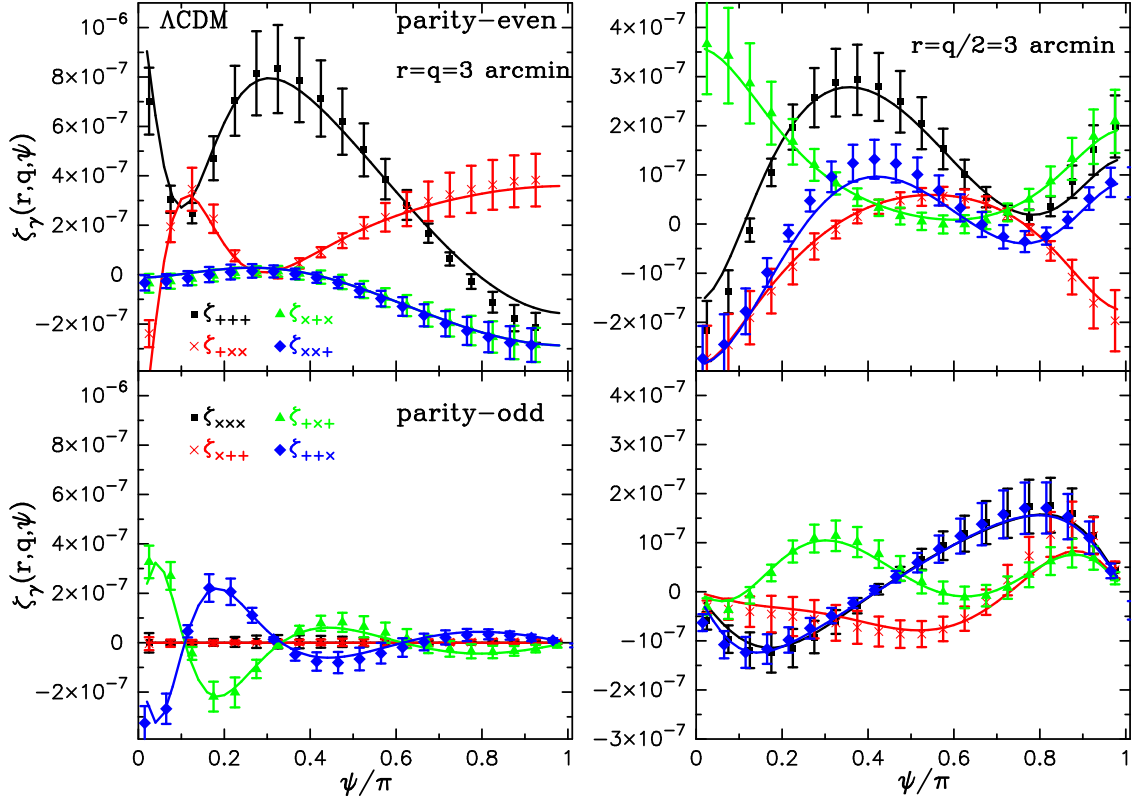}
  \end{center}
\caption{The eight shear 3PCFs for the \LCDM model against triangle
configurations, as in Figure \ref{fig:c3ptlcdm}.  
The upper and lower plots show the results for the
parity-even and -odd functions, respectively. Note that range on the
y-axis for the right panel is about two times smaller than in the left panel.
The solid curves show the halo model predictions for the eight shear
3PCFs, while the symbols are the simulation results as indicated.  
}
\label{fig:s3ptlcdm}
\end{figure}
%%%%%%%%%%%%%%%%%%%%%%%%%%%%%%%%%%%%%%%%%%%%%%%%%%%%%%%%%%%%%%%%%%%%%%
The accuracy of the halo model for the shear 3PCFs is tested in greater
detail in Figure \ref{fig:s3ptlcdm}, which compares model
predictions with simulations for varying triangle configurations
for the \LCDM model.  The upper and lower panels show the parity-even and -odd
functions, respectively. 
As can be seen from the lower left panel, both the halo model predictions and
the simulation results verify that two parity-odd functions vanish:
$\zeta_{\times\times\times}=\zeta_{\times++}=0$ for
isosceles triangles (see \S \ref{parity}).  However, the
other two parity-odd functions do carry lensing
information as pointed out by SL03 (also see TJ03a), as the
$\times$ component is at the vertex bounded by unequal sides.  
For $\psi=\pi/3$ the triangle is equilateral and these two functions 
also vanish.  

The right panel of Figure \ref{fig:s3ptlcdm} shows that all the 
eight functions are non-zero for general triangle configurations 
(SL03; TJ03a).  In contrast to the convergence 3PCF shown in 
Figure \ref{fig:c3ptlcdm}, the shear 3PCFs display complex configuration 
dependences and change sign with varying $\psi$. 
These features reflect the detailed structure of the underlying mass
distribution as well as properties of the spin-2 field generated by
an $E$-mode signal. $\zeta_{+++}$ peaks around $\psi=\pi/3$, where the triangle
configuration is close to equilateral. For the
general triangles shown in the right panel, all the eight functions have
roughly comparable amplitude.  Comparing the left and right
panels shows that more elongated triangles lead to smaller amplitudes
of the shear 3PCFs.  
These results are to be contrasted with the
expectation that the 3D mass 3PCF has higher amplitude for 
elongated triangles on large scales ($\simgt 10$Mpc),
reflecting the dominance of anisotropic structures in the perturbative
regime (Scoccimarro \& Frieman 1999; Bernardeau et al. 2002b; also see Figure 5
in TJ03b).  The range $\psi=[0,\pi]$ is shown in our figures; 
one should keep in mind the relation
$\zeta(\psi)=\pm\zeta(2\pi-\psi)$ for the lensing $E$-mode, where $+$ or
$-$ sign is taken for the parity-even or -odd functions, respectively
(\S \ref{parity}; also see Figure 3 in TJ03a ). 

%%%%%%%%%%%%%%%%%%%%%%%%%%%%%%%%%%%%%%%%%%%%%%%%%%%%%%%%%%%%%%%%%%%%%%
\begin{figure}
  \begin{center}
    \leavevmode\epsfxsize=15.cm \epsfbox{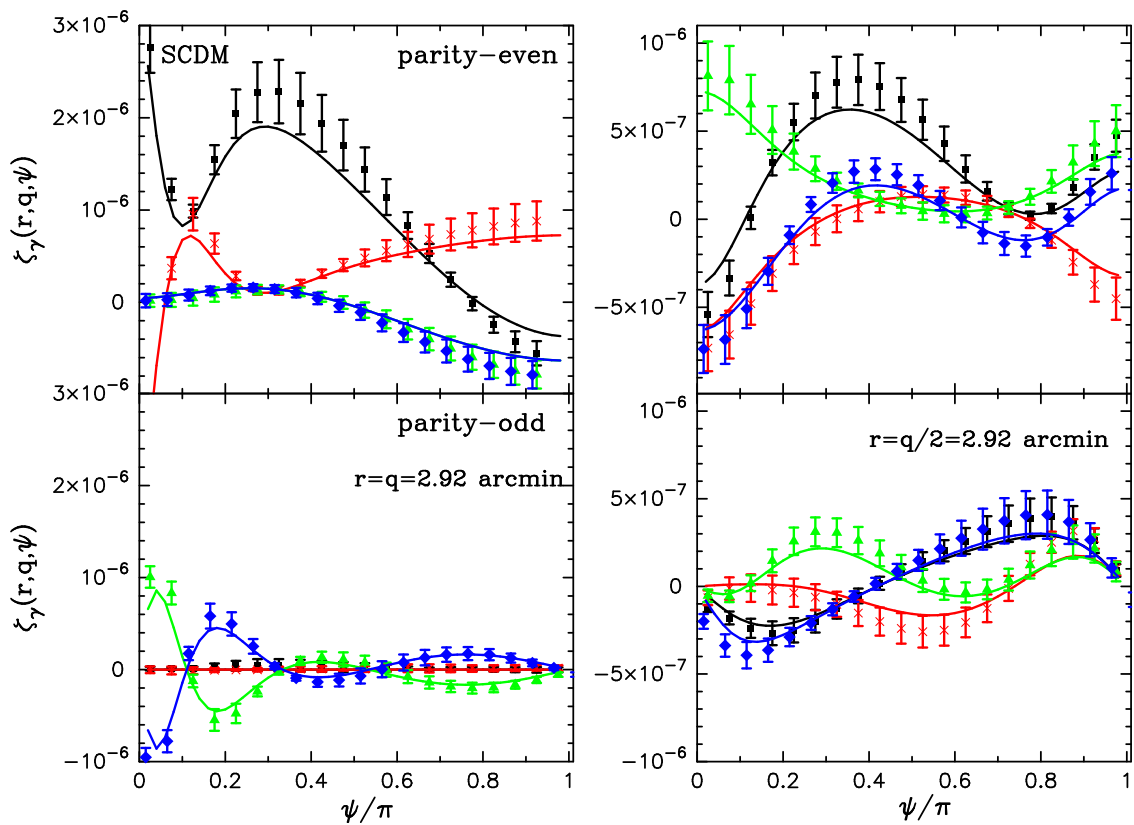}
  \end{center}
\caption{As in the previous figure, for the SCDM model.
}  \label{fig:s3ptscdm}
\end{figure}
%%%%%%%%%%%%%%%%%%%%%%%%%%%%%%%%%%%%%%%%%%%%%%%%%%%%%%%%%%%%%%%%%%%
\begin{figure}
  \begin{center}
    \leavevmode\epsfxsize=15.cm \epsfbox{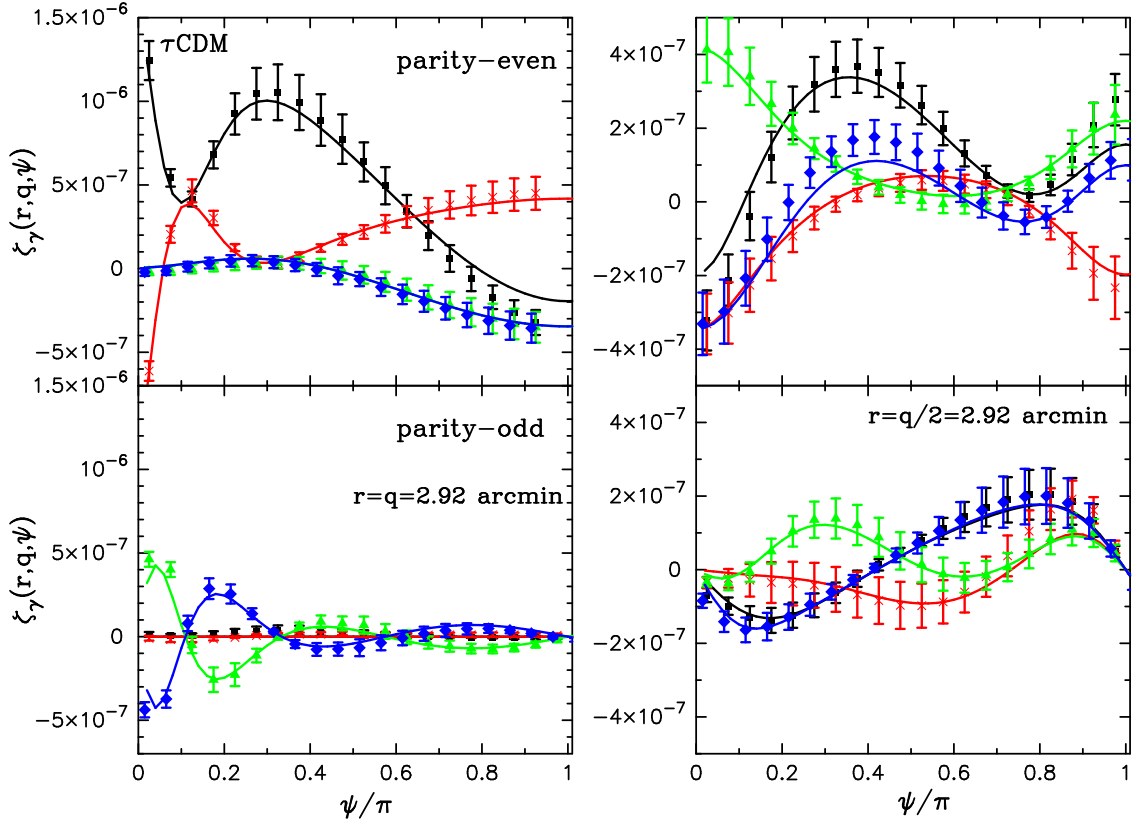}
  \end{center}
\caption{As in the previous figure, for the $\tau$CDM model. 
The cosmological parameters for the $\tau$CDM model are the same as
for the SCDM model, except for the shape parameter $\Gamma$ which
is $0.21$, while it is $0.5$ for SCDM. 
}  \label{fig:s3pttcdm}
\end{figure}
%%%%%%%%%%%%%%%%%%%%%%%%%%%%%%%%%%%%%%%%%%%%%%%%%%%%%%%%%%%%%%%%%%%%
To further check the accuracy of the halo model for different cosmological 
models, Figures \ref{fig:s3ptscdm} and \ref{fig:s3pttcdm} show the results for
the SCDM and $\tau$CDM models, respectively.  Note that the cosmological
parameters for the $\tau$CDM model are the same for the SCDM model except
for the shape parameter $\Gamma=0.21$ ($\Gamma=0.5$ for the SCDM model). 
The angular resolution of the two models is the same
(Jain et al. 2000).  The halo model predictions again
match the simulation results for all eight functions.  
A comparison of Figure \ref{fig:s3ptscdm} and $\ref{fig:s3pttcdm}$
shows that the amplitude of the shear 3PCF depends on the shape of
the input linear mass power spectrum. In the context of the halo model
picture, this is captured by the dependence of the halo mass function on
the power spectrum shape on the angular scales we considered.  In
summary, from Figures {\ref{fig:s3ptlcdm}}-\ref{fig:s3pttcdm}, one can
see that the shear 3PCF amplitudes are sensitive to the cosmological
models, but the oscillatory shape of each function is quite similar for
the three CDM models.

\subsection{Summary:  the halo model accuracy for predicting the lensing 
statistics} \label{summodel}

Here we summarize the results shown in Figures
\ref{fig:2ptlcdm}-\ref{fig:s3pttcdm}, which investigated the
2PCFs and the 3PCFs of the convergence and shear fields.  It has
been shown that the halo model predictions are in remarkable agreement
with the simulation results for all these statistical quantities over
the angular scales we have considered.  In particular, the halo model
reproduces the amplitudes and the complex configuration
dependences for the eight shear 3PCFs.  
We have used only the 1-halo term and not adjusted any model parameters 
to get this agreement.  This implies
that the shear 3PCFs result from correlations between
the tangential shear pattern around a single  NFW profile,
as pointed out by TJ03b and ZS03.  The agreement is striking, since the
halo model rests on simplified assumptions of smooth, spherical halos 
while the halos in CDM simulations are aspherical and contain substructure
(e.g., Jing \& Suto 2002). The projection of halos
oriented in different directions and the nonlocal properties of the 
shear appear to dilute the effect of some of the detailed structure of
halos and make the spherically symmetric profile
a good approximation in the statistical sense.  

It is not clear whether the
agreement remains on smaller angular scales ($\simlt 1'$), due to the
resolution limitation of the simulations we have used. It  is of 
great interest to address this issue using higher resolution simulations. 
On these scales new ingredients may be needed for the halo model as well, 
such as the inclusion of substructure (Sheth \& Jain 2002). 
The agreement we have shown holds for different
cosmological models (\LCDM, SCDM and $\tau$CDM models).  This implies
that the cosmological model dependences can be captured through the
spherical collapse model and the mass function used in the halo
model.  These results lead us to conclude that the halo model provides
an analytical method for predicting higher order 
lensing statistics with sufficient accuracy for our purposes.

\section{Measurement of halo profile parameters from shear correlations}
\label{profile}

In the following we address the question: how can
measurements of the lensing 2PCF and 3PCFs be used to constrain halo
model parameters? In particular, we focus on the halo profile
parameters: the inner slope $\alpha$ for the generalized NFW profile
in equation (\ref{eqn:nfw}) and the halo concentration $c$ in equation
(\ref{eqn:conc}). We will show that forthcoming lensing surveys can put
stringent constraints on these parameters.  For this study we will use
the convergence 2PCF and 3PCF for simplicity, although the shear
correlations are the direct observable. To use the shear 3PCFs, it is
necessary to combine all the eight 3PCFs. Since the lensing information
obtained combining the eight shear 3PCFs are related to 
the convergence 3PCF, the results we obtain should hold as a first
approximation. We also note that 
the statistics of the convergence field may be directly measured
from future space based surveys such as the one proposed for 
the SNAP satellite (Jain 2002).

\subsection{Sensitivity of the lensing 2PCF and 3PCF to profile parameters} 

In TJ03b, we showed that the 2PCF and 3PCF of the 3D mass distribution
at small scales are sensitive to halo profile parameters
(see Figures 12-14 in TJ03b). It was shown that the 3PCF is more
sensitive to the halo profile than the 2PCF. This
is expected to hold for lensing statistics also, since the lensing
fields are projections of the 3D mass distribution.

We derive analytical expressions in Appendix \ref{conv} 
for the halo convergence 
for $\alpha=0, 1$ and $2$ in the generalized NFW profile of 
equation (\ref{eqn:nfw}).  To compute the
convergence 2PCF and 3PCF for general $\alpha$ ($0\le\alpha\le2$) we
simply interpolate from the 1-halo term predictions for $\alpha=0, 1$ and $2$,
using formulae similar to equation (42) in TJ03b; the interpolation 
is expected to work to better than $10\%$. 

%%%%%%%%%%%%%%%%%%%%%%%%%%%%%%%%%%%%%%%%%%%%%%%%%%%%%%%%%%%%%%%%%%%
\begin{figure}
  \begin{center}
    \leavevmode\epsfxsize=17.cm \epsfbox{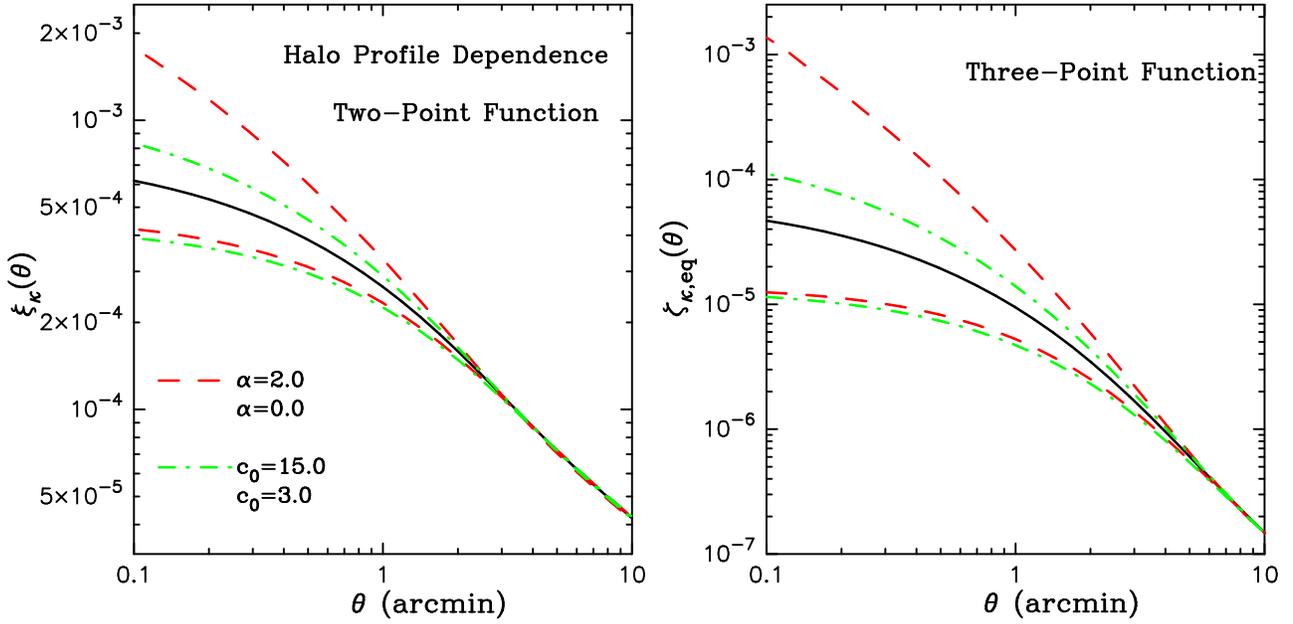}
  \end{center}
\caption{The dependences of the 2PCF
(left panel) and 3PCF (right panel) of the convergence field
on halo profile parameters.  The upper
and lower dashed curves show the halo model predictions with
$\alpha=2$ and $0$ for the inner slope parameter of the generalized NFW
profile of equation (\ref{eqn:nfw}). The upper and lower dot-dashed
curves are the results for the concentration parameter $c_0=15$ and $3$.
The solid curve is the halo
 model result for our reference model, 
the NFW profile with $\alpha=1, c_0=9$. 
}  \label{fig:3ptalpha}
\end{figure}
%%%%%%%%%%%%%%%%%%%%%%%%%%%%%%%%%%%%%%%%%%%%%%%%%%%%%%%%%%%%%%%%%%%%
%%%%%%%%%%%%%%%%%%%%%%%%%%%%%%%%%%%%%%%%%%%%%%%%%%%%%%%%%%%%%%%%%%%
\begin{figure}
  \begin{center}
    \leavevmode\epsfxsize=17.cm \epsfbox{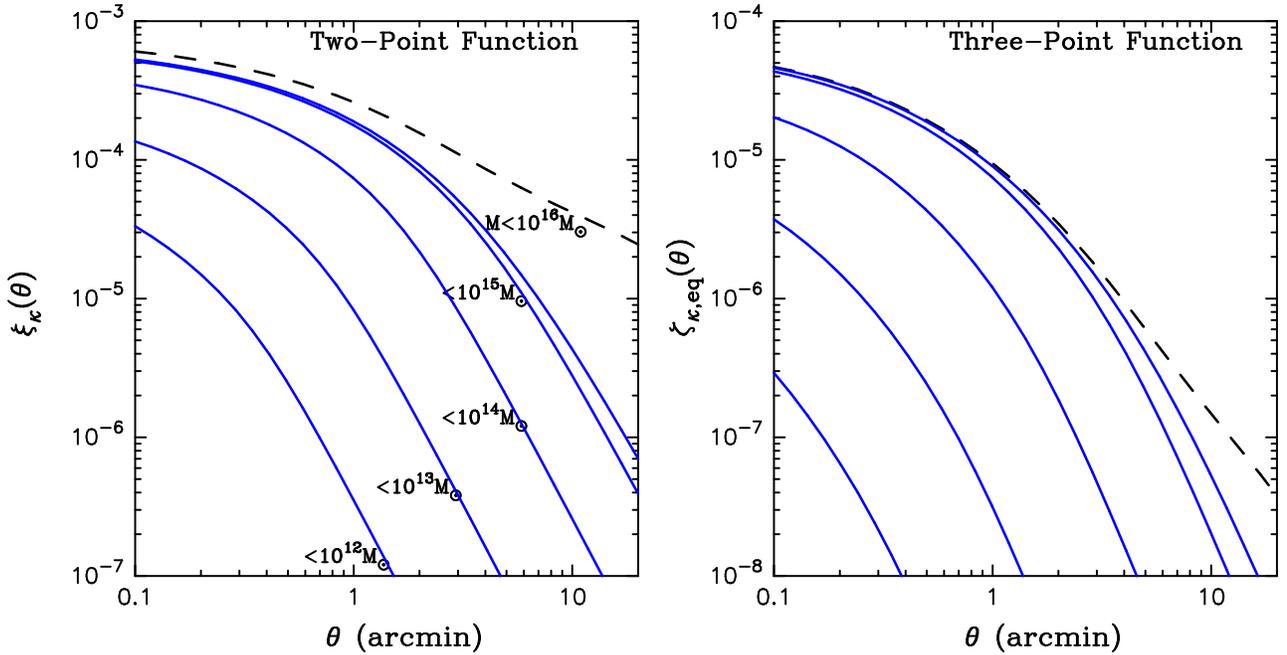}
  \end{center}
\caption{The dependences of the 1-halo term contribution to the
convergence 2PCF (left panel) and 3PCF (right panel) on the maximum mass
cutoff used in the calculation. From top to bottom, the five solid
curves are the results for a maximum mass of $10^{16}$, $10^{15}$,
$10^{14}$, $10^{13}$ and $10^{12}M_\odot$, as indicated.  The dashed
curve shows the total halo model prediction. Most contributions arise
from halos of $M>10^{13}M_\odot$ on the scales we have considered, and
the 3PCF is more sensitive to massive halos than the 2PCF.  }
\label{fig:3ptMmax}
\end{figure}
%%%%%%%%%%%%%%%%%%%%%%%%%%%%%%%%%%%%%%%%%%%%%%%%%%%%%%%%%%%%%%%%%%%%

Figure \ref{fig:3ptalpha} shows the sensitivity of the convergence 2PCF
(left panel) and 3PCF (right panel) to the inner slope parameter
$\alpha$ and to the concentration parameter $c_0$; in our
parameterization $c=c_0(1+z)^{-1}(M/M_\ast)^{-\beta}$ with $\beta=0.13$.
Increasing $\alpha$ or $c_0$ steepens the 2PCF and 3PCF for scales
$\theta\simlt 3'$, since it increases the density profile
in the inner region, $r\le r_{\rm vir}/c$.  
The curves coincide with each other at large scales, because the
outer region has the slope $r^{-3}$ for all $\alpha$.  
The 3PCF is more sensitive to modifications of the halo profile than
the 2PCF; this is because of the extra power of the halo profile in the
1-halo term.

Figure \ref{fig:3ptMmax} shows the mass range of halos that contribute
to the lensing statistics on scales of our interest. The figure shows 
the dependences on the maximum mass cutoff in the halo model calculation
on the convergence 2PCF and 3PCF. Massive halos with
$M>10^{13}M_\odot$ provide more than $80\%$ of the
contribution over the scales we have considered. At
smaller angular scales, less massive halos are more
relevant.  One can also see that the 3PCF is more sensitive to 
massive halos than the 2PCF. 

\subsection{Covariance of the 2PCF and 3PCF estimators}

To rigorously extract parameter information from cosmic shear measurements,  
it is crucial to consider the covariance between the lensing statistics 
in different bins. This issue for the shear 2PCFs has been investigated
by Schneider et al. (2002b).  It was shown that there are strong
correlations between the shear 2PCFs in different bins on small scales. 
Extending their method, we present
analytic expressions for the covariance of the convergence 3PCF and
2PCF. Note that the method described below can be extended to compute
the covariance for the shear 3PCFs, if one accounts for the spin-2 phase
factors of the shear fields and the projection operators.

Following Schneider et al. (2002b), an estimator of the convergence 3PCF
from realistic data can be expressed as
\begin{equation}
\zeta^{\rm est}_{\kappa}(r,q,\psi)=\frac{1}{N_{\rm trip}}
\sum_{ijk}\kappa_i\kappa_j\kappa_k \Delta_{ijk}(r,q,\psi), \hspace{2em}
N_{\rm trip}=\sum_{ijk}\Delta_{ijk}(r,q,\psi),
\end{equation}
where index $i$ denotes source galaxies and $N_{\rm trip}$ is the
number of triplets of galaxies that form a given triangle configuration
within the bin width. 
The function $\Delta_{ijk}(r,q,\psi)$ is the selection function for
triplets: it is unity if the three points with indices $i,j$ and $k$ are
in the triangle configuration and zero otherwise.  In the weak
lensing limit, the observed convergence field can be expressed as the 
sum of the lensing signal and the noise contamination due to intrinsic
ellipticities: $\kappa =\kappa^{\rm lens}+n_\epsilon$.  
The expectation value of the estimator above
is obtained by averaging over source ellipticities and performing
an ensemble average of the convergence field, denoted by $\skaco{\cdots}$. 
We will use the notation 
$\skaco{\zeta^{\rm est}_{\kappa}}\equiv \zeta_{\kappa}$.

\begin{figure}
  \begin{center}
    \leavevmode\epsfxsize=17.cm \epsfbox{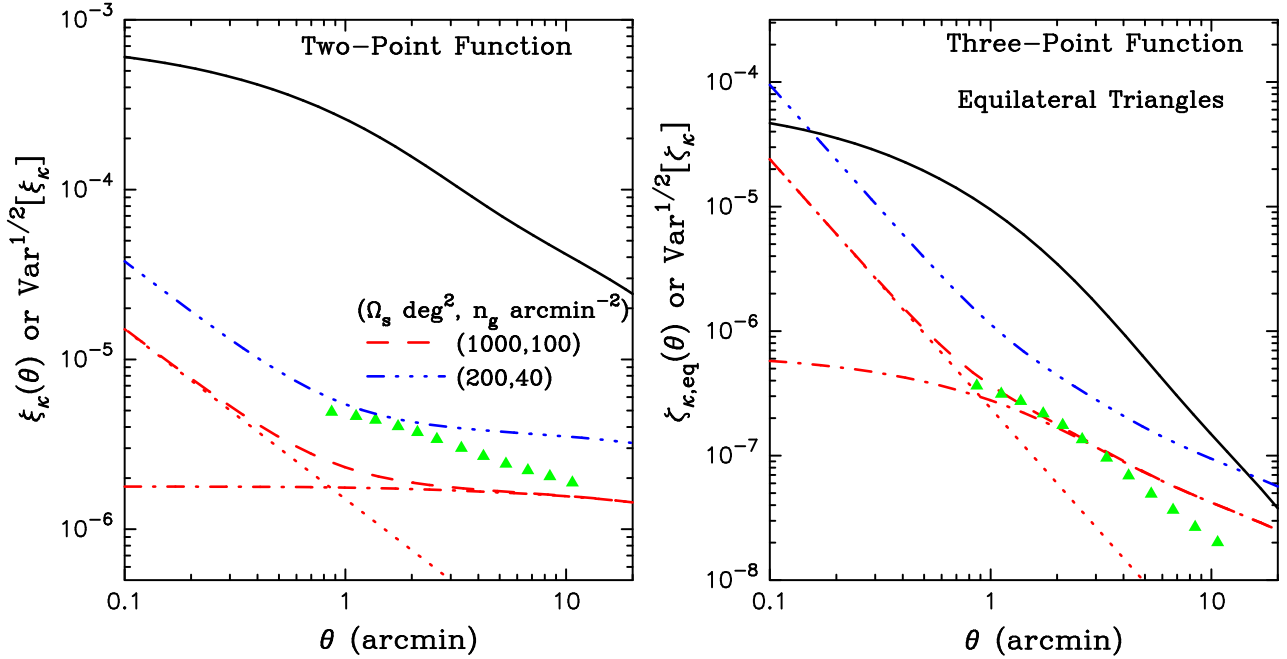}
  \end{center}
\caption{Error estimates for measurements of the convergence 2PCF
(left panel) and 3PCF (right panel) for the \LCDM model.
The error estimate includes both the shot noise due
to intrinsic ellipticities, with rms $\sigma_{\epsilon}=0.4$, 
and sample variance.
Two cases for survey area $\Omega_{\rm s}$ and
number density of source galaxies $n_g$ are shown. 
For $(\Omega_{\rm s},n_g)=(1000~ {\rm
deg}^2, 100~ {\rm arcmin}^{-2})$, the dotted and dot-dashed curves show
the shot noise contribution and the sample variance separately: 
shot-noise  dominates on sub-arcminute scales. The solid
curve shows the halo model prediction for the 2PCF or the 3PCF.  The
triangular symbols show the sample variance estimated from simulations
for $\Omega_{\rm s}=1000$.
}
\label{fig:sncov}
\end{figure}

The covariance of the convergence 3PCF is defined as
\begin{equation}
{\rm Cov}[\zeta_{\kappa},\zeta'_{\kappa}]
=\skaco{(\zeta^{\rm est}_{\kappa}-\zeta_\kappa)
(\zeta^{{\rm est}\prime}_\kappa-\zeta_{\kappa}^\prime)}
=\skaco{\zeta^{\rm est}_{\kappa}\zeta^{{\rm est}\prime}_\kappa}-
\zeta_\kappa\zeta_\kappa^\prime. 
\end{equation}
where we have simplified the notation so that $\zeta_\kappa$ and
$\zeta'_\kappa$ denote $\zeta_{\kappa}(r,q,\psi)$ and
$\zeta_{\kappa}(r',q',\psi')$, respectively. Similarly as done in
Schneider et al. (2002b),
the first term on the r.h.s. of the above equation can be rewritten as
\begin{equation}
\skaco{\zeta_{\kappa}^{\rm est}\zeta^{{\rm est}\prime}_\kappa}
=\frac{1}{N_{\rm trip}(r,q,\psi)N_{\rm trip}(r',q',\psi')}
\sum_{ijkabc}\Delta_{ijk}(r,q,\psi)\Delta_{abc}(r',q',\psi')
\skaco{\kappa_i\kappa_j\kappa_k\kappa_a\kappa_b\kappa_c}.
\label{eqn:cov}
\end{equation}
The covariance estimate thus requires an evaluation of the 6-point
correlation function of the observed convergence field. If the noise
field and the convergence field are statistically uncorrelated, the
6-point correlation function can be expressed as
\begin{eqnarray}
\skaco{\kappa_i\kappa_j\kappa_k\kappa_a\kappa_b\kappa_c}
\approx \sigma_\epsilon^6(\delta_{ia}\delta_{jb}\delta_{kc}+5~ {\rm perm.})
+\left[
  \xi_\kappa(\theta_{ij})\xi_{\kappa}(\theta_{ka})\xi_{\kappa}(\theta_{bc})
+14~ {\rm perm.}\right],
\label{eqn:6pt}
\end{eqnarray}
where $\sigma_\epsilon$ is the rms of the intrinsic ellipticities
\footnote{Van Waerbeke (2000) discussed a more accurate 
description of the noise field for the convergence, taking into account 
the smoothing kernel used for the reconstruction. } 
and we have ignored terms of $O(\sigma_\epsilon^4 \xi_\kappa)$ and
$O(\sigma_\epsilon^2\xi^{2}_\kappa)$, which are relevant only in	
the transition regime between those in which the first
or second term on the r.h.s of equation (\ref{eqn:6pt}) dominate. 
In addition, the second term ignores
the non-Gaussian contribution that is due to the connected parts of the
three-, four- and six-point correlation functions, and thus
underestimates the sample variance.  For a more conservative estimate
of the non-Gaussian contribution, one 
might replace the connected parts of the three-, four- and six-point
functions with their unconnected parts, providing
$\skaco{\kappa^{\rm cs}_i\cdots\kappa^{\rm cs}_c}
=8\left[\xi_\kappa(\theta_{ij})\xi_{\kappa}(\theta_{ka})
\xi_{\kappa}(\theta_{bc}) + 14~ {\rm perms.}\right]$. However, in this
paper we use the above equation for simplicity. This is likely to
be a good approximation for two reasons. One, our
estimates are consistent with the results from simulations which
contain the full non-Gaussian contribution, as shown in Figure
\ref{fig:sncov}.  Moreover the shot noise contribution dominates
the covariance on sub-arcminute scales, which provide the 
main constraints on halo profiles. 

Inserting equation (\ref{eqn:6pt}) into equation (\ref{eqn:cov}) and
performing an ensemble average over source galaxy positions yields
\begin{equation}
{\rm Cov}[\zeta(r,q,\psi),\zeta(r',q',\psi')]=
\left[\Delta\zeta_N(r,q,\psi)
\right]^2\delta(\{r,q,\psi\}-
\{r',q',\psi'\})+{\cal R}(r,q,\psi,r',q',\psi'),
\label{eqn:covresult}
\end{equation}
with 
\begin{eqnarray}
\Delta\zeta_N(r,q,\psi)&\equiv&
\frac{\sigma_\epsilon^3}{\sqrt{N_{\rm trip}}}\nonumber\\
&=&2.40\times 10^{-7}
\left(\frac{\sigma_e}{0.4}\right)^{3}\left(\frac{\Omega_{\rm s}}
{10^3 ~ {\rm deg}^2}\right)^{-1/2}\left(\frac{n_g}{10^2 ~ 
{\rm arcmin}^{-2}}\right)^{-3/2}\nonumber\\
&&\times\left(\frac{r}{1'}\right)^{-1}
\left(\frac{\Delta \ln r}{0.1}\right)^{-1/2}
\left(\frac{q}{1'}\right)^{-1}
\left(\frac{\Delta \ln q}{0.1}\right)^{-1/2}
\left(\frac{\Delta\psi/\pi}{0.2}\right)^{-1/2},
\label{eqn:3ptnoise}\\
{\cal R}(r,q,\psi,r',q',\psi')&\equiv& 
\frac{1}{2\pi \Omega_{\rm s}}\int_0^{s_{{\rm max}}} 
\! \! sds \int^{2\pi}_0\! d\varphi_r
\int^{2\pi}_0 \! d\varphi_r^\prime 
\left[\xi(r)\xi(\bm{s}-\bm{r}-\bm{q})
\xi(\bm{q}') + \mbox{ 14 perms.}
\right](s,\varphi_r,\varphi_r'; r,q,\psi,r',q',\psi'),
\label{eqn:cscov}
\end{eqnarray}
where $\Omega_{\rm s}$ and $n_g$ are the survey area and the number
density of source galaxies, respectively, and $\Delta r$, $\Delta q$ and
$\Delta \psi$ denote the bin widths for the three parameters
$(r,q,\psi)$ that specify the triangle configuration. The notation used
in equation (\ref{eqn:cscov}) is $s_{\rm max}=\sqrt{\Omega_{\rm s}/\pi}$,
$\bm{r}=r(\cos\varphi_r,\sin\varphi_r)$,
$\bm{q}=q(-\cos(\psi-\varphi_r),\sin(\psi-\varphi_r))$ and so on.  We
have assumed that the survey geometry does not affect covariance
estimation -- a good approximation as long as we consider
sufficiently small scales compared to the survey size.  The first term
on the r.h.s of equation (\ref{eqn:covresult}) denotes the shot noise
contribution due to the intrinsic ellipticities, where the function
$\delta(\{r,q,\psi\}-\{r',q',\psi'\})$ is defined to be unity if
$r=r'$, $q=q'$ and $\psi=\psi'$ within the bin widths, and
zero otherwise. The derivation of this term requires an estimate of the
triplet number for a given triangle configuration, which we estimate as
$N_{\rm trip}=n_g \Omega_{\rm s} \times n_g \pi r(\Delta r)\times n_g
q(\Delta q)(\Delta \psi) $, where the first, second and third factors
denote the number of galaxies at the
vertices, 1, 2 and 3, respectively (see \S \ref{3ptalg}).  The second
term ${\cal R}$ in equation (\ref{eqn:covresult}) denotes the sample
variance.  Several interesting points made by Schneider
et al. (2002b) for the 2PCF hold for the convergence
3PCF as well.  (1) All the terms are proportional to $\Omega_{\rm s}^{-1}$, and
therefore the relative contribution of the terms is independent of 
survey area, if the area is sufficiently large.  (2) The sample variance
${\cal R}$ does not depend on the survey particulars such as $n_{g}$
and $\sigma_\epsilon$. Further, ${\cal R}$ is independent of the bin
widths.  This implies that combining the 3PCFs in different bins cannot
reduce the sample variance.  This is indeed verified by ray-tracing
simulations. (3) The off-diagonal components of the covariance arise
only from the sample variance ${\cal R}$.

To compute the sample variance ${\cal R}$ for a given cosmological model
using equation (\ref{eqn:covresult}), we first make a table of the model
predictions of the convergence 2PCF as a function of the separation
angle. Then, we can use the same table to compute the sample variance
between the two convergence 3PCFs of {\em any} triangle configurations.
Therefore, this method is much more tractable than a direct
implementation including contributions from the three-, four- and six-point
functions, where we have to account for their configuration dependences
in the integration of equation (\ref{eqn:cscov}). An alternative way to
estimate the covariance is to use ray-tracing simulations. However,
to do this rigorously requires an adequate number of independent
realizations, since sample variance is large for the higher-order moments.  
The PTHalos method recently proposed by Scoccimarro \& Sheth (2002) can be a 
powerful tool for such an approach.

Likewise, we can derive the covariance of the convergence 2PCF as
\begin{equation}
{\rm Cov}[\xi(r),\xi(r')]=[\Delta\xi_N(r)]^2\delta(r-r')
+{\cal R}_{\rm 2pt}(r,r')
\label{eqn:2ptcov}
\end{equation}
with 
\begin{eqnarray}
\Delta \xi_N&\equiv&\frac{\sigma^2_\epsilon}{\sqrt{N_{\rm pair}}}
= 1.5\times 10^{-6}
\left(\frac{\sigma_e}{0.4}\right)^{2}\left(\frac{\Omega_{\rm s}}
{10^3 ~ {\rm deg}^2}\right)^{-1/2}\left(\frac{n_g}{10^2 ~ 
{\rm arcmin}^{-2}}\right)^{-1}\left(\frac{r}{1'}\right)^{-1}
\left(\frac{\Delta \ln r}{0.1}\right)^{-1/2},\label{eqn:2ptnoise}\\
{\cal R}_{\rm 2pt}(r,r')&\equiv&\frac{1}{2\pi\Omega_{\rm s}}
\int_0^{s_{\rm max}}\!\! sds\int^{2\pi}_0\!\! d\varphi_r
\int^{2\pi}_0\!\!d\varphi_r'\left[
\xi(\bm{s})\xi(|\bm{s}+\bm{r}'-\bm{r}|)+\xi(|\bm{s}+\bm{r}'|)\xi(|\bm{s}-\bm{r}|)
\right],
\end{eqnarray}
where the number of pairs are estimated as $N_{\rm pair}=n_g\Omega_{\rm
s}\times n_g\pi r(\Delta r)$.  

%%%%%%%%%%%%%%%%%%%%%%%%%%%%%%%%%%%%%%%%%%%%%%%%%%%%%%%%%%%%%%%%%%%
%%%%%%%%%%%%%%%%%%%%%%%%%%%%%%%%%%%%%%%%%%%%%%%%%%%%%%%%%%%%%%%%%%%%%
Figure \ref{fig:sncov} plots the square root of the diagonal component
of the covariance matrix for the convergence 2PCF (left panel) and 3PCF
(right panel) for the \LCDM model. This is an estimate of the error
on the 2PCF and 3PCF measurements from a lensing survey.  We
consider two cases,  specified by the survey area
$\Omega_{\rm s}$ and the number density of source galaxies $n_g$. The
dashed and broken curves show the results for
$(\Omega_{s},n_g)=(1000,100)$ and $(200,40)$ in units of ${\rm
degree}^2$ and ${\rm arcmin}^{-2}$, respectively. The former is expected
from future imaging survey, while the latter applies
for a survey like the CFHT Legacy Survey, which has just begun.  
For $(\Omega_{s},n_g)=(1000,100)$, the dotted and dot-dashed curves show the
shot noise contamination and the sample variance separately, implying
that the shot noise provides the dominant contribution at small scales
$\simlt 1'$.  The triangle symbols denotes the simulation results for the
sample variance for $\Omega_{\rm s}=1000$ degree$^2$. Note that the
simulation result is computed from 36 realizations of 
a 11.7 degree$^2$ simulated map, and is then scaled 
as $\propto \Omega_{\rm s}^{-1/2}$. Out analytic estimates are 
consistent with the simulation results, within the rather large 
error bars of the latter (not plotted to preserve clarity).  

The solid curve in each panel of Figure \ref{fig:sncov} shows 
the halo model prediction for the
2PCF or the 3PCF.  Comparing the solid curves with the error estimate
gives the signal-to-noise ($S/N$) ratio for
measuring the 2PCF and 3PCF. Clearly these survey parameters will allow 
for measurements of shear correlations at high significance, even on
sub-arcminute scales.  We note that the $S/N$ estimate for the 3PCF is
only for one triangle configuration -- we can combine the
3PCFs from different configurations to improve the $S/N$ at these scales.

\subsection{Constraints on $\alpha$ and $c_0$}

Next we apply the covariances computed above to demonstrate how
combined measurements of the 2PCF and 3PCF can be used to
constrain parameters of the halo profile. We use the standard $\chi^2$
statistic, expressed for our case as:
\begin{equation}
\chi^2=\chi_{\rm 2pt}^2+\chi_{\rm 3pt}^2,
\label{eqn:chi2}
\end{equation}
where
\begin{eqnarray}
\chi^2_{\rm 2pt}&\equiv&\sum_{i\le j}(\hat{\xi}_i-\xi_i)
[{\rm Cov}(\xi)]^{-1}_{ij}(\hat{\xi}_j-\xi_j),
\nonumber\\
\chi^2_{\rm 3pt}&\equiv&\sum_{i\le j}
(\hat{\zeta}_i-\zeta_i)[{\rm Cov}(\zeta)]^{-1}_{ij}(\hat{\zeta}_j-\zeta_j),
\end{eqnarray}
where $\hat{\xi}$ and $\hat{\zeta}$ denote the 2PCF and 3PCF for the
fiducial model and $[{\rm Cov}]^{-1}_{ij}$ denotes the inverse
of the covariance matrix.  The index $i$ runs among
different bins for $\xi$ and $\zeta$.  In equation (\ref{eqn:chi2}), we
have ignored the covariance between the 2PCF and 3PCF for simplicity. 
We make this approximation because only the sample variance 
contributes to this covariance (the shot noise contribution vanishes as 
it involves the fifth power of the intrinsic ellipticities), which
is small on sub-arcminute scales as discussed above. 

So far we have used the parameters $r$, $q$ and $\psi$ to describe 
triangle configurations.  To perform the $\chi^2$ fitting
we employ an alternative set of parameters used in the
literature (e.g., Peebles 1980):
\begin{equation}
r\equiv \theta_{12},\hspace{2em}
u\equiv \frac{\theta_{23}}{\theta_{12}},\hspace{2em}
v\equiv \frac{\theta_{31}-\theta_{23}}{\theta_{12}}, 
\end{equation}
with the condition $\theta_{12}\le\theta_{23}\le\theta_{31}$, which
imposes the constraints $u\ge 1$ and $0\le v\le 1$.  
Different sets of $r$, $u$ and $v$ correspond to different triangles, 
so we do not have to worry about double-counting. 

We treat the inner slope parameter
$\alpha$ and the halo concentration normalization $c_0$ as free
parameters. We fix $\beta=0.13$ (the mass dependence of the concentration)
and we set the other model parameters to be those for the \LCDM
model.  We discuss below why this choice is a good first attempt at
the use of correlation statistics to measure halo profiles. As shown 
in Figure \ref{fig:3ptalpha}, these profile
parameters are sensitive to the lensing statistics at small scales
$\simlt 2'$.  We therefore restrict ourselves to these scales.  We
consider 10 logarithmic bins in $r=[0.\!\!'12,2']$ with the bin width $\Delta
r/r=0.1$. For the 3PCF, we consider 5 bins each for $u$ and $v$: 
$u=1,2,3,4,5$ and $v=0.1,0.3,0.5,0.7,0.9$.  Thus we use 10 bins for
the 2PCFs and 250 for the 3PCF.  How the binning
affects the fitting of model parameters must be carefully examined, to
avoid over- or under-estimating the constraints (e.g, Scoccimarro \&
Frieman 1999). Since we correctly take into account the covariance for the
2PCF and 3PCF, including the off-diagonal components, 
we avoid over-constraining the parameters. 
The triplet number used for the shot noise evaluation
is estimated for the binning we consider as $N_{\rm
trip}=n_g^3\Omega_{\rm s}\times \pi r\Delta r \times qr\Delta u\Delta v$
with $\Delta u=1$ and $\Delta v=0.2$. To save computational expense for
the sample variance of the 3PCF, we ignore the dependence on the $v$
parameter -- we compute the sample variance for different bins of $r$
and $u$, but with $v=0.5$ fixed, resulting in $50\times 50$ computations
for the sample variance. This is adequate, since the
configuration dependence is weak as shown in Figure \ref{fig:c3ptlcdm}.

%%%%%%%%%%%%%%%%%%%%%%%%%%%%%%%%%%%%%%%%%%%%%%%%%%%%%%%%%%%%%%%%%%%
\begin{figure}
  \begin{center}
    \leavevmode\epsfxsize=16.cm \epsfbox{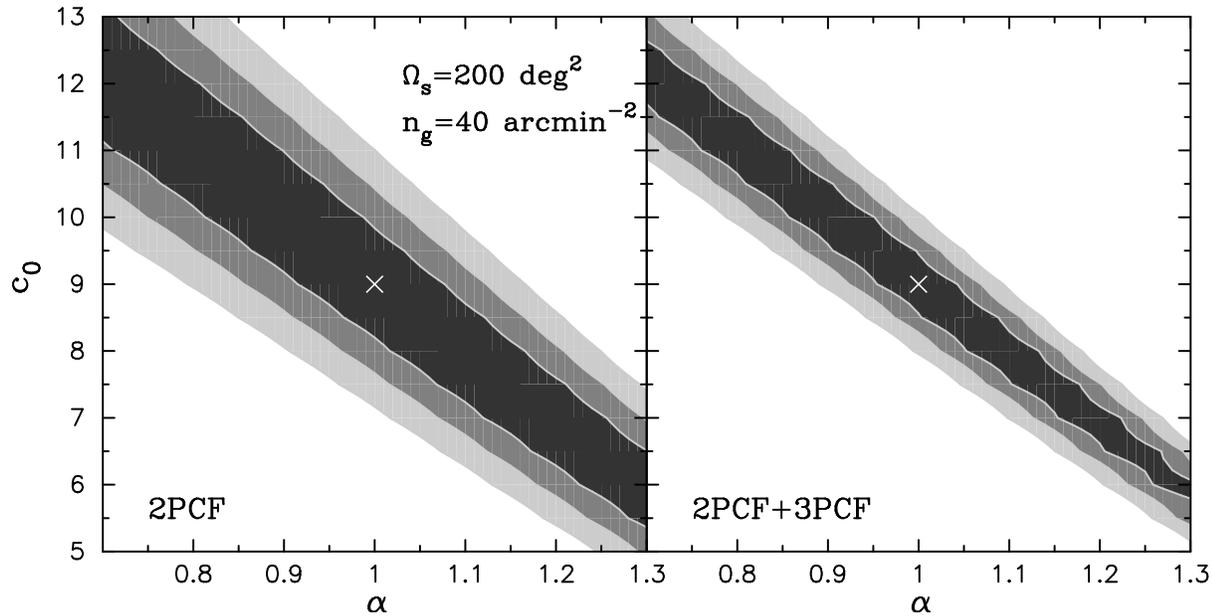}
  \end{center}
\caption{Contour plots of the constraints on the inner slope parameter
$\alpha$ and the halo concentration $c_0$ expected from a survey 
with area $\Omega_{\rm s}=200$ degree$^2$ and $n_g=40$ arcmin$^{-2}$.  
The left panel shows the result if we use only the 2PCF measurement,
while the right panel shows the result from combined measurements of the 2PCF
and 3PCF. The cross symbol denotes input fiducial model of
$(\alpha,c_0)=(1.0,9.0)$.  To perform the fitting, we have used 
measurements on scales $[0.\!\!'2, 2']$ and
combined $250$ triangle configurations for the 3PCF (see text for
details).  The three shaded regions represent $\Delta\chi^2=2.30 (68.3\%)$,
$6.17(95.4\%)$ and $11.8 (99.7\%)$, respectively.  The 3PCF measurement
can tighten the constraints, reflecting the fact that it
carries lensing information complementary to the 2PCF. 
} \label{fig:chi}
\end{figure}
%%%%%%%%%%%%%%%%%%%%%%%%%%%%%%%%%%%%%%%%%%%%%%%%%%%%%%%%%%%%%%%%%%%%%
%%%%%%%%%%%%%%%%%%%%%%%%%%%%%%%%%%%%%%%%%%%%%%%%%%%%%%%%%%%%%%%%%%%
\begin{figure}
  \begin{center}
    \leavevmode\epsfxsize=16.cm \epsfbox{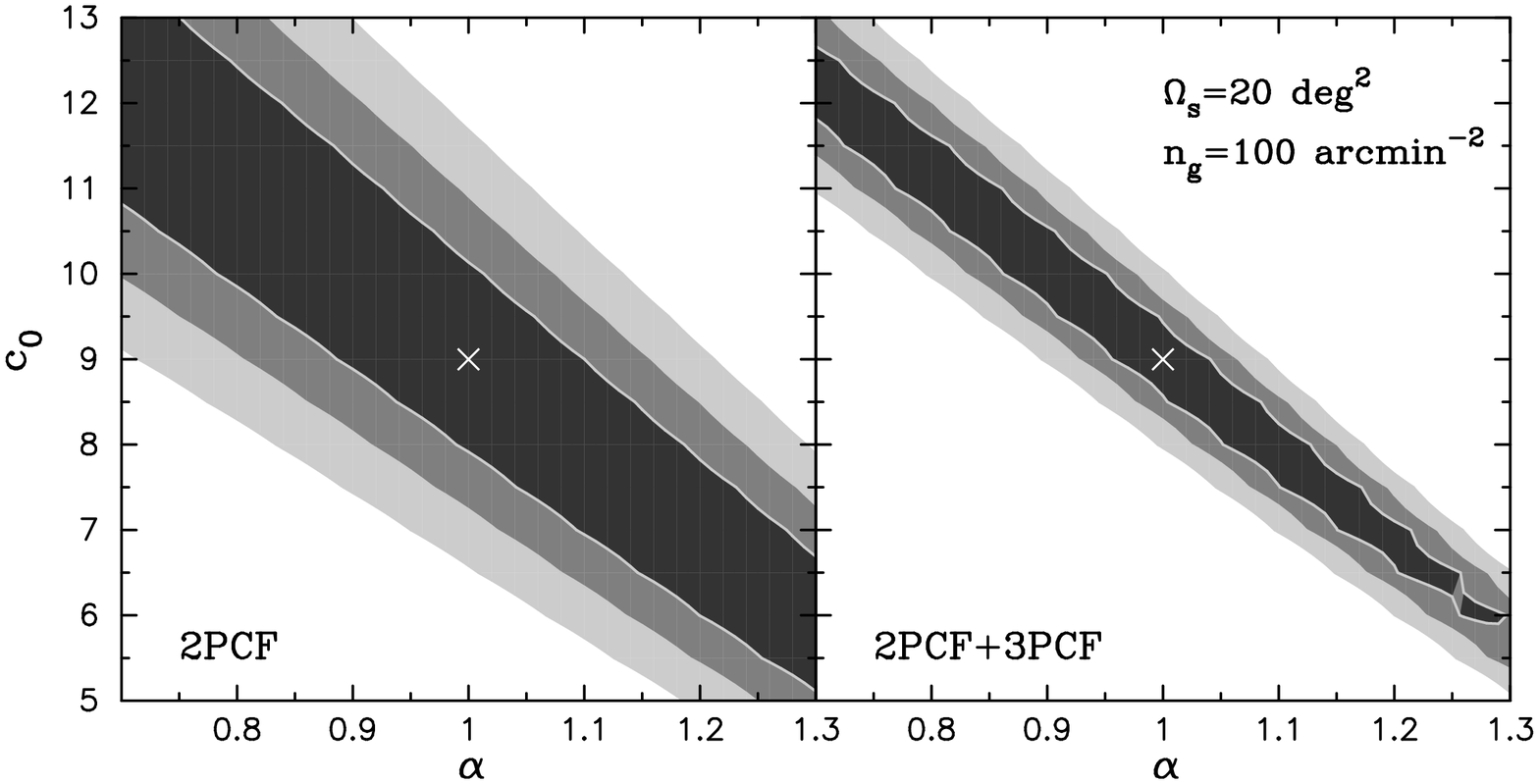}
  \end{center}
\caption{As in the previous figure, but for a survey with
$\Omega_{\rm s}=20$ degree$^2$ and $n_g=100$
 arcmin$^{-2}$. The results show the value of increased number density
of galaxies that can be achieved with a deeper survey. For such a survey 
the 3PCF provides a much larger improvement in the constraints.
} \label{fig:ching100}
\end{figure}
%%%%%%%%%%%%%%%%%%%%%%%%%%%%%%%%%%%%%%%%%%%%%%%%%%%%%%%%%%%%%%%%%%%%%
Figure \ref{fig:chi} shows contour plots of the constraints on the inner
slope parameter $\alpha$ and the halo concentration $c_0$. We consider
$\Omega_{\rm s}=200$ degree$^2$ and $n_g=40$ arcmin$^{-2}$ for the
survey area and the number density of source galaxies, respectively. The
left panel shows the constraints if we use the measurement of the 2PCF
only, while the right panel shows the constraints from combined
measurements of the 2PCF and the 3PCF. The cross symbol denotes our
fiducial model of $(\alpha,c_0)=(1,9)$, which is the NFW profile with
the concentration given by Bullock et al. (2001).  For a 
Gaussian probability distribution function for the 2PCF and 3PCF,  
the three shaded regions correspond to $\Delta\chi^2=2.30$, $6.17$ and
$11.8$ corresponding to $68.3\%$, $95.4\%$ and $99.73\%$ confidence
levels (C.L.), respectively.  One can see that the
constraints on $\alpha$ and $c_0$ are degenerate: the effect of increasing
(decreasing) $\alpha$ on the 2PCF and 3PCF is compensated by decreasing
(increasing) $c_0$.  Nevertheless, the right panel
shows that the 3PCF measurement can tighten the constraints, implying
that the 3PCF provides additional information on these parameters,
although the parameter degeneracy is not broken. As a
result, this type of survey can be used to put stringent constraints on
$\alpha$ and $c_0$ within $5\%$ level ($95\%$ C.L.), if one of the
parameters is fixed, and systematic uncertainties in the measurements
and the theoretical model do not dominate. 

Figure \ref{fig:ching100} shows the result expected from a different
type of survey from the one in the previous figure: a deep, 
small area survey, with $(\Omega_{\rm s},n_g)=(20,
100)$. For this case, the 3PCF can substantially tighten the 
constraints provided from the 2PCF.  
However, the degeneracy between $\alpha$ and $c_0$ remains.
Recently, Takada \& Hamana (2003) proposed that a joint measurement of
the magnification statistics and the cosmic shear could be used to break
the degeneracy or at least put upper bounds on $\alpha$ and $c_0$,
since the amplitude of the magnification correlation is more sensitive
to an increase of $\alpha$ and $c_0$ than the cosmic shear correlation. 
This arises from the non-linear relation between the magnification 
and shear, given by $\mu=|(1-\kappa)^2-\gamma^2|^{-1}$. 
%Specific combinations of the 2PCF and the 3PCF, such as the skewness
%parameter $Q_\kappa$, may do better at breaking degeneracies, 
%analogous to the expectation that $Q_\kappa$ can break the degeneracies in
%$\Omega_{\rm m0}$ and $\sigma_8$ in the large-scale regime ($\simgt 1'$)
%(Bernardeau
%et al. 1997; Jain \& Seljak 1997; TJ02).

\section{Discussion}
\label{disc}

In this paper, we have developed the halo model for computing the
higher order correlations of the cosmic shear field, extending the
real-space dark matter halo approach developed in TJ03b. 
A detailed investigation of the three-point correlation
functions (3PCF) of the convergence and shear fields with respect to the
size and configuration dependence of triangles has been presented.  
Our method provides an accurate, analytical way of computing
the 3PCFs with little computational expense.  We have
focused on the eight shear 3PCFs defined from combination of
the $+/\times$ projections of the shear fields at each vertex of a
given triangle (SL03, ZS03 and TJ03a).  The shear 3PCFs 
defined in this way have characteristic properties that help
disentangle the lensing $E$-mode from the the $B$-mode due to possible
systematic errors and other non-linear effects (see \S \ref{parity}).
They are a direct probe of the gravitational clustering of the mass
distribution and can provide an independent 
test of the CDM paradigm of structure formation.

We have carefully checked the accuracy of our model by comparing the
predictions with ray-tracing simulation results. We paid particular
attention to the triangle configuration dependences of the 3PCFs, 
since our halo model uses simple spherically symmetric profiles. 
We find excellent agreement over the angular scales and models 
we have considered, as shown in 
Figures \ref{fig:2ptlcdm}-{\ref{fig:s3pttcdm}}. 
The halo model reproduces the complex configuration
dependences for the eight shear 3PCFs, as well as their amplitudes. 
The agreement is found for plausible model ingredients:
mass function (Sheth \& Tormen 1999) and
NFW halo profiles with recent prescriptions for the  halo 
concentration (Bullock et al. 2001).
 We chose the best parameter values identified
in the literature and did not adjust any parameters. 

On scales $\simgt 1'$ the 3PCF 
can be used to break degeneracies in cosmological parameters, in 
particular in the $\sigma_8$-$\Omega_{\rm m0}$
determinations so far made from the 2PCF measurement. 
Figures \ref{fig:c3ptlcdmeq}-\ref{fig:s3pttcdm} show a clear
dependence of the lensing 3PCFs on the cosmological models. In addition, 
on large scales the 3PCF  can constrain primordial 
non-Gaussianity, which can be separated from the gravitationally induced
signal using its dependence on scale (R. Scoccimarro, private communication). 
In practice, measuring the shear 3PCFs is more feasible than the
convergence 3PCF, since obtaining the convergence requires
a non-local reconstruction from the observed shear field. 
Detections of shear three-point moments have recently been reported
 (Bernardeau et al. 2002a; Pen et al. 2003). Thus, current survey 
data are likely to already allow for shear 3PCF measurements (see
Figure \ref{fig:sncov} and Figure 4 in TJ03a 
for theoretical justification).  
Our method provides the only well-tested analytical approach to
interpret the measured signals 
in terms of cosmological parameters. 

A second application we have proposed is to use 
shear statistics on sub-arcminute scales to constrain
halo profile properties.   Forthcoming
lensing surveys  promise to measure the
sub-arcminute signals with high significance (see Figure
\ref{fig:sncov}; also see Figures 17 and 18 in Van Waerbeke et al. 2001b
for a measurement of the shear 2PCFs).  
This will open a new window in the use of shear correlation functions, 
beyond the determination of cosmological parameters. The $n$-point
correlation functions on sub-arcminute scales arise mainly from
correlations between the lensing fields around a single halo of
$M>10^{13}M_\odot$ (see Figure \ref{fig:3ptMmax}). The halo model allows
us to interpret the measured signals in terms of the halo
profile properties. The inner slope of the generalized NFW profile and
the halo concentration are sensitive to the amplitudes of the lensing
2PCF and 3PCF on these scales (see Figure \ref{fig:3ptalpha}). 

We have demonstrated how combined measurements of the 2PCF and 3PCF can put
stringent constraints on halo profile properties from forthcoming
lensing data. This was done by taking into account the
covariance for the 2PCF and 3PCF measurements, which contains 
contributions from the shot noise due to the intrinsic ellipticities and
sample variance.  For example, Figure \ref{fig:chi} shows that a
survey with parameters similar to those of the CFHT Legacy survey can constrain
the inner slope parameter with $5\%$ accuracy ($95\%$ C.L.) if the
halo concentration is fixed. Figure \ref{fig:ching100} shows the
dramatic improvements possible with a deeper survey; on the basis
of these figures, it follows that a deep survey with area 
$\simgt 200$ square degrees and source number density $\sim 100$ per
square arcminutes can achieve an accuracy of $1\%$ in density profile
parameters. The use of the 3PCF is critical in being able to achieve
this accuracy, and the use of the four-point function would also be 
valuable if it could be measured. 
In this paper we have ignored
the effect of uncertainties in the mass
function and its possible 
degeneracy with the inner slope and concentration.
White (2002) suggested adjustments to the parameters in 
the Sheth-Tormen mass function from fitting to 
simulations. We confirmed that this modification alters the
halo model predictions for lensing correlations
over angular scales ($\simlt 5'$)
where the 1-halo term is relevant. Therefore, measuring the lensing
2PCF and 3PCF can be similarly used to constrain  
the shape of halo mass function over a mass range of
$10^{13}-10^{15}M_\odot$. 
This would be complementary to other methods such as cluster counts. 
Breaking the degeneracy in the halo model parameters
will require either using the different 3PCFs
of the shear, which we have not done here, or input from other 
methods such as the convergence reconstruction approach discussed below, 
or a statistical study of strong lensing, $X$-ray and SZ effect
on cluster/group scales.
These are important issues to be  addressed.  

A fundamental result from CDM simulations is that 
the density profiles of halos are universal
across a wide range of mass scales (e.g., NFW). 
Therefore, applying our method to different length scales would
sample halo profiles on different mass scales and 
offer a powerful test of the CDM paradigm. The halo model formalism
can also be extended to include the effects of substructure, 
currently a subject of study due to a possible conflict between
CDM theory and observations (e.g. Sheth \& Jain 2002; 
Dalal \& Kochanek 2002). Substructure
and triaxiality of halos, discussed further below, 
would increase the amplitude of correlation functions on the 
smallest scales. Since the two-, three-, and four-point functions scale
differently with these effects, a careful study is merited to develop
the correlation function approach as a probe of different small scale effects. 
Using lensing correlations on scales of $0.1-2$
arcminutes, halos with masses of $10^{13}-M^{15}M_\odot$ will be probed.  

The method of constraining halo profile properties from 
 shear correlations is complementary to the approaches of
Reblinsky \& Bartelmann (1999), Dahle, Hannested \& Sommer-Larsen (2003),
White et al. (2002),
Miyazaki et al. (2002) and Padmanabhan et al. (2003).  They rely
on the reconstruction of the convergence field from observed
ellipticities (see Dahle et al. 2003 and
Miyazaki et al. 2002 for
the  implementation to actual data).  Halo profiles are constrained
by looking at convergence profiles
around individual peaks in the convergence map.  However, 
there are some limitations to this
method. The first is lensing projection effects -- we cannot
accurately measure properties of the primary halo due
to superposition with a void region or less massive halos along the
same line of sight, even if the redshift of the lens halo is available
 (White et al. 2002; Padmanabhan et al. 2003).  The
other limitation is the angular resolution of the convergence
reconstruction. In practice, reconstructing the convergence field on a
given patch of the sky requires an averaging of the observed
ellipticities over an adequate number of source galaxies to reduce the
noise contamination as well as enhance the contrast of the signals
arising from halos.  For plausible survey parameters
the reconstruction resolution is larger than an arcminute. 
The reconstructed convergence around a halo 
is thus smoothed, which makes it difficult to see the inner region
in the halo profile. Padmanabhan et al. (2003) concluded that the inner
slope of NFW profiles cannot be constrained using convergence maps.  

The resolution limitation of convergence maps can be offset by stacking 
clusters and by follow-up deeper weak lensing observations and other
methods such as the SZ effect, $X$-ray observations and strong
lensing. It is not an easy task however, and is subject to biases
associated with identifying centers and mass scales with peaks. 
The strength of
the correlation function method we have discussed is that it
allows for constraints
on the inner halo profile statistically, even though individual halos 
are not resolved on these scales. Our approach treats the data 
objectively, while taking a parameterized approach to the modeling. 
This appears a sensible approach at a time when cosmological parameters
are well constrained and we have a broad understanding of halo properties. 

The real-space halo model formulation, developed in TJ03b and this
paper, does not rely on a model of the 3D non-linear power spectrum. 
This fact leads to interesting applications for lensing statistics. We can
directly compute any $n$-point correlation functions of lensing
observables, such as the reduced shear field $g=\gamma/(1-\kappa)$ and
the lensing magnification $\mu=1/[(1-\kappa)^2+\gamma^2]$ 
(Takada \& Hamana 2003). 
This can be
done merely by replacing $\gamma_M$ in equation (\ref{eqn:shear2pt}) 
with $g_M$ or $\mu_M$ for a given halo of mass $M$, respectively.  The 
resulting prediction is exact in the sense that it fully accounts for the
non-linear contribution of lensing.  So far, the cosmological 
interpretation of lensing two-point statistics has been made 
using theoretical predictions computed from a model of the 3D power 
spectrum, which requires a
perturbative approach, such as setting $\bm{g}\approx \bm{\gamma}$ or
$\mu\approx 1+2\kappa$ (e.g., Van Waerbeke et al. 2001b; McKay et al. 2001; 
Guzik \& Seljak 2002; Benitez \&
Martinez-Gonzalez 1997; Gazta\~naga 2003).  

There are some uncertainties we have ignored in the halo model
formulation.  We have employed a spherically symmetric halo profile, 
while CDM simulations show that halos are triaxial (Jing \& Suto 2002).  
In addition, high-resolution simulations have also shown that about
$10\%$ of mass distribution in a halo is in small sub-clumps
(e.g, Ghigna et al. 2000). 
We have shown that the halo model computed from the NFW profile
matches the simulation results for all the lensing statistics we
have considered. However, it is unclear whether this agreement holds
on sub-arcminute scales. The 3PCF can be the lowest order statistical
quantity to probe the detailed mass distribution of halos via its 
dependence on triangle configurations.  Substructure and triaxiality
are expected to have different scale and configuration dependences
than changes to the halo parameters that we have considered. It is
an interesting problem for future work to work out in detail the different
effects that emerge on scales of order $\sim 0.1$ arcminute. 
For the applications described above, it is important to test the 
analytical predictions with ray-tracing simulation with higher
resolution. This is a pressing need and a challenge for future 
numerical work. The PTHalos method developed by Scoccimarro \& Sheth (2002) 
could be a powerful alternate  tool for such a study.

\bigskip

We would like to thank T. Hamana for kindly providing his ray-tracing
simulation data and for discussions.  We also thank R. Scoccimarro,
R. Sheth, M. Jarvis  and G. Bernstein for valuable comments and
discussions and 
thank P. Schneider for valuable discussions on the parity
transformation properties of the shear three-point functions. 
This work is supported by NASA grants NAG5-10923, NAG5-10924 
and a Keck foundation grant.
\appendix

\section{Validity of the real-space halo approach for shear statistics}
\label{valid}

%%%%%%%%%%%%%%%%%%%%%%%%%%%%%%%%%%%%%%%%%%%%%%%%%%%%%%%%%%%%%%%%%%%%%%
\begin{figure}
  \begin{center}
    \leavevmode\epsfxsize=10.cm \epsfbox{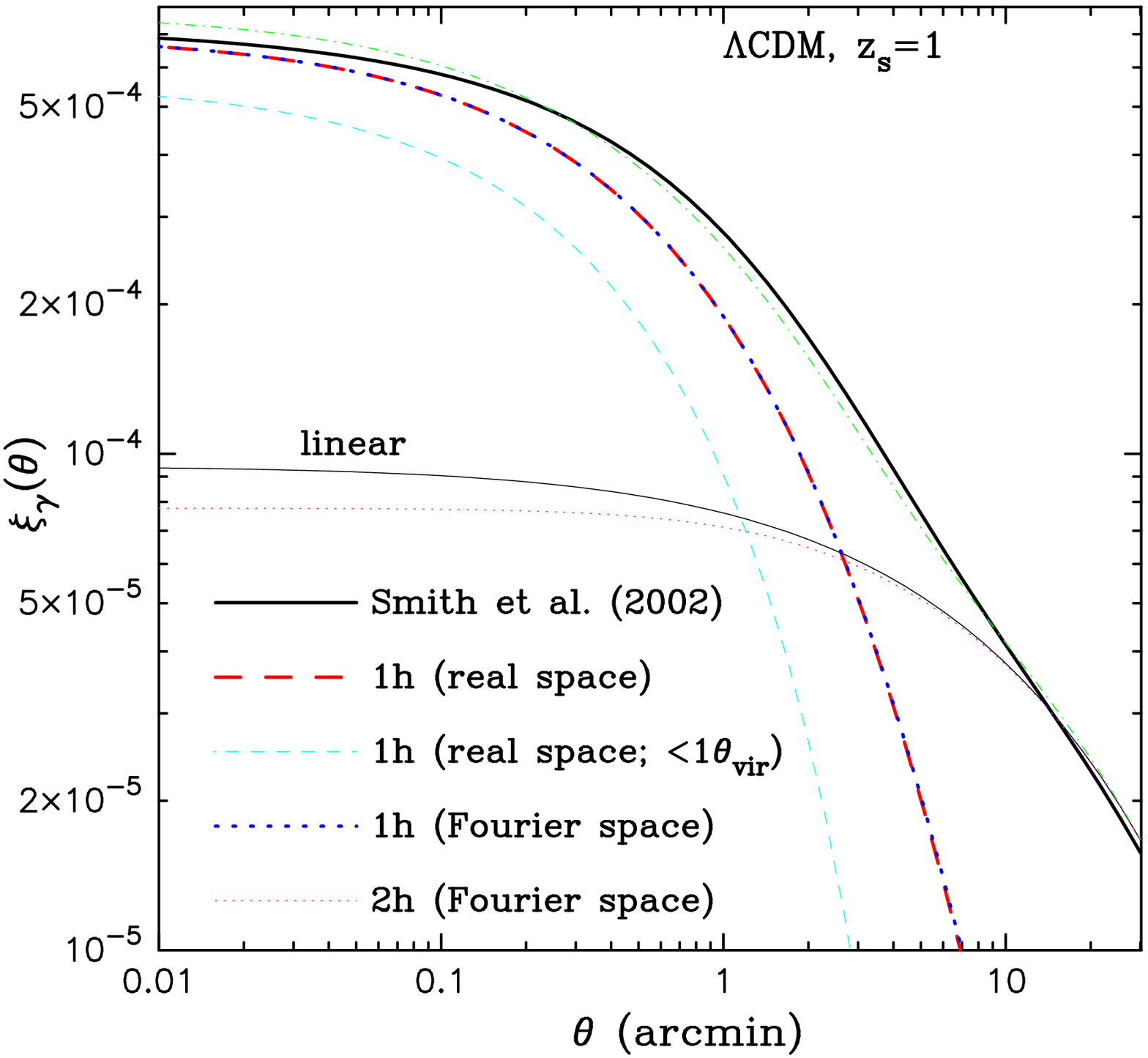}
  \end{center}
\caption{ The two-point correlation function of the shear field,
$\xi_{\gamma}(\theta)=\skaco{\bm{\gamma}\cdot\bm{\gamma}^\ast}$, vs.
separation angle $\theta$. The thick dashed curve shows the
real-space halo model prediction for the 1-halo term, computed using
equation (\ref{eqn:shear2pt}), while the thick dotted curve shows the
Fourier-space halo model prediction. 
The sum with the 2-halo term, denoted by the thin dotted curve, leads to the halo model
prediction for the total power (1h+2h terms) shown by the 
dot-dashed curve.  For comparison, the thick and thin solid curves show
the predictions computed from the Smith02 fitting formula for the
non-linear mass power spectrum and the linear power spectrum,
respectively. The thin dashed curve is the real-space halo model
prediction when the integration range is confined to the virial radius. 
} \label{fig:shear2pt}
\end{figure}
%%%%%%%%%%%%%%%%%%%%%%%%%%%%%%%%%%%%%%%%%%%%%%%%%%%%%%%%%%%%%%%%%%%%%%
In this appendix, we  address the validity of the real-space halo
model developed in \S \ref{halomodel}.

Figure \ref{fig:shear2pt} plots the shear 2PCF,
$\xi_{\gamma}(\theta)=\skaco{\bm{\gamma}\cdot\bm{\gamma}^\ast}$, against
separation angle $\theta$ for the $\Lambda$CDM model and source
redshift $z_s=1$.  The thick dashed and dotted curves
are the real-space and Fourier-space halo model predictions for the
1-halo term, which are computed using equations (\ref{eqn:shear2pt}) and
(\ref{eqn:fshear2pt}), respectively. Note that we have used the
NFW profile.  There is excellent agreement over the
angular scales we have considered, verifying that the real-space model
formulated in \S \ref{model} is equivalent to the Fourier-space model.
This agreement is quite encouraging, since the real-space halo model for
the shear field contains an infinite integration range
to account for the non-local property of the shear fields. To
more explicitly demonstrate the importance of the integration range, the
thin dashed curve is the real-space halo model prediction when the
integration range $\int\!\!  d^2\bm{s}$ is confined to the virial
region, as is done for the convergence field (see equations
(\ref{eqn:conv2pt})). It significantly underestimates the 1-halo term.  
Figure 3 in TJ03b also demonstrates that the
real-space halo model for the convergence field is equivalent to the
Fourier-space halo model.  These results lead us to conclude
that the real-space halo model formulation for the shear and convergence
fields has been made in a self-consistent way, in agreement 
with the Fourier-space
model well studied in the literature (e.g., Cooray \& Sheth 2002).

The thin dotted curve denotes the 2-halo term, which provides the dominant
contribution to the 2PCF for $\theta\simgt 4'$.  For comparison,
the thin solid curve is the result for $\xi_\gamma$ computed from the
linear mass power spectrum.  It very slightly overestimates the 2-halo
term which includes the biasing of halos. The 2-halo term thus captures the 
clustering properties in the linear regime.

The accuracy of the halo model for predicting the lensing 2PCFs can be
seen by comparing the dot-dashed and thick solid curves (the
detailed comparison with ray-tracing simulation results will be
presented below).  The dot-dashed curve shows the total halo
model prediction (1- plus 2-halo terms), while the thick solid curve is
the result computed from the fitting formula proposed in Smith et
al. (2002) (hereafter Smith02).  The Smith02 formula was calibrated
to reproduce the non-linear mass power spectra from high
resolution $N$-body simulations for various cosmological models.  As can
be seen, the halo model prediction matches the Smith02 result within
$10\%$ at $0.\!\!'1\le \theta\le 30'$. 
The reliable range of the Smith02 formula, is $k \simlt 10~
h\mbox{Mpc}^{-1}$ in the mass power spectrum.  Since the lensing field
is the projected field of the 3D mass distribution, the valid range
roughly corresponds to the angular scale
$\theta\simgt 2\pi/[k_{\rm max}d_A(z=0.4)]\approx 0'\!  \!  .6$, where
$z=0.4$ is close to the peak redshift of the lensing projection function
for the source redshift $z_s=1$ (e.g., see Figure 4 in TJ02). Therefore,
it is unclear whether or not the discrepancy between the halo model and
the Smith02 result at $\theta \simlt 0.\!\! ' 1$ is genuine.

It is worth mentioning why we use the fiducial model
$(c_0,\beta)=(9,0.13)$ for the halo concentration in this paper,
since we used a steeper mass slope $\beta=0.2$ in TJ02 and
TJ03b.  The Smith02 formula predicts a steeper $k$ slope of the
non-linear power spectrum $P(k)$ than predicted from the fitting formula
of Peacock \& Dodds (1996) (hereafter PD), which has been widely used in
the literature. 
\footnote{The PD formula is based in part on 
the physically motivated {\em stable clustering hypothesis},
which allows one to analytically predict the behavior of strongly
non-linear clustering. 
The Smith02 results display a weak violation of the stable clustering
hypothesis. This violation can occur for the halo model as well
(Ma \& Fry 2000b,c; also see Scoccimarro et
al. 2001; TJ03b).}  For the halo model, the $k$-slope is
determined by the halo profile parameters,
the slope of the mass function, and the input linear power spectrum
(Seljak 2000; Ma \& Fry 2000a,b,c; TJ03b).  The halo model with
$\beta=0.13$ gives a slightly better fit to the $k$-slope of the Smith02
power spectrum than with $\beta=0.2$.  
It is reassuring that we find agreement with N-body simulations 
with plausible ingredients for the halo model, each of which is supported
by independent studies: the halo profile (NFW), the halo
concentration (Bullock et al. 2001) and the halo mass function (Sheth \&
Tormen 1999).  These studies explored halo properties
by resimulating regions containing halos in a larger scale
simulation with a higher resolution simulation.  Hence, one advantage
of the halo model is that it can be easily refined by incorporating 
results from different $N$-body simulations with different sizes. 

%%%%%%%%%%%%%%%%%%%%%%%%%%%%%%%%%%%%%%%%%%%%%%%%%%%%%%%%%%%%%%%%%%%%%%
\begin{figure}
  \begin{center}
    \leavevmode\epsfxsize=16.cm \epsfbox{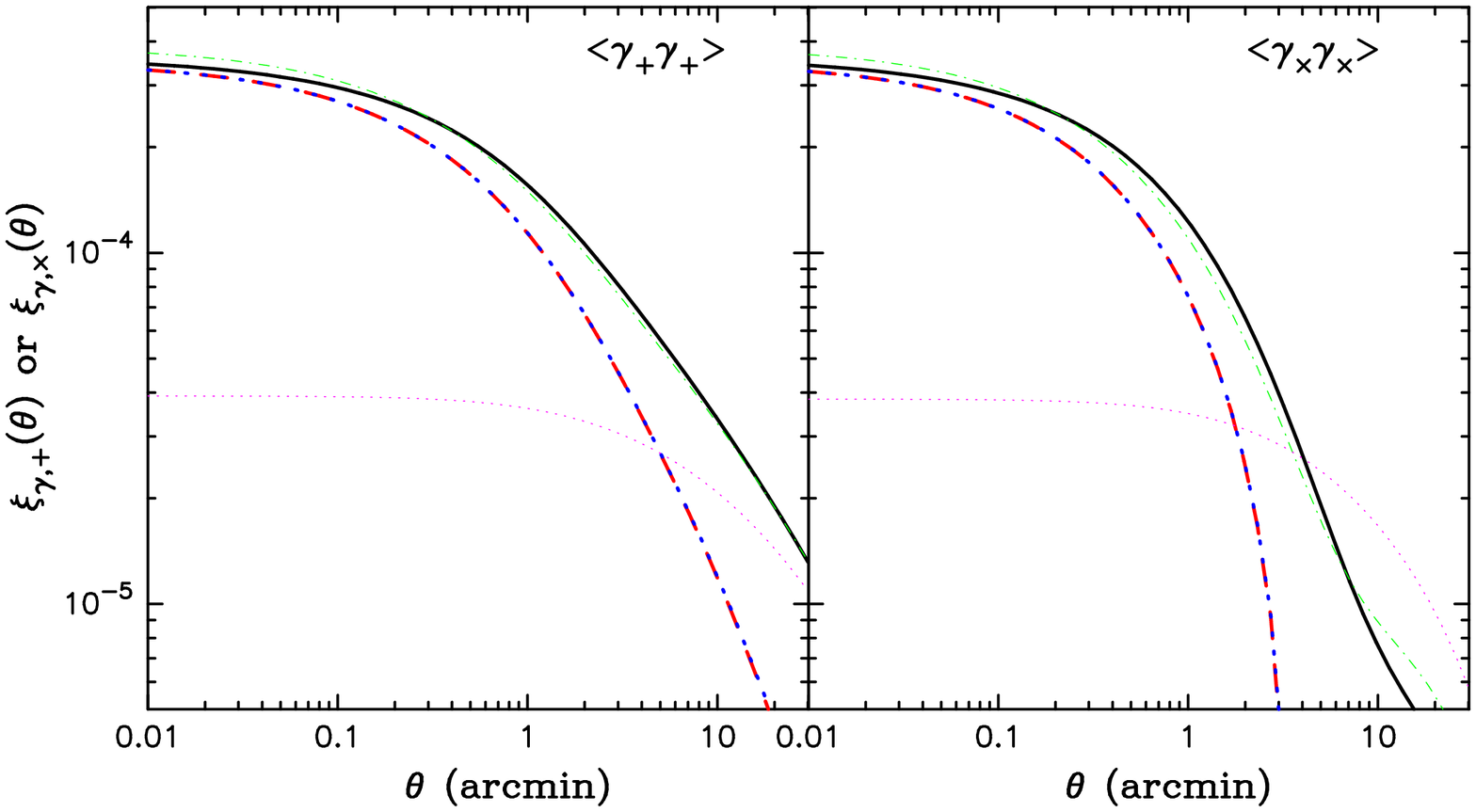}
  \end{center}
\caption{The two-point correlation functions of the shear fields,
 $\xi_{\gamma,+}$ (left panel) and $\xi_{\gamma,\times}$ (right panel),
 as in Figure \ref{fig:shear2pt}. 
 Although the 2-halo term for $\xi_{\gamma,\times}$ 
 overestimates the amplitude at large scales $\theta\simgt 5'$,
 the halo model reproduces the Smith02 result because the
 1-halo term is negative.}  \label{fig:tshear2pt}
\end{figure}
%%%%%%%%%%%%%%%%%%%%%%%%%%%%%%%%%%%%%%%%%%%%%%%%%%%%%%%%%%%%%%%%%%%%%%
Figure \ref{fig:tshear2pt} compares the
theoretical predictions for $\xi_{\gamma,+}(\theta)$ (left panel) and
$\xi_\times(\theta)$ (right panel), as done in the previous figure. 
The real-space halo model result
for the 1-halo term (thick dashed curve) again 
agrees with the Fourier-space result (thick dotted curve) for
both $\xi_{\gamma,+}$ and $\xi_{\gamma,\times}$.  
The agreement is as a result of the integration of the
phase factors $\epsilon_\mu\epsilon_\nu$ in equation
(\ref{eqn:deftshear2pt}) for the real-space case. 
Comparison of the dot-dashed curve with the solid curve in each panel
shows that the halo model prediction for the total 2PCF 
matches the Smith02 result.
Note that, although the 2-halo term (or the linear theory prediction) 
for $\xi_{\gamma,\times}$ overestimates the amplitude at large
scales $\theta\simgt 5'$, the agreement of the halo model with the
Smith02 result at these scales results from sum of the negative 1-halo term
plus the 2-halo term.

%%%%%%%%%%%%%%%%%%%%%%%%%%%%%%%%%%%%%%%%%%%%%%%%%%%%%%%%%%%%%%%%%%%%%%
\begin{figure}
  \begin{center}
    \leavevmode\epsfxsize=9.cm \epsfbox{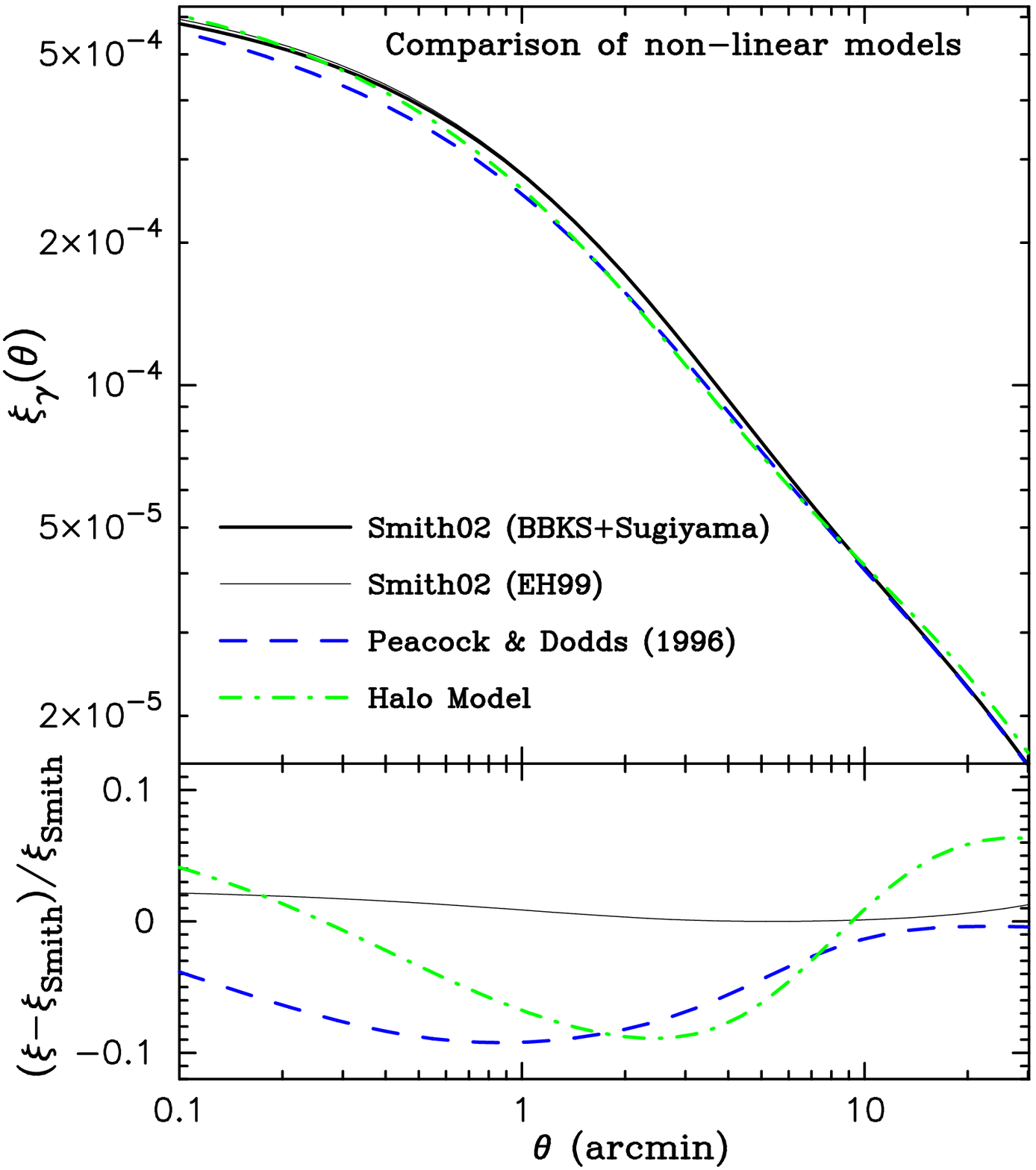}
  \end{center}
\caption{Comparison of model predictions for
$\xi_\gamma$.  The solid, dashed and dot-dashed curves are the results
from the Smith02 formula, the PD formula and the halo model,
respectively.  The lower panel explicitly shows the deviation of each
model prediction from the Smith02 model.  Although all the results agree
with each other within $10\%$, the Smith02 result 
for $\xi_{\gamma}$ is higher than the PD result by  $0-10\%$ at
$0'\!\!.1\simlt \theta\simlt 10'$ (see text for details).  Note that the
Smith02 result is computed using the BBKS
transfer function. To
demonstrate the effect of the input linear power spectrum, the thin
solid curve is the Smith02 result when we use the fitting formula of
Eisenstein \& Hu (1999), which is more accurate. 
} \label{fig:smithpd}
\end{figure}
%%%%%%%%%%%%%%%%%%%%%%%%%%%%%%%%%%%%%%%%%%%%%%%%%%%%%%%%%%%%%%%%%%%%%%
Figure \ref{fig:smithpd} compares theoretical predictions for
$\xi_\gamma$ from various models of non-linear gravitational
clustering. The thick solid, dashed and dot-dashed curves are the
results from the Smith02 formula, the PD formula and the halo model,
respectively.  The lower panel shows the deviation relative to the
Smith02 result (thick solid curve) and shows agreement among these
models within 10$\%$ over the scales we have considered. The 
Smith02 prediction is higher than PD
 by $0-10\%$ at $1\le \theta \le 10'$. This discrepancy also
exists for the aperture mass variance (Schneider et al. 1998), which has
been used to disentangle the $E/B$-modes from the actual measurement
(e.g, Van Waerbeke et al. 2001b; Jarvis et al. 2003).  Most of the
cosmic shear analysis to constrain cosmological models has been made by
comparing the measured signals with the theoretical predictions computed
from the PD formula.  Hence, the discrepancy shown implies that 
$\sigma_8$ might be overestimated by up to $\sim 5\%$, if
the constraint is obtained from 
$\theta<10'$.  On the other hand, the halo model slightly overestimates
the Smith02 and PD results on $\theta>10'$, where the 2-halo term yields
the dominant contribution.  This is
because the standard implementation of the halo model imposes the
condition that the 2-halo term reproduce the shear 2PCF computed from
the {\em linear} power spectrum on large scales -- this cannot reproduce
the suppression seen in the realistic non-linear power spectrum over
the transition scales between the non-linear and linear
regimes.  Finally, the sensitivity of the Smith02 result to the input 
linear power spectrum is shown by the thin solid curve, which 
uses the transfer function proposed in Eisenstein \&
Hu (1999), which is more accurate than the BBKS plus Sugiyama model.  The
comparison shows a  difference less than $3\%$.  This sensitivity 
also holds for the halo model prediction.  Hence, we can safely use the 
BBKS plus Sugiyama model to compute the halo model predictions. 

%%%%%%%%%%%%%%%%%%%%%%%%%%%%%%%%%%%%%%%%%%%%%%%%%%%%%%%%%%%%%%%%%%%%%%
\begin{figure}
  \begin{center}
    \leavevmode\epsfxsize=9.cm \epsfbox{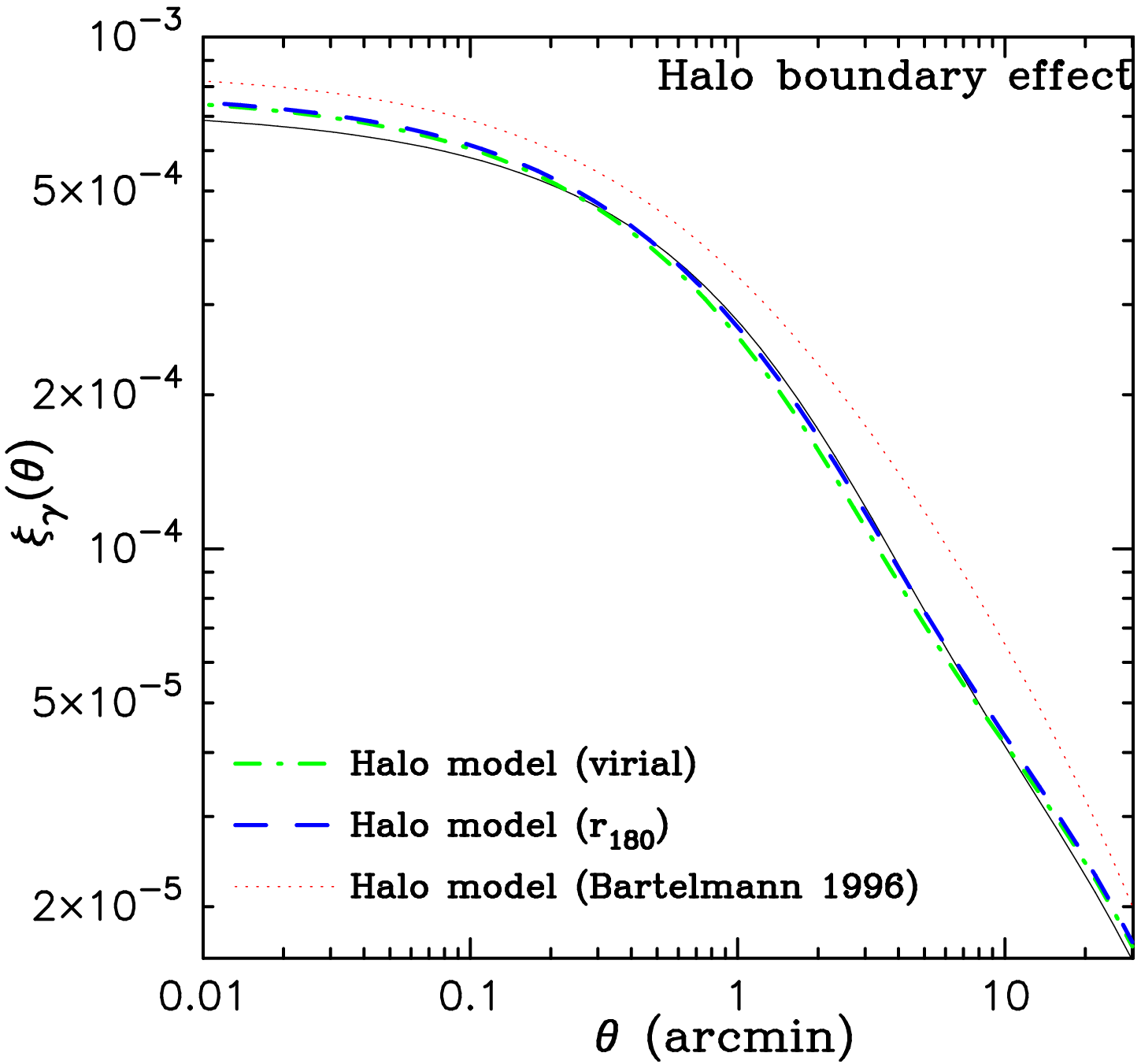}
  \end{center}
\caption{The figure shows how the halo boundary condition employed in
 the halo model calculation affects the prediction for $\xi_{\gamma}$.
 The thick dot-dashed and dashed curves are the results when we employ
 the NFW profile truncated at the virial radius and at the radius
 $r_{180}$, respectively.  The radius $r_{180}$ is defined so that mean
 density enclosed by a sphere with radius $r_{180}$  is $180$
 times the background density.  
 The upper dotted curve shows the halo model prediction
 if one uses the expression in Bartelmann (1996) for the shear profile
 to compute the 1-halo term.  For reference, the thin solid curve shows
 the Smith02 result in Figure \ref{fig:shear2pt}.}  \label{fig:2ptbound}
\end{figure}
%%%%%%%%%%%%%%%%%%%%%%%%%%%%%%%%%%%%%%%%%%%%%%%%%%%%%%%%%%%%%%%%%%%%%%
Figure \ref{fig:2ptbound} explores how a modification of the halo
boundary condition used in the halo model calculation affects the model
prediction.  As stated below equation (\ref{eqn:massfun}), the
Sheth-Tormen mass function (\ref{eqn:massfun}) tends to better fit the
mass function measured from $N$-body simulations, if one employs the
halo mass estimator, $M_{180}$, enclosed within a region of the
overdensity $\Delta_{180}(=180)$, than the virial mass estimator (e.g.,
White 2002).  Therefore, one possible modification of the halo model is
to employ the halo profile truncated at the radius $r_{180}$
\footnote{Note that, as long as the halo mass function is then considered
a function of $M_{180}$, mass conservation and the
normalization of the mass function are not violated}. If we assume that
the mass distribution in a halo follows an NFW profile up to $r_{180}$, 
we can obtain the relation between $M$ and
$M_{180}$, which allows us to re-express all the relevant quantities in
terms of the new mass $M_{180}$, as demonstrated in Hu \& Kravtsov (2003).
In addition, for consistency with the simulation results in White
(2002), we employ the parameters of $a=0.67$ and $p=0.3$ for the
Sheth-Tormen mass function (\ref{eqn:massfun}) in the halo model
calculation (see Table 2 in White 2002).
%%%
The thick dashed curve plots the halo model prediction for
$\xi_{\gamma}$ for the halo boundary $r_{180}$, while the thick
dot-dashed curve is the result for the virial boundary.  As can be seen, 
the two results are close, although the halo model of
$r_{180}$ better matches the Smith02 result denoted by the thin solid
curve over a range of the scale $0.\!\!'5\simlt \theta\simlt 5'$.  
This is mainly due to the enhancement of the 1-halo term
in the halo model with $r_{180}$, since
$r_{180}>r_{\rm vir}$ ($\Delta_v=334>180$) for the \LCDM model, hence
the 1-halo term covers a larger range than the virial region.  
The dotted curve shows the halo model result when one employs the
expression for the shear profile $\gamma_M$ in Bartelmann (1996) to
compute the 1-halo term. The expression is derived by the
line-of-sight projection of the NFW profile, allowing it to extend to 
infinite radius. 
It substantially overestimates the amplitude for $\xi_\gamma$ over
the scales we have considered.  Hence, if one intends to account
for the mass contribution outside the virial region, it is
necessary to first modify the mass defined within the new
halo boundary and then to modify the halo model ingredients 
in terms of the new halo mass. This could improve the halo
model accuracy over the transition scales between the non-linear
and linear regimes, since the mass distribution outside the virial
region is relevant for the quasi-linear regime ($\delta\sim
1$; also see the discussions around Figure 8 in TJ03b).  We
have confirmed that the boundary condition $r_{180}$ slightly
improves the agreement between the halo model prediction for the lensing
3PCFs and the simulation results. However, in this paper, we implement
the virial boundary for simplicity.

To summarize the results shown in Figures
\ref{fig:shear2pt}-\ref{fig:tshear2pt}, we have shown that our
real-space halo model formulation for cosmic shear statistics is
self-consistent with the Fourier-space halo model.
A great advantage of the real-space halo model is that it 
enables us to analytically compute any $n$-point correlation functions
for both the convergence and shear fields on small angular scales, without
additional computational effort compared to the two-point
function.

\section{Convergence field for a generalized NFW profile}
\label{conv}

In this appendix, we present the convergence field around a generalized NFW
profile with $\alpha\ne 1$ given in equation (\ref{eqn:nfw}). 

For general $\alpha$ the convergence field cannot be analytically
computed. However,  if $\alpha=0$ and $1$, we obtain  analytical
expressions to evaluate $\kappa_M$ in 
equation (\ref{eqn:convm}) in terms of 
$\Sigma_M$, which is given by
\begin{eqnarray}
\Sigma_M(\theta)=\frac{Mc^2f_0}{2\pi \rvir^2}F(c\theta/\theta_{\rm vir}), 
\end{eqnarray}
with
\begin{eqnarray}
F(x)=
\left\{
\begin{array}{ll}
\displaystyle
\frac{\sqrt{c^2-x^2}(2+c+x^2+2cx^2)}{2(1+c)^2(1-x^2)^2}
-
\frac{3}{2}\frac{x^2}{(1-x^2)^{5/2}}{\rm arccosh}\frac{x^2+c}{x(1+c)},
& (x<1)\\
\displaystyle 
\frac{\sqrt{c^{ 2}-1}}{5(1+c)^3}\left\{1+c(c+3)\right\},
& (x=1)\\
\displaystyle
\frac{\sqrt{c^2-x^2}(2+c+x^2+2cx^2)}{2(1+c)^2(1-x^2)^2}
-
\frac{3}{2}\frac{x^2}{(x^2-1)^{5/2}}{\rm arccos}\frac{x^2+c}{x(1+c)},
& (1<x\le c),
\end{array} 
\right.
\end{eqnarray}
for $\alpha=0$ and 
\begin{eqnarray}
\Sigma_M(\theta)=\frac{Mc^2f_2}{2\pi \rvir^2}F(c\theta/\theta_{\rm vir}), 
\end{eqnarray}
with
\begin{eqnarray}
F(x)=
\left\{
\begin{array}{ll}
\displaystyle
\frac{1}{x}{\rm arctan}\frac{\sqrt{c^2-x^2}}{x}
-
\frac{1}{\sqrt{1-x^2}}{\rm arccosh}\frac{x^2+c}{x(1+c)},
& (x<1)\\
\displaystyle -\sqrt{\frac{c-1}{c+1}}+{\rm arctan}\sqrt{c^2-1},
& (x=1)\\
\displaystyle
\frac{1}{x}{\rm arctan}\frac{\sqrt{c^2-x^2}}{x}
-
\frac{1}{\sqrt{x^2-1}}{\rm arccos}\frac{x^2+c}{x(1+c)},
& (1<x\le c),
\end{array} 
\right.
\end{eqnarray}
for $\alpha=2$, respectively. The factors $f_0$ and $f_2$ in the above
equations 
are $f_0^{-1}=-c(2+3c)/(2(1+c)^2)+\ln(1+c)$ and
$f_2^{-1}=\ln(1+c)$. We again note that the convergence fields above
are defined from the generalized NFW profile truncated at the virial
radius, leading to $\Sigma_M(\theta)=0$ for $\theta>\theta_{\rm vir}$.
These expressions are used to address the sensitivity of the lensing
statistics to the inner slope parameter $\alpha$. 
Similarly, we can derive analytical expressions for the shear
profiles for $\alpha=0$ and $2$, as in equation (\ref{eqn:gammam}).

%%%%%%%%%%%%%%%%%%%%%%

%\twocolumn

\label{lastpage}
\end{document}